\documentclass[a4paper,12pt]{article}
\usepackage{amsfonts,amsthm,amsmath,amssymb,graphicx,hyperref,cite}
\usepackage{color}

\begin{document}

\newtheorem{definition}{Definition}[section]
\newcommand{\be}{\begin{equation}}
\newcommand{\ee}{\end{equation}}
\newcommand{\bea}{\begin{eqnarray}}
\newcommand{\eea}{\end{eqnarray}}
\newcommand{\LE}{\left[}
\newcommand{\R}{\right]}
\newcommand{\nn}{\nonumber}
\newcommand{\Tr}{\text{Tr}}
\newcommand{\N}{\mathcal{N}}
\newcommand{\G}{\Gamma}
\newcommand{\vf}{\varphi}
\newcommand{\LL}{\mathcal{L}}
\newcommand{\HH}{\mathcal{H}}
\newcommand{\arctanh}{\text{arctanh}}
\newcommand{\up}{\uparrow}
\newcommand{\down}{\downarrow}
\newcommand{\ket}[1]{\left| #1 \right>}
\newcommand{\bra}[1]{\left< #1 \right|}
\newcommand{\ketbra}[1]{\left|#1\right>\left<#1\right|}
\newcommand{\rd}{\partial}
\newcommand{\de}{\partial}
\newcommand{\ba}{\begin{eqnarray}}
\newcommand{\ea}{\end{eqnarray}}
\newcommand{\db}{\bar{\partial}}
\newcommand{\we}{\wedge}
\newcommand{\ca}{\mathcal}
\newcommand{\lr}{\leftrightarrow}
\newcommand{\f}{\frac}
\newcommand{\s}{\sqrt}
\newcommand{\vp}{\varphi}
\newcommand{\hvp}{\hat{\varphi}}
\newcommand{\tvp}{\tilde{\varphi}}
\newcommand{\tp}{\tilde{\phi}}
\newcommand{\ti}{\tilde}
\newcommand{\ap}{\alpha}
\newcommand{\pr}{\propto}
\newcommand{\mb}{\mathbf}
\newcommand{\ddd}{\cdot\cdot\cdot}
\newcommand{\no}{\nonumber \\}
\newcommand{\la}{\langle}
\newcommand{\lb}{\rangle}
\newcommand{\ep}{\epsilon}
 \def\we{\wedge}
 \def\lr{\leftrightarrow}
 \def\f {\frac}
 \def\ti{\tilde}
 \def\ap{\alpha}
 \def\pr{\propto}
 \def\mb{\mathbf}
 \def\ddd{\cdot\cdot\cdot}
 \def\no{\nonumber \\}
 \def\la{\langle}
 \def\lb{\rangle}
 \def\ep{\epsilon}

\numberwithin{equation}{section}

\begin{titlepage}
\thispagestyle{empty}

\begin{flushright}
YITP-21-25\\
IPMU21-0022\\
\end{flushright}

\vspace{.4cm}
\begin{center}
\noindent{\Large \textbf{Holographic Path-Integral Optimization}}\\
\vspace{2cm}

Jan Boruch$^{a}$, Pawel Caputa$^{a}$, Dongsheng Ge$^{a}$ 
and Tadashi Takayanagi$^{b,c,d}$
\vspace{1cm}

{\it
 $^{a}$Faculty of Physics, University of Warsaw, ul. Pasteura 5, 02-093 Warsaw, Poland.\\
$^b$Yukawa Institute for Theoretical Physics, Kyoto University, \\
Kitashirakawa Oiwakecho, Sakyo-ku, Kyoto 606-8502, Japan\\
$^{c}$Inamori Research Institute for Science, \\ 620 Suiginya-cho, Shimogyo-ku, Kyoto 600-8411 Japan\\
$^{d}$Kavli Institute for the Physics and Mathematics of the Universe (Kavli IPMU),\\
University of Tokyo, Kashiwa, Chiba 277-8582, Japan\\
}

\vskip 2em
\end{center}

\vspace{.5cm}
\begin{abstract}

In this work we elaborate on holographic description of the path-integral optimization in conformal field theories (CFT) using Hartle-Hawking wave functions in Anti-de Sitter spacetimes. We argue that the maximization of the Hartle-Hawking wave function is equivalent to  the path-integral optimization procedure in CFT. In particular, we show that metrics that  maximize gravity wave functions computed in particular holographic geometries, precisely match those derived in the path-integral optimization procedure for their dual CFT states.  The present work is a detailed version of  \cite{Boruch:2020wax} and contains many new results such as analysis of excited states in various dimensions including JT gravity, and a new way of estimating holographic path-integral complexity from Hartle-Hawking wave functions. Finally, we generalize the analysis to Lorentzian Anti-de Sitter and de Sitter geometries and use it to shed light on path-integral optimization in Lorentzian CFTs. 
\end{abstract}

\end{titlepage}

\tableofcontents

\section{Introduction}

Holography, or the so-called Anti-de Sitter space/Conformal Field Theory (AdS/CFT) correspondence \cite{Ma}, is one of the most promising approaches to quantum gravity. The main reason is that it relates quantum theories of gravity on Anti-de Sitter spacetimes to a class of CFTs, where there is no gravity present. Therefore, in AdS/CFT, gravitational spacetime is emergent and, as it is manifest in the holographic entanglement entropy \cite{RT,HRT,Ra}, it is encoded in the geometry of quantum entanglement in CFT. Thus we expect that the dynamics of quantum entanglement in CFTs is directly related to that of quantum gravity. 

Surprisingly, some tensor networks provide useful playgrounds where holographic correspondence is realized in discretized lattice models. This direction of research started from the pioneering work \cite{Swingle}, which conjectured the relation between AdS gravity and a particular tensor network for CFT states called MERA (multi-scale entanglement renormalization ansatz) \cite{MERA,TNR,cMERA}. Among others, this tensor network setup beautifully explains the geometric calculation of entanglement entropy in AdS/CFT \cite{RT,HRT}. Refer to e.g. \cite{Qi:2013caa,HAPPY,HQ,Beny,NRT,MT,Cz,Czech:2015kbp,Milsted:2018yur,Milsted:2018san,Bhattacharyya:2016hbx,Hung:2018mcn,Chen:2021ipv,Jahn:2020ukq,Jahn:2021uqr} for further progress and improvements in this direction. 

Even though the connection between tensor networks and AdS geometry leads to an interesting progress in qualitative understanding of the mechanism behind AdS/CFT, it still remains among toy models for holography (however, refer to a few recent attempts to improve that \cite{Takayanagi:2018pml,VanRaamsdonk:2018zws,Bao:2018pvs,Bao:2019fpq,Caputa:2020fbc}). The main obstacle of tensor network approaches to AdS/CFT is the manifest presence of lattice regularizations, which prohibits us from understanding analytical results in the continuum limit. On the other hand current approaches to continuous tensor networks such as e.g. cMERA (continuous MERA) are limited to free theories (see however \cite{Cotler:2016dha,Cotler:2018ufx,Cotler:2018ehb,Fernandez-Melgarejo:2019sjo,Fernandez-Melgarejo:2020fzw} for recent progress beyond free CFTs). Therefore, in order to draw conclusions about a genuine holography, we need to develop a tensor network-like formalism applicable to continuous and interacting, holographic CFTs.

The path-integral optimization, which was proposed in \cite{Caputa:2017urj} and is reviewed in the next section, is a promising formalism for this purpose. The construction starts with a wave function representation of a CFT state defined by Feynman's path-integral on a flat Euclidean space with prescribed boundary condition for quantum fields.  
Then, utilizing the conformal symmetry, we perform a Weyl re-scaling of the Euclidean metric while keeping the boundary condition fixed, so that the final quantum state prepared by the path-integral remains unchanged. Intuitively, we can interpret path-integrals on different Weyl-rescaled geometries as different continuous (non-unitary) tensor networks preparing the same quantum state (see e.g. \cite{Milsted:2018yur} for relation between tensor networks and geometry of path-integrals). 

The ``lattice structure" of the path-integral network is specified by the metric such that there is a unit lattice site per unit area. This way, we can simplify such continuous tensor networks by a coarse-graining procedure implemented by choosing an appropriate metric (Weyl factor). The maximization of this coarse-graining is called the ``path-integral optimization". Quantitatively, the path-integral optimization corresponds to minimizing the ``path-integral complexity", a functional of the background metric, given by e.g. the Liouville action in two-dimensional CFTs.\footnote{A close connection between the computational complexity and the AdS/CFT already started from \cite{Susskind}.} More generally, ``complexity" gives us a notion of how hard it is to prepare a given quantum state and in this context we can regard it as a size of the path-integral tensor network.  

After this procedure it was found that optimal metrics for CFT path-integrals are indeed hyperbolic ($AdS$) \cite{Caputa:2017urj}. This is a strong evidence that the path-integral optimization provides a successful, continuum version of the AdS/tensor network correspondence. Nevertheless, the emergence of $AdS$ geometries from this approach has been mysterious and precise connection (i.e. the gravity dual) to the AdS/CFT has not been well understood. Moreover, up to date, we have been able to fully analyze only two-dimensional CFT examples while higher-dimensional generalization of this procedure relied only on a conjectured path-integral complexity action.

In this article, we present a more direct relation between the path-integral optimization and the AdS/CFT in the light of the Hartle-Hawking wave function. We will argue that the path-integral optimization corresponds to the maximization of the Hartle-Hawking wave function in gravity on a Euclidean AdS background. Our construction will be based on generalized Hartle-Hawking wave functions that require imposing a specific boundary condition on the asymptotic boundary of AdS, dual to a specific CFT state. This maximization naturally arises in the saddle point approximation when we calculate physical observables using the Hartle-Hawking wave function. From this perspective, we can also understand how to generalize the path-integral optimization to a quantum regime beyond the semi-classical approximation. 

The arguments below will not only explain the origin of the optimization procedure from the viewpoint of the AdS/CFT but also shed new light on several interesting directions in the path-integral optimization program. Firstly, the Hartle-Hawking wave function method works well in higher dimensions and we use it to extract information about the ``correct" path-integral optimization in any dimensions. Secondly, we extend the analysis to Lorentzian signature including both Anti de-Sitter and de-Sitter spacetimes. This is a starting point for exploring holography in general spacetimes using gravitational path-integrals. Finally, in Euclidean as well as Lorentzian setups, we propose a natural definition of holographic counterpart of the path-integral complexity in terms of gravity action with the Hayward term. Below, we compute it in all examples and discuss some of its interesting properties.

This paper is organized as follows. In section \ref{Sec:2 Review}, we give a review of path-integral optimization in two as well as in higher dimensions. In section \ref{Sec: 3 AdSP}, we provide a detailed derivation of path-integral optimization geometry from Hartle-Hawking wave function in $AdS_{d+1}$ using Poincare coordinates, define the Euclidean holographic path-integral complexity and discuss a few important points of our proposal. In section \ref{Sec:4 MoreEx} we analyze more examples in Euclidean $AdS$ geometries and in section \ref{Sec:5 JT} we test our proposal in the context of JT gravity. In section \ref{Sec:6 Lor} we study Lorentzian $AdS_{d+1}$ and $dS_{d+1}$ geometries where we derive Hartle-Hawking wave functions and use them as a guidance for Lorentzian path-integral optimization in CFT. 
In section \ref{Sec:Tension}, we discuss foliations of $AdS$ and $dS$ spacetimes by the slices that maximize Hartle-Hawking wave functions and interpret their tension as an emergent time. Finally, in section \ref{Sec:7 Conclusions} we conclude and leave several technical details into appendices.

\section{Review of the Path-Integral Optimization}\label{Sec:2 Review}
We start with a brief review of the path-integral optimization \cite{Caputa:2017urj} that provides a framework for constructing continuous tensor networks for CFT wave functions. We will discuss only some of its important properties and new insights that will be particularly relevant for the holographic interpretation in later sections (see also \cite{MTW,Bhattacharyya:2018wym,Caputa:2020mgb,Takayanagi:2018pml} for more details of the formalism). We will separately discuss CFTs in two and in higher dimensions. 

\subsection{Two-Dimensional CFTs}
The main idea of the path-integral optimization is as follows. We start with a two-dimensional CFTs on the Euclidean plane $\mathbb{R}^2$ with coordinates $(\tau,x)$ and denote all the fields in the CFT by $\Phi(\tau,x)$. The ground state wave function $\Psi_{CFT}[\Phi(x)]$ of the CFT, that is formally a functional of the boundary conditions imposed for the CFT fields at the time slice $\tau=0$: $\Phi(x)\equiv \Phi(x,\tau=0)$, is defined by the Euclidean path-integral as
\ba
\Psi_{CFT}[\Phi(x)] =\int \prod_{-\infty<\tau\leq 0,x}[D\ti{\Phi}(\tau,x)] e^{-S_{CFT}[\ti{\Phi}]}\delta(\ti{\Phi}(0,x)-\Phi(x)),
\ea
where $S_{CFT}$ is the Euclidean action of the 2d CFT under consideration.\\
This exact definition of the wave functions is usually only a formal expression and can be rarely used to derive the state beyond free CFTs. On the other hand, this path-integral representation contains all the information about the state, including $AdS$ dual geometry for holographic CFTs, structure of entanglement, as well as data required to estimate state's complexity. The main problem is to extract this information directly from the path-integral. The path-integral optimization is then an attempt to solve this problem and, in particular, to extract information about holographic dual geometry and complexity from the path-integral representation of CFT states.

The key step in the path-integral optimization is to deform the metric on the two-dimensional Euclidean plane on which we perform the path-integral in the following way
\ba
ds^2=e^{2\phi(\tau,x)}(d\tau^2+dx^2).  \label{curm}
\ea
Then, the original path-integral computing  $\Psi_{CFT}[\Phi(x)]$ is evaluated on a flat  $\mathbb{R}^2$ metric with $e^{2\phi(\tau,x)}=1/\ep^2$, where $\epsilon$ is a UV regularization scale (i.e. lattice constant) introduced once we discretize path-integrals of quantum fields into those on a lattice. On the other hand, general Weyl factors $e^{2\phi(\tau,x)}$ can be interpreted as a position-dependent choices of the lattice discretization. For this interpretation we need to impose an extra rule, namely, that there is a single lattice site per unit area. This way we can relate a given metric to a (non-unitary) tensor network, which is a discretization of Euclidean path-integral, and volume of the above geometry to the number of tensors.

Next, let us write the wave function obtained from the path-integral on the curved space with metric (\ref{curm})
as $\Psi_{CFT}^{\phi}[\Phi(x)]$. If we impose the boundary condition
\be
e^{2\phi(0,x)}=\frac{1}{\ep^2}\equiv e^{2\phi_0},  \label{bcon}
\ee
then this wave function is proportional to the original one $\Psi_{CFT}[\Phi(x)](=\Psi_{CFT}^{\phi_0}[\Phi(x)])$ 
computed by path-integral on flat space $\mathbb{R}^2$. The reason for this is that CFTs are invariant under Weyl rescalings up to the Weyl anomaly, so we  have
\ba
\Psi_{CFT}^{\phi}[\Phi(x)]=e^{S^{(e)}_L[\phi]-S^{(e)}_L[\phi_0]}\cdot \Psi_{CFT}^{\phi_0}[\Phi(x)], \label{pathinf}
\ea
where the functional in the exponent $S^{(e)}_L[\phi]$ is the Liouville action on a Euclidean flat space \cite{Po}
\ba
S^{(e)}_L[\phi]=\frac{c}{24\pi}\int^\infty_{-\infty} \!dx\! \int^{0}_{-\infty}\! d\tau\! \left[(\de_x\phi)^2+(\de_\tau\phi)^2
+\mu e^{2\phi}\right]. ~~ \label{LVac}
\ea
Notice that Liouville action universally depends on the number of degrees of freedom denoted by the central charge $c$ of the 2d CFT that appears as a pre-factor. From the relation (\ref{pathinf}), we see that the quantum state still remains the same CFT vacuum $|0\lb$ for any choice of the metric (\ref{curm}) as long as the boundary condition (\ref{bcon}) at $\tau=0$ is satisfied. Moreover, the assumption about the discretization, i.e. that one unit area of the metric (\ref{curm}) corresponds to a single lattice site, fixes the value of the ``cosmological constant" $\mu$ to $\mu=1$ as in \cite{Caputa:2017urj}. Nevertheless, we will keep this constant parameter for later convenience and will also provide an interpretation for $\mu\neq 1$ in the gravity dual setup. 

One way to derive the Liouville action is to examine the expectation value of the trace of energy stress tensor in CFT, which is given by 
(we write $i=\tau,x$)
\ba
\la T^i_i \lb=-\frac{c}{12\pi}\de_i\de^i \phi+\mu_0 e^{2\phi}. \label{emtrp}
\ea
The first term is proportional to the Ricci scalar curvature and describes the Weyl anomaly that is universal for 2d CFTs. The second term simply comes from the UV divergence (see e.g. App. \ref{LorentzianDet}). Since the expectation value of the trace $T^i_i$ is the derivative of the partition function with respect to the Weyl scaling factor $\phi$, we can reproduce (\ref{LVac}) with setting $\mu=12\pi\mu_0/c$. Note that here, we are interested in the bare action (before renormalization) such as the Liouville function $S^{(e)}_L[\phi]$. Indeed, the Liouville potential term gets quadratically divergent as $e^{2\phi}\sim \ep^{-2}$. Nevertheless, we keep it and, as we will momentarily see, it plays an important role in the path-integral optimization story. Finally, since the Liouville potential term in (\ref{LVac}) arises from the UV divergence,  it should dominate over the kinetic term when we take the UV limit $\ep\to 0$. This condition is satisfied when
\ba
(\de_i \phi)^2 \ll e^{2\phi}\ \  \ (i=x,\tau).  \label{contlim}
\ea
We will come back to this limit in the gravitational construction in later sections.

Next, we describe how the optimal metrics for Euclidean path-integrals are chosen. The basic strategy of path-integral optimization is to coarse-grain the discretization of path-integral as much as possible, such that e.g. the cost of numerical computation of path-integral is minimal, while keeping the final boundary condition that specifies the correct wave function fixed.  For this we need a measure of computational cost of a path-integration i.e. an estimate of its computational complexity. In the original path-integral optimization, the actual Liouville action $S^{(e)}_L[\phi]$ (with $\mu=1$)  was proposed as the complexity functional \cite{Caputa:2017urj}, which should be minimized in the optimization procedure. This was motivated by the minimization of the overall factor of the wave function, which is proportional to $e^{S^{(e)}_L[\phi]}$ as in (\ref{pathinf}). Although, as usually, the overall factor does not affect physical quantities in quantum mechanics, this quantity can be interpreted as a measure of the number of repetitions of numerical integrals when we discretise path-integrals into lattice calculations.\footnote{Remember that our choice of lattice 
regularization is specified by the metric (\ref{curm}).} 

It was shown \cite{Caputa:2017urj} that such optimization (minimization) procedure properly selects the most efficient discretization of path-integral which produces the correct vacuum state. More precisely, the minimization of $S^{(e)}_L[\phi]$ is achieved by solving the Liouville equation 
\be
(\de_x^2+\de^2_\tau)\phi=\mu e^{2\phi},
\ee
which is equivalent to the condition for constant negative (hyperbolic) Ricci scalar of the 2-dimensional metric \eqref{curm}
\be
R=-2\mu.
\ee
Recall that this conclusion is a consequence of the fact that we work with the bare Liouville action with the potential term ($\mu\neq 0$).\\
The most general solution of this equation is given in terms of two functions: $A(w)$ and $B(\bar{w})$, of $w=\tau+ix$ and its complex conjugate $\bar{w}=\tau-ix$ as
\be
e^{2\phi}=\frac{4A'(w)B'(\bar{w})}{\mu(1-A(w)B(\bar{w}))^2}.
\ee 
For example, for the path-integral preparing the vacuum state of the CFT on a line, with the boundary condition (\ref{bcon}), the minimization leads to the solution 
\ba
e^{2\phi(\tau,x)}=\frac{1}{(\sqrt{\mu}\tau-\ep)^2}, \ \ \ (-\infty<\tau\leq 0).\label{SolVP}
\ea
We regard the solution with $\mu=1$ as the genuine optimized metric, which minimized the original Liouville action (remember we set $\mu=1$ for our complexity functional \cite{Caputa:2017urj}). In addition, in this work, we propose to interpret solutions with $\mu<1$, for which the metric gets larger than the original with $\mu=1$, as only partially optimized  where the UV cut-off length scale is taken to be larger. On the other hand, solutions with $\mu>1$, for which the metric is smaller than the original with $\mu=1$, intuitively, correspond to fine-graining the cut-off scale more than the current lattice spacing (i.e. $\ep$). We would like to suggest that these solutions with $\mu>1$ may not be allowed. One of the arguments is that, as $\mu$ gets larger, the coarse-graining gets too fast (i.e. the curvature $R$ is ``too big") to maintain the Weyl invariance which is only guaranteed in the continuum limit and which leads to our basic equivalence relation \eqref{pathinf}.

The observation that  (\ref{SolVP}) coincides with the time slice of a three-dimensional AdS (AdS$_3$)  \cite{Caputa:2017urj}, offered the main implication that the path-integral optimization, as a version of continuous tensor-network, can explain the emergence of the AdS geometry purely from CFT. Indeed, the discretized path-integral can be regarded as a (non-unitary) tensor network and its relation to AdS provides a path-integral version of the conjectured interpretation of AdS/CFT as a tensor network. Moreover, as a product of the above procedure, the Liouville action $S^{(e)}_L[\phi]$ (the ``path-integral complexity" \cite{Caputa:2017urj}) became the first workable definition of computational complexity in continuous, interacting CFTs, that could be computed in genuinely holographic models \cite{Susskind} (refer also to \cite{Czech:2017ryf,Caputa:2018kdj,Camargo:2019isp,Erdmenger:2020sup,Chagnet:2021uvi} for further remarkable connections between path-integral and circuit complexities). This method has been further generalized to various CFTs and QFTs (see e.g. \cite{Bhattacharyya:2018wym,Sato:2019kik,Caputa:2020mgb,Molina-Vilaplana:2018sfn,Jafari:2019qns,Ghodrati:2019bzz,Ahmadain:2020jfm,Yang:2020tna,Chandra:2021kdv}) and also used to successfully compute entanglement of purification in 2d CFTs \cite{Caputa:2018xuf}, which was recently verified numerically in \cite{Camargo:2020yfv}.

Nevertheless, despite this emergence of hyperbolic geometries, a direct relation between the path-integral optimization and AdS/CFT has remained an open problem. Moreover, it has not been clear how to promote the classical Liouville theory which appeared in the path-integral optimization to the quantum regime. The need for that was found in \cite{Caputa:2017urj} where, in order reproduce the correct gravity metric dual to a primary state with $1/c$ corrections, one would have to replace a classical Liouville result with a quantum one. The main problem with ``quantum" path-integral optimization is that integral $\int [D\phi] e^{S^{(e)}_L[\phi]}$, which is expected to arise from (\ref{pathinf}) in the quantized version, does not make sense as it is not bounded from below. Indeed, the standard quantum Liouville is defined by the path-integral  $\int [D\phi] e^{-S^{(e)}_L[\phi]}$, but it has not been clear how such well-defined expression arises from the above optimization. In other words, we cannot get the minimization as a saddle point approximation of path-integrals and its derivation within the AdS/CFT may appear too challenging. As we will see, the new approach \cite{Boruch:2020wax} further developed in this paper will naturally resolve these two problems in the light of Hartle-Hawking wave functions.

Last but not least, we would like to point an important subtlety that, in the solution (\ref{SolVP}), $(\de_i\phi)^2$ and $e^{2\phi}$ are of the same order, so the UV criterion (\ref{contlim}) is not satisfied. This actually implies 
that the path-integral optimization using the Liouville action is only qualitative. Indeed, its derivation (\ref{emtrp}) involves the continuum limit and, generally, we expect finite cut-off corrections to the Liouville action. As we will see later, the gravity approach will naturally come with such corrections.

\subsection{Higher-dimensional CFTs}
In higher dimensions, the status of path-integral optimization is still not entirely clear. At least in even dimensions, it is tempting to follow the intuition from the Liouville action that, as discussed above, can be defined as integrated anomaly. For example in four dimensions, this would lead us to the so-called $Q$-curvature actions (see e.g. \cite{Levy:2018bdc}) that naturally generalises the Polyakov action from 2d. However, such actions, even though very interesting, are genuinely higher order in derivatives (or the curvature) and less intuitive from the tensor-network perspective.
Instead, in \cite{Caputa:2017urj}, we proposed a phenomenological generalisation to an ``effective" path-integral complexity action for higher-dimensional CFTs in the form
\be
I[\phi,\hat{g}]=N\int_{\Sigma}d^dx\sqrt{\hat{g}}\left[e^{(d-2)\phi}\hat{g}^{ab}\partial_a\phi\partial_b\phi+\frac{e^{(d-2)\phi}}{(d-1)(d-2)}R_{\hat{g}}+\mu e^{d\phi}\right],\label{HDimA}
\ee
where $R_{\hat{g}}$ is the Ricci scalar for the reference metric $\hat{g}$ and the normalization
\be
N=\frac{d-1}{16\pi G_N},
\ee
was fixed in order to reproduce entanglement entropy for spherical regions \cite{Caputa:2017urj}.\\
We have argued that path-integral complexity computed by \eqref{HDimA}, as well as improved Liouville complexity\footnote{with the volume term subtracted.} are good relative measures of complexity between two continuous tensor networks described by metrics $g_1$ and $g_2$. Moreover, both of these complexity actions satisfy the so-called ``co-cycle conditions" in terms of three metrics 
\be
I[g_1,g_2]=-I[g_2,g_1],\qquad I[g_1,g_2]+I[g_2,g_3]=I[g_1,g_3].\label{co-cycle}
\ee
Interestingly, from the above properties it is clear that $I[g_1,g_2]$ can be negative (unlike measures of complexity based on e.g. distance in the Hilbert space). \\
One more comment we should make at this point is that in \eqref{HDimA} we generalized the action from \cite{Caputa:2017urj} by including coupling $\mu$ in front of the potential. To get a better intuition regarding this action and the meaning of $\mu$, let us now rewrite it in a more transparent way. Namely, consider a general $d$-dimensional metric of the form
 \be
 ds^2=e^{2\phi}\hat{g}_{ab}dx^adx^b,
 \ee 
and its Ricci scalar curvature given by
\be
R^{(d)}=e^{-2\phi}(R_{\hat{g}}-2(d-1)\hat{\Box}\phi-(d-2)(d-1)\hat{g}^{ab}\partial_a\phi\partial_b\phi).
\ee
We can check that, after integrating $\hat{\Box}\phi$ by parts, \eqref{HDimA} can be actually rewritten as
\be
I[\phi,\hat{g}]=\frac{1}{16\pi G^{(d)}_N (d-2)}\int d^dx\sqrt{g}\left[R^{(d)}-2\Lambda^{(d)}\mu\right],
\ee
where
\be
\Lambda^{(d)}=-\frac{(d-1)(d-2)}{2}.
\ee
This is obviously the Einstein-Hilbert action with negative cosmological constant $\Lambda^{(d)}\mu$ in $d$-dimensions, i.e. the same number of dimensions as the CFT. Comparing with the standard gravity action, $\mu$ plays  the role of the $AdS$ radius $1/l^2$ that sits in the denominator of the cosmological constant. Let us stress that this resemblance is only at the formal level and our effective complexity action, by generalizing the Liouville action, was aimed to describe the geometry of the continuous tensor networks from path-integrals. Derivation of this action from the optimization procedure in CFTs is still an open problem.
Nevertheless, the holographic approach \cite{Boruch:2020wax}, in principle, provides the full answer to the correct path-integral complexity functional (including finite cut-off corrections) in higher dimensions and will reproduce and justify \eqref{HDimA} in the UV limit.

Setting aside the CFT derivation for the moment, this form of the action has obvious consequences for optimal path-integral geometries. Clearly, by varying the action with respect to $\phi$ we end up with the trace of Einstein's equations in $d$-dimensions
\be
\left(R^{(d)}_{ab}-\frac{1}{2}R^{(d)}g_{ab}+\Lambda^{(d)}\mu g_{ab}\right)g^{ab}\delta\phi=0,
\ee
which is of course equivalent to the constant negative Ricci curvature condition
\be
R^{(d)}=-d(d-1)\mu=2\Lambda^{(d+1)}\mu.\label{HDOE}
\ee
This provides a generalization of the optimization constraint for the path-integral geometries (Liouville equation in 2d) to higher dimensions.\\
It is also interesting to see that the negative cosmological constant that appears on the right hand side of this equation is $d+1$-dimensional. As we will see later, this fact naturally comes from the Hamiltonian constraint in the gravity proposal. Finally, the parameter $\mu$ that fixes the value of the curvature will also find a holographic counterpart in terms of tension $T$ in the Hartle-Hawking construction. After this review, we can proceed with the gravity dual description.

\section{Optimization from Hartle-Hawking Wave Functions}\label{Sec: 3 AdSP}
In this section we begin by explaining the construction reported in a short article \cite{Boruch:2020wax}, then we solve the vacuum example in detail and discuss various subtleties and properties of this new proposal. Our goal is to understand the path-integral optimization procedure, reviewed above, in holographic CFTs that are equivalently described by gravity in the one-dimension higher Anti-de Sitter spacetime \cite{Ma}. This correspondence, or the  AdS/CFT for short, usually involves super-conformal field theories and string theory or supergravity in AdS, however, in this work, we will limit the discussion to large-$N$ conformal field theories and the pure gravity sector.
\subsection{Definition of Hartle-Hawking Wave Functions}
\begin{figure}[t!]
  \centering
  \includegraphics[width=10cm]{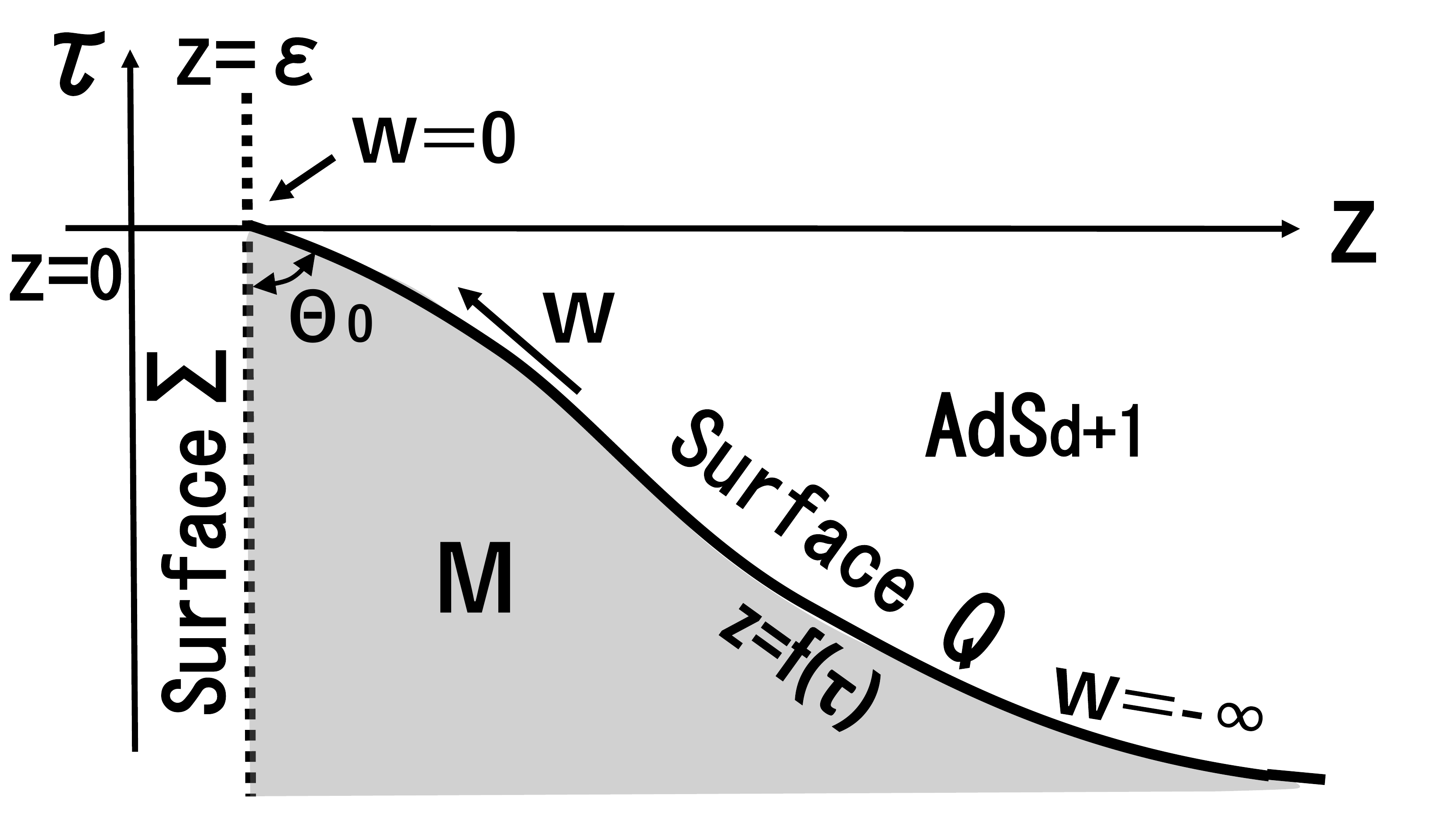}
  \caption{The on-shell action in the shaded region $M$
computes the Hartle-Hawking wave function, related to the path-integral optimization for holographic CFTs.}
\label{gravityfig}
\end{figure}

In what follows, we proceed with the analysis of the vacuum of a CFT on $\mathbb{R}^{d}$ that, on the gravity side, corresponds to Euclidean Poincare $AdS_{d+1}$ geometry\footnote{In Euclidean examples we will set the $AdS$ radius to $l=1$.}
\ba
ds^2=\frac{dz^2+d\tau^2+\sum_{i=1}^{d-1}  dx_i^2}{z^2}.  \label{metx}
\ea
In this geometry, we focus on the shaded region $M$ in Fig.\ref{gravityfig} between surfaces $\Sigma$ and $Q$ in the bulk of $AdS$. In this region, we will be interested in a gravitational ``Hartle-Hawking" wave function \cite{Hartle:1983ai}\footnote{See e.g. \cite{Heemskerk:2010hk,Harlow:2011ke} for previous important applications of gravity wave functions in the AdS/CFT correspondence.}, denoted by $\Psi_{HH}[g_{ab}]$, which is a functional of the induced metric $g_{ab}$ on $Q$. More precisely, we consider a path-integral of Euclidean gravity from the asymptotic boundary $\Sigma$ of AdS, described by the cut-off surface $z=\ep$ with $\tau<0$, up to the surface $Q$, specified by $z=f(\tau)$ that extends from $z=\ep$ and $\tau=0$ (where it meets $\Sigma$ at an angle $\theta_0$) towards the bulk as depicted in Fig.\ref{gravityfig}. 
\\
Note that, for simplicity, we focus on translationally-invariant surfaces $Q$. Moreover, we assume the following form of the metric on $Q$ 
\ba
ds^2=e^{2\phi(w)}(dw^2+\sum_{i=1}^{d-1}dx_i^2),  \label{diago}
\ea
where $-\infty<w<0$ (that is a function of $\tau$) renders the induced metric on $Q$ conformally flat.\footnote{Note that the above construction assumes  that a gravity solution which calculates the Hartle-Hawking wave function, is given by a sub-region in the Poincare AdS. The metric (\ref{diago}) obtained in this way covers all possible metrics on $Q$ for $d=2$ with the condition (\ref{bcon}) because all solutions to the vacuum Einstein equation are locally equivalent to the Poincare AdS$_3$. However, for $d>2$, the above construction covers only a sub-class of the metrics on the $d$-dimensional surface $Q$. Nevertheless, as we will see below, this ansatz includes a large class of metrics of our direct interest. More generally, we can extend target metrics to completely general ones by directly solving the Einstein's equations.}  In this setup, we now define the Hartle-Hawking wave function $\Psi_{HH}[\phi]$ as follows
\ba
\Psi_{HH}[\phi]=\int [Dg_{\mu\nu}] e^{-I_G[g]}\delta(g_{ab}|_Q-e^{2\phi}\delta_{ab}), \label{HwH}
\ea
where $I_G[g]$ is the $d+1$-dimensional Einstein-Hilbert action with negative cosmological constant $\Lambda$ and the Gibbons-Hawking boundary term
\ba
I_G\!=\!-\frac{1}{2\kappa^2}\!\int_{M}\! \s{g}(R\!-\!2\Lambda)\!-\!\frac{1}{\kappa^2}\!\int_{Q\cup \Sigma}\!\s{h} K,~
 \label{actionM}
\ea
where we also denoted $\kappa^2=8\pi G_N$.\footnote{One of the defining features of the Hartle-Hawking wave functions in canonical quantum gravity is the fact that they solve the Wheeler-de Witt equation, i.e. the quantum Hamiltonian constraint. In this work we will be mostly concerned with the semi-classical regime and will not worry about subtleties of the quantum definition of \eqref{HwH} such as e.g. sum over topologies etc.} More generally, as we will see below, one should also carefully add a Hayward term in regions with non-smooth boundaries like our $M$ above. \\
Cautious reader will notice that we also (implicitly) imposed an initial condition (i.e. fixed the metric) on $\Sigma$ that corresponds to the pure AdS dual to the CFT vacuum. In principle, we can consider more general boundary conditions on $\Sigma$ as well as $\Psi_{HH}[\phi(w,x)]$ corresponding to excited states of a CFT (see following sections). 

Given this definition, we are now ready to make the connection with the CFT path-integral optimization.
Namely, the key step of the proposal \cite{Boruch:2020wax} is to identify the metric (\ref{diago}) on $Q$ as the ``holographic dual" of the metric (\ref{curm}) in the path-integral optimization for holographic CFTs. Moreover, we propose that the gravity dual of the optimization procedure corresponds to the maximization of  the Hartle-Hawking wave function $\Psi_{HH}[\phi]$ with respect to $\phi$ (more generally, with respect to the metric on $Q$). This maximization arises naturally when we evaluate e.g. generic gravity correlation functions using the Hartle-Hawking wave functional
\ba
\la O_1O_2\ldots\lb =\int [D\phi] |\Psi_{HH}[\phi]|^2 O_1O_2\ldots,  \label{corhh}
\ea
by applying the saddle point approximation.\footnote{In fact, in AdS/CFT, this computation could also naturally appear in the correlator of local bulk operators where operators $O_i$ would be related to the CFT ones by the standard HKLL prescription.} The main advantage of this dual description is that the maximization of Hartle-Hawking wave function works well even in the presence of quantum fluctuations of $\phi$, as opposed to the minimization of $e^{S^{(e)}_L[\phi]}$ as discussed before. Indeed, we will compute explicitly in the next section that $\Psi_{HH}[\phi]$ is proportional to $e^{-S^{(e)}_L[\phi]}$ up to finite cut-off corrections.
 
As we already mentioned, in this work, we will be able to discuss a generalization of the path-integral optimization based on the Liouville action (and its higher dimensional version \eqref{HDimA}) with a non-trivial coefficient $\mu$ of the exponential potential. For this new construction, on the gravity side, it will be useful to introduce a tension term on the surface $Q$, which also plays an important role in the AdS/BCFT \cite{Ta} (we assume $T<0$ below)
\ba
I_T[h]=\frac{T}{\kappa^2}\int_Q \s{h}. \label{tensiont}
\ea
This allows us to define a one-parameter family of ``deformed" Hartle-Hawking wave functions in the following way
\ba
\Psi^{(T)}_{HH}[\phi]\!=\!\int\![Dg_{\mu\nu}\!] e^{-I_G[g]-I_T[e^{2\phi}]}\,\delta(g_{ab}|_Q\!-\!e^{2\phi}\delta_{ab}).~~\label{HHWFT}
\ea
The standard Hartle-Hawking wave functions are obtained by setting $T=0$. We can regard $T$ as a 
chemical potential or the Legendre transformation for the area of $Q$, acting on Hartle-Hawking wave functions.
As we will see, the tension term in gravity will be responsible for shifting the value of the cosmological constant $\mu$ in the Liouville action.\footnote{Later, we will set $T=0$ while evaluating the holographic path-integral complexity. Adding $T$  should then be thought of as just a trick for selecting partially optimized configurations as solutions to the problem of maximization of the Hartle-Hawking wave functions.} More importantly, since the maximum of $\Psi^{(T)}_{HH}[\phi]$ will correspond to a family of surfaces $Q$ in AdS parameterized by the tension $T$, we will observe that $T$ plays the role of an emergent holographic time. 

\subsection{Evaluation of $\Psi^{(T)}_{HH}[\phi]$ }
Computation of the wave function \eqref{HHWFT} in full quantum gravity is clearly beyond the present technology but we can nevertheless make progress using saddle point analysis. In the remaining discussions we will focus on evaluating $\Psi^{(T)}_{HH}[\phi]$ using the semi-classical approximation i.e. as an on-shell gravity action with tension term added
\ba
\Psi^{(T)}_{HH}[\phi]\simeq e^{-I_G[g]-I_T[h]}|_{\mbox{on-shell}},  \label{HHwv}
\ea  
in region $M$. In this limit, the evaluation of the Hartle-Hawking wave function boils down to a standard computation from the AdS/BCFT, of a classical gravity action in a portion of $AdS_{d+1}$ up to an end-of-the-world brane $Q$ with tension $T$. However, the important technical assumption in this computation will be the fact that the (fictitious) ``brane" that describes surface $Q$ does not back-react on the bulk geometry. This allows us to evaluate the wave function using the Poincare AdS metric \eqref{metx} in a tractable way. This assumption is sufficient to find solutions whose metrics take the conformally flat form (\ref{diago}). For more general metrics we may need to solve the back-reactions in the presence of the brane, which is also a well-defined problem.

Let us now compute the gravity action step by step. First, since we are using the on-shell solution of Einstein's equations \eqref{metx}, we have $R-2\Lambda=-2d$ and the bulk action contributes
\bea
-\frac{1}{2\kappa^2}\int_{M} d^{d+1}x\sqrt{g}\left(R-2\Lambda\right)&=&\frac{d}{\kappa^2}\int d^{d-1}x\int d\tau\int^{f(\tau)}_\epsilon\frac{dz}{z^{d+1}},\nn\\
&=&\frac{V_x}{\kappa^2}\left(\frac{L_\tau}{\epsilon^{d}}-\int\frac{d\tau}{f(\tau)^{d}}\right),
\eea
where we denoted the divergent volumes and the length of $\tau$ direction as
\be
V_x\equiv \int d^{d-1}x,\qquad L_\tau\equiv \int^0_{-\infty} d\tau.
\ee
Clearly, the bulk action not only gives rise to a divergent part from the UV limit but also contributes a part with $f(\tau)$ to the effective action that will describe metric on $Q$. Moreover, notice that the bulk action vanishes when we take $Q\to\Sigma$ i.e. $f(\tau)=\epsilon$.\\
Next, from the UV boundary surface $\Sigma: z=\epsilon$ with normal $n^\mu=-z\delta^{\mu z}$, we have $K|_\Sigma= d,$ and the contribution from the Gibbons-Hawking term
\be
-\frac{1}{\kappa^2}\int_{\Sigma} d^{d}x\sqrt{h}K|_\Sigma=-\frac{d}{\kappa^2}\frac{V_x L_\tau}{\epsilon^{d}}.
\ee
Finally, we carefully evaluate the contribution from $Q$, specified by function $z=f(\tau)$. Firstly, the induced metric on $Q$ is given by
\ba
ds^2=\frac{(1+f'^2)d\tau^2+dx^2_i}{f^2}\equiv e^{2\phi(w)}\left(dw^2+dx^2_i\right),\label{METQ}
\ea
where in the second step we defined
\be
e^{2\phi}\equiv\frac{1}{f^2},
\ee
and, to make the metric diagonal, we introduced ``conformal" coordinate $w$ with the range $-\infty<w<0$, satisfying
\be
\frac{dw}{d\tau}=\sqrt{1+f'^2}=\left(1-\dot{f}^2\right)^{-1/2}.\label{wprim}
\ee
We also introduced a short-hand notation $f'=f'(\tau)=\de_\tau f(\tau)$, $\dot{f}=\dot{f}(w)=\de_w f(w)$ and the relation $f'(\tau)=w'(\tau)\partial_wf(w)=w'(\tau)\dot{f}$ was used to solve for the second relation. In order to rewrite formulas in terms of $w$ coordinate, we will later use iteratively $f''(\tau)=w'(\tau)\partial_w f'(\tau)$ and from the definition
\be
\dot{f}=-e^{-\phi(w)}\dot{\phi}(w),\qquad \ddot{f}=-e^{-\phi(w)}\left(\ddot{\phi}(w)-\dot{\phi}(w)^2\right).\label{fphi}
\ee
Secondly, we can compute extrinsic curvature and its trace on $Q$. With coordinates on $Q$: $y^i=\{\tau,x^a\}$, and projectors
\be
e^\alpha_i\equiv \frac{\partial x^\alpha}{\partial y^i},\qquad e^\alpha_\tau=\{1,\vec{0},f'(\tau)\},\qquad e^\alpha_{x^a}=\delta^{\alpha}_a,
\ee 
as well as the outward-pointing normal vector 
\be
n^{\mu}=\frac{z}{\sqrt{1+f'(\tau)^2}}\left\{-f'(\tau),\vec{0},1\right\},
\ee
we can compute the components of the extrinsic curvature
\be
K_{ij}=e^{\alpha}_i e^\beta_j \nabla_\alpha n_\beta.
\ee
Expressed in terms of $f(\tau)$ they read 
\bea
K_{\tau\tau}&=&-\frac{1+f'(\tau)^2+f(\tau)f''(\tau)}{f(\tau)^2\sqrt{1+f'(\tau)^2}},\qquad K_{\tau a}=0,\nn\\
K_{ab}&=&-\frac{1}{f(\tau)^2\sqrt{1+f'(\tau)^2}}\delta_{ab},\label{Kabs}
\eea
and after taking the trace we obtain
\be
K|_Q=-\frac{f(\tau)f''(\tau)+d(1+f'(\tau)^2)}{(1+f'(\tau)^2)^{3/2}}.\label{KfP}
\ee
Note here that when $Q\to\Sigma$ ($f(\tau)\to\epsilon$), $K|_Q$ has the opposite sign to $K|_\Sigma$ hence, by construction, the sum of the two Gibbons-Hawking terms also vanishes when $Q\to\Sigma$. We will use this fact in discussion on holographic path-integral complexity.\\
Then, using relations between $\tau$ and $w$, we can write \eqref{KfP} in terms of $\phi(w)$ as
\be
K|_Q=\frac{e^{-2\phi}\left(\ddot{\phi}+(d-1)\dot{\phi}^2\right)-d}{\sqrt{1-e^{-2\phi}\dot{\phi}^2}}.\label{Kphi}
\ee
Putting all these ingredients together, we have the on-shell gravity action with the tension term expressed in terms of $f(\tau)$
\ba
I_G+I_T=-\frac{d-1}{\kappa^2}\frac{V_xL_\tau}{\epsilon^{d}}-\frac{V_x}{\kappa^2}\int \frac{d\tau}{f^{d}}w'(\tau)\left[K|_Q+\frac{1}{w'(\tau)}-T\right],\label{Actf}
\ea
with $w'(\tau)$ given by the first relation in \eqref{wprim} and $K|_Q$ given by \eqref{KfP}.\\
We can also rewrite this action in terms of $w$ and $\phi(w)$ field as 
\bea
I_G+I_T=-\frac{d-1}{\kappa^2}\frac{V_xL_\tau}{\epsilon^{d}}+\frac{V_x}{\kappa^2}\int dw\,e^{d\phi}\left[\frac{d-1-e^{-2\phi}\left(\ddot{\phi}+(d-2)\dot{\phi}^2\right)}{\sqrt{1-e^{-2\phi}\dot{\phi}^2}}+T\right].
\eea
Furthermore, we can integrate by parts and rewrite this action as first-order in derivatives 
\ba
I_G+I_T&=&-\frac{(d-1)V_xL_\tau}{\kappa^2\ep^{d}}+\frac{(d-1)V_x}{\kappa^2}\int dw e^{d\phi}G(\dot{\phi})+\frac{V_x}{\kappa^2}\int dw e^{d\phi}T\no
&& -\frac{V_x}{\kappa^2}\left[e^{(d-1)\phi}\arcsin(\dot{\phi}e^{-\phi})\right]^0_{-\infty},  \label{actgrv}
\ea
where $G$ is the following function bounded from below
\ba
G(\dot{\phi})\!=\!\s{1-e^{-2\phi}{\dot{\phi}}^2}\!+\!\dot{\phi}e^{-\phi}\arcsin(\dot{\phi}e^{-\phi}). \label{gphki}
\ea
This is our main result in this subsection and a general algorithm for computing the Hartle-Hawking wave functions semi-classically. In order to derive it, it was crucial that we treated $Q$ with tension $T$ as a probe in Poincare $AdS_{d+1}$. With this action, we are now ready to maximize the semi-classical wave function \eqref{HHwv}.
\subsection{Maximization}
The next step in our procedure is to maximize the Hartle-Hawking wave function (\ref{HHwv}) by choosing the appropriate metric on $Q$. This step should be understood as the gravity counterpart of the path-integral optimization on the CFT side and we will justify this claim below. The maximization condition is obtained by taking a variation of the on-shell action (\ref{actgrv}) with respect to $\phi$, leading to
\ba
\frac{e^{-2\phi}(\ddot{\phi}+(d-1)\dot{\phi}^2)-d}{\s{1-e^{-2\phi}{\dot{\phi}}^2}}=\frac{d}{d-1}T. \label{eoma}
\ea
Comparing with \eqref{Kphi}, we find that the left-hand side is given by $K|_Q$, therefore \eqref{eoma} becomes
\be
K|_Q=\frac{d}{d-1}T.\label{ourEqn}
\ee
In other words, the maximization implies that $Q$ should be a constant mean-curvature (CMC) slice of the $AdS_{d+1}$ metric \eqref{metx} dual to the CFT vacuum.\\
More generally, the maximization procedure requires variation of the on-shell gravity action with tension term with respect to the induced metric $h_{ij}$ on $Q$. On general grounds this gives 
\be
(K_{ij}-Kh_{ij}+Th_{ij})\delta h^{ij}=0,\label{VarNeu}
\ee
i.e. the Neumann boundary condition on $Q$. In the ``conformal" gauge and with $\phi(w)$ being the only component of the metric, this variation is proportional to the metric itself and gives the trace of the Neumann condition \eqref{ourEqn}. We will see explicitly below that our metrics will solve \eqref{ourEqn} as well as all the components of \eqref{VarNeu} so the maximization procedure is equivalent to imposing the full Neumann boundary condition on $Q$
\ba
K_{ij}-Kh_{ij}=-Th_{ij},  \label{bdycond}
\ea
which is also usually imposed in the standard AdS/BCFT construction \cite{Ta}. See more in App. \ref{InhomSlices}. \\
Now, by definition, $f(w=0)=\epsilon$ (see Fig. \ref{gravityfig}), and we search for solutions of \eqref{eoma} with the boundary condition
\ba
e^{2\phi}|_{w=0}=\frac{1}{f^2(w=0)}=\frac{1}{\ep^2}. \label{bccg}
\ea
This condition also matches the two metrics, the one on $Q$ and the one on $\Sigma$, at the corner $\tau=w=0$ where the two surfaces meet. The relevant solutions can be found when $T<0$ and we derive
\ba
e^{2\phi}=\frac{1}{f^2(w)}=\frac{1}{\left(-\sqrt{1-\frac{T^2}{(d-1)^2}}w+\ep\right)^2}.\label{solta}
\ea
Note that for $T=0$, metric \eqref{solta}, in the ``conformal coordinate" $w$, precisely matches the CFT metric from the path-integral optimization \eqref{SolVP} written in $\tau$ with $\mu=1$. Identifying the two solutions for all values of the parameters suggests the dictionary between the two parameters
\be
\mu=1-\frac{T^2}{(d-1)^2},\label{mutr}
\ee
and we will elaborate on it below and check it in all the Euclidean examples.\\
By definition, our solution $\phi$ determines function $f(\tau)$. Indeed using \eqref{wprim} we can rewrite \eqref{solta} in terms of the original coordinate $\tau$ that from integrating \eqref{wprim} is given by
\be
\tau=\frac{|T|}{d-1}w,\label{dictaumu}
\ee
so that surface $Q$ is parametrized by
\ba
z=f(\tau)=\s{1-\frac{T^2}{(d-1)^2}}\frac{(d-1)}{T}\tau +\epsilon .  \label{surfacep}
\ea
\begin{figure}[b!]
  \centering
  \includegraphics[width=7cm]{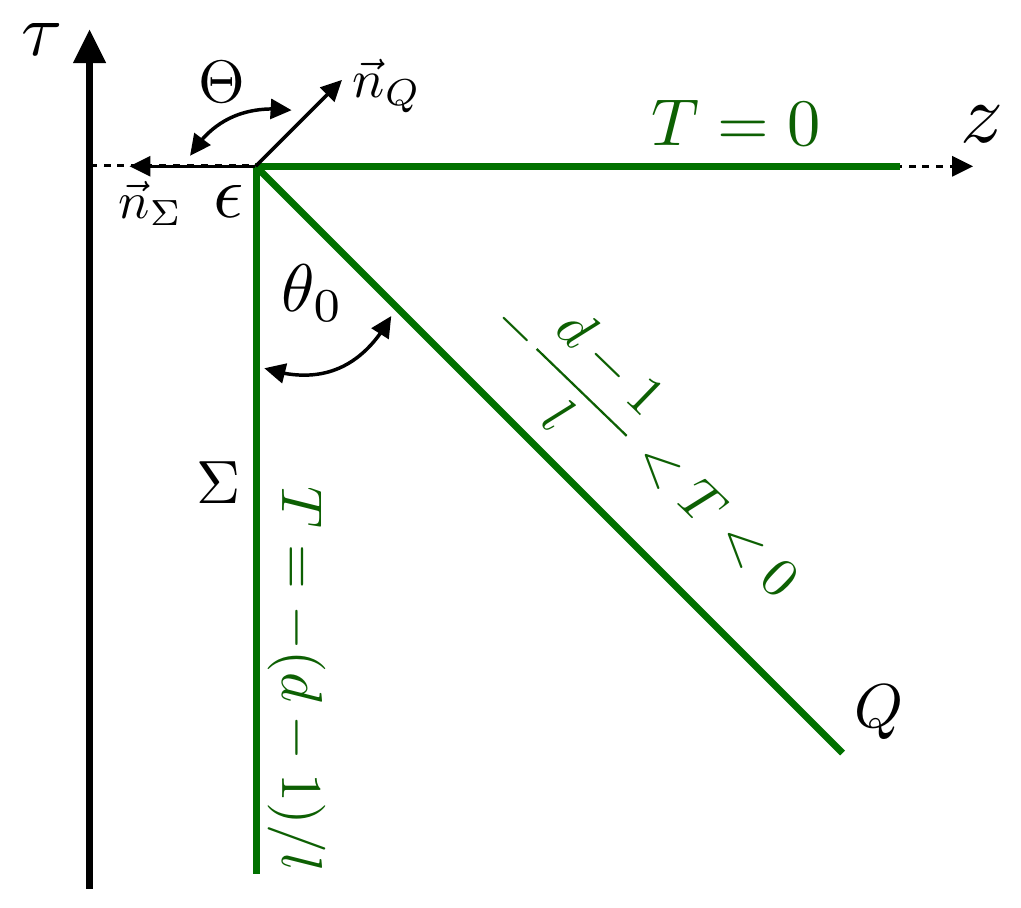}
  \caption{Solutions \eqref{surfacep} are half planes interpolating between the boundary at $T=-(d-1)/l$ and the $\tau=0$ slice for $T=0$ (in the text we used $l=1$). The inner angle $\theta_0$ is related to $\Theta$ and $T$ via \eqref{anglet}.}
\label{SlicesTE}
\end{figure}\\
Surfaces $Q$ that maximize Hartle-Hawking wave functions, shown on Fig. \ref{SlicesTE}, are then half-planes that interpolate between $z=\epsilon$ at $T=-(d-1)$ and the constant time-slice of $AdS_{d+1}$ for $T=0$ given by $\tau(z)=0$.\\
The tension parameter $T$ is naturally related to the angle between $\Sigma$ and $Q$. More precisely, we have
\be
\cos\Theta=n_\Sigma\cdot n_{Q}|_{\tau=0}=\frac{T}{d-1},\qquad \sin\theta_0=\sqrt{1-\frac{T^2}{(d-1)^2}},
\label{anglet}
\ee
where angle $\Theta$ is simply related to the inner  angle $\theta_0$ between $\Sigma$ and $Q$ by $\Theta=\pi-\theta_0$ (see Fig \ref{SlicesTE}). The allowed values of $\theta_0\in [0,\pi/2]$ are reflected in the range of $T\in [-(d-1),0]$.\\
If we recall our interpretation of parameter $0<\mu\le1$ in the path-integral complexity action and take the relation \eqref{mutr}, these gravity solutions provide an elegant geometrization of sub-optimal continuous tensor networks from the path-integral optimization. Indeed for $T\to -(d-1)$, surface $Q\to\Sigma$ represents the original, flat (most un-optimal) tensor network. On the other hand, for $T\to0$ the $\tau=0$ slice of the bulk corresponds to the fully optimized path-integral geometry with $\mu=1$. All the CMC surfaces in between with $-(d-1)<T\le 0$ naturally correspond to partially optimized path-integral tensor networks from the path-integral complexity action with $0<\mu<1$.

\subsection{Holographic Path-Integral Complexity}\label{Sec:Hayward}
The important part of the CFT analysis was focused on the path-integral complexity actions as relative measures of computational complexity. Now, after the optimization we obtained a family of slices $Q$ labeled by tension $T$ and we argued that they interpolate between the boundary $\Sigma$ and the optimal slice for $T=0$. In this section we make this statement more quantitative and propose how to estimate complexity of these sub-optimal networks in the gravitational construction.\\
After what we have already discussed, it is natural to expect that the gravity action in region M is a good candidate for the relative measure of complexity between $\Sigma$ and $Q$. However, the full story is slightly more involved. In order to prepare for our final quantity, let us then first evaluate the classical action \eqref{Actf} or \eqref{actgrv} that computes the Hartle-Hawking wave function $\Psi^{(T)}_{HH}[\phi]$ on the solution  (\ref{solta}). We can verify that the boundary term, part of the action with $G(\dot{\phi})$ and the tension term precisely cancel and we are only left with the divergent contribution
\ba
I_G+I_T=-\frac{(d-1)V_xL_\tau}{\kappa^2\ep^{d}},
\ea
so the on-shell action with the tension term added is independent of the tension parameter $T$. This way, in order to measure computational complexity from the Hartle-Hawking wave function construction, we propose to consider the value of the action without the tension term added. This is natural since the role of the tension term is only to fix the intermediate slices and, by definition, it also doesn't contribute to the complexity of the $\tau=0$ slice.\\
 The on-shell action without the tension term is then evaluated as follows
\ba
I_G=-\frac{(d-1)V_xL_\tau}{\kappa^2\ep^{d}}+
\frac{V_x}{\kappa^2}\frac{|T|}{(d-1)\s{1-\frac{T^2}{(d-1)^2}}}\frac{1}{\ep^{d-1}}.\label{IGTC}
\ea
Still, this is not the end of the story. Recall that for the correct variational principle, gravitational action in regions with non-smooth boundaries should be supplemented by the so-called Hayward term \cite{Hayward:1993my} defined as
\be
I_H=-\frac{1}{\kappa^2}\int \sqrt{\gamma}(\Theta+\vp_0),\label{Hay0}
\ee
where as shown on Fig. \ref{SlicesTE}, $\Theta$ is defined in terms of the normal vectors to the surfaces that meet at the corner (see \eqref{anglet}) and $\gamma_{ij}$ is the induced metric on the corner. By adding $\varphi_0$ in \eqref{Hay0} we wanted to stress that, from the point of view of the variational problem, any constant value of $\vp_0$ works sensibly in this definition. In order to estimate the complexity of partially optimized slices, we will argue that the appropriate choice of $\vp_0$ is given by $\vp_0=-\pi$, which leads to the following expression for the corner contribution
\be
I'_H\equiv\frac{1}{\kappa^2}\int \sqrt{\gamma}\,\theta_0,  \label{haymod}
\ee
that is determined entirely by the data in region $M$. We will refer to this term as ``modified Hayward term" expressed by the inner angle $\theta_0$ between $\Sigma$ and $Q$.\\
A clear motivation for this way of fixing the ambiguity in the Hayward term comes in fact from the path-integral complexity. Namely, as we pointed in the careful derivation of the Hartle-Hawking wave functions, the gravity action (bulk as well as the two Gibbons-Hawking contributions) vanish in the limit of $Q\to \Sigma$. This is clearly a desired property of a relative complexity between metrics on $\Sigma$ and $Q$. If we added the standard Hayward term to the gravitational action on M, since it vanishes only for smooth boundaries (in our setup $\theta_0=\pi$) it would spoil this property. This way, in order to keep a well-defined variational principle in the cusped region M as well as have a sensible holographic definition of a relative complexity functional, we propose to fix the constant ambiguity in the Hayward term to $\varphi_0=-\pi$ and use \eqref{haymod} in addition to the gravity action \eqref{IGTC}.

Finally, we propose to estimate the holographic path-integral complexity of the partially optimized network $Q$ (relative to $\Sigma$) corresponding to the Euclidean Hartle-Hawking wave function with fixed value of the tension $T$ as the sum of the contribution of the on-shell gravity action in region $M$ (without the tension term) and the modified Hayward term\footnote{In fact a similar definition was used before in \cite{Takayanagi:2018pml} with the overall sign difference since it was evaluated in the bulk region up to the surface $Q$ that represented a cut-off in that work.}
\be
\mathcal{C}^{(e)}_T\equiv I_G+I'_H. \label{euclcomp}
\ee
In the following sections we will study this quantity in various explicit examples of Euclidean geometries. Interestingly, before going to any specific examples, we can verify that $\mathcal{C}^{(e)}_T$ indeed satisfies the same co-cycle properties \eqref{co-cycle} as the CFT path-integral complexity (see App. \ref{GRaction}). \\
Evaluating $\mathcal{C}^{(e)}_T$ in the vacuum example gives 
\ba
\mathcal{C}^{(e)}_T=-\frac{(d-1)V_xL_\tau}{\kappa^2\ep^{d}}+\frac{V_x}{\kappa^2\ep^{d-1}}
(\theta_0+\cot\theta_0), \label{compigh}
\ea
where $\theta_0$ and its relation to $T$ is computed in \eqref{anglet}. This shows that $\mathcal{C}^{(e)}_T$ is indeed a monotonically decreasing function of 
$\theta_0$ or equally of $T$. Thus, it is minimized when $T=0$ ($\tau=0$ slice), as expected from the optimization procedure. The minimal value is given by
\be
\mathcal{C}^{(e)}_{T=0}=-\frac{(d-1)V_xL_\tau}{\kappa^2\ep^{d}}+\frac{V_x}{\kappa^2\ep^{d-1}}
\frac{\pi}{2}.  \label{euccompvacc}
\ee
Notice that the first negatively divergent term in (\ref{compigh}) can be regarded as the subtraction of the complexity for the reference state before the optimization (i.e. $e^{\phi(w)}=1/\ep$ at $\Sigma$). \\
Let us also point a subtlety that, by construction, \eqref{compigh} should vanish when $\theta_0\to0$ or equivalently $T\to-(d-1)$. This was clear from the definition of $\mathcal{C}^{(e)}_{T}$ but after evaluating it on-shell (and doing the $\tau$-integral) in order to reproduce this fact we should take a careful (formal) limit of
\be
\theta_0\to\cot^{-1}\left(\frac{(d-1)L_\tau}{\epsilon}\right),\qquad T\to\frac{-(d-1)}{\sqrt{1+\frac{\epsilon^2}{(d-1)^2L^2_\tau}}},\label{ForLim}
\ee
with $L_\tau/\epsilon\to\infty$.\\
This concludes our analysis of the holographic path-integral complexity from the Hartle-Hawking wave function and in section \ref{Sec:4 MoreEx} we will provide further evidence and support for our definition in various Euclidean examples.

\subsection{Maximization and Gravity Hamiltonian Constraint}
In this subsection we give further comments on the maximization procedure. In particular, we elaborate on the fact that path-integral geometries in CFT are found from the constant Ricci scalar constraint (e.g. Liouville equation) whereas surfaces $Q$ are obtained from the Neumann boundary condition. We will see that it is in fact the gravity Hamiltonian constraint that elegantly connects the two procedures.\\ 
As we argued on general grounds, as in AdS/BCFT \cite{Ta}, the maximization of the Hartle-Hawking wave function is equivalent to imposing Neumann boundary condition on $Q$. We can check this explicitly using components of the extrinsic curvature tensor derived in \eqref{Kabs}. After writing \eqref{bdycond} in components we get
\be
\frac{1}{\sqrt{1+f'(\tau)^2}}=-\frac{T}{d-1},\quad -\frac{f(\tau)f''(\tau)+(d-1)(1+f'(\tau)^2)}{(1+f'(\tau)^2)^{3/2}}\delta_{ab}=T\delta_{ab},\label{NBCt}
\ee
or in terms of $w$ and $\phi(w)$ 
\be
\sqrt{1-\dot{\phi}^2e^{-2\phi}}=-\frac{T}{d-1},\qquad \frac{e^{-2\phi}\left(\ddot{\phi}+(d-2)\dot{\phi}^2\right)-(d-1)}{\sqrt{1-\dot{\phi}^2e^{-2\phi}}}\delta_{ab}=T\delta_{ab}.\label{NBCw}
\ee
Plugging \eqref{surfacep} to the first and \eqref{solta} to the second pair of constraints, we can verify that full Neumann boundary conditions are satisfied for $T<0$ straightforwardly.\footnote{Note that to find $Q$ (i.e. $f(\tau)$) that maximizes the Hartle-Hawking wave function, we could also vary the action \eqref{Actf} with respect to $f(\tau)$ and get the constraint
\be
\frac{2ff''+d(1+f'^2)}{(1+f'^2)^2}=-\frac{T}{d-1}\frac{ff''+d(1+f'^2)}{(1+f'^2)^{3/2}}.
\ee
We  can easily check that \eqref{surfacep} is a solution of this equation with $T<0$.} See also App. \ref{InhomSlices} for more discussion on solving the Neumann boundary condition.\\
Given the induced metric on \eqref{METQ} with maximizing solution \eqref{solta}, we can compute the Ricci scalar curvature (intrinsic) of our bulk slices 
\be
R^{(d)}=e^{-2\phi}(-2(d-1)\partial^2_w\phi-(d-2)(d-1)(\partial_w\phi)^2)=2\Lambda^{(d+1)}\left(1-\frac{T^2}{(d-1)^2}\right),\label{Riccid}
\ee
where $\Lambda^{(d+1)}=-d(d-1)/2$. Clearly, slices $Q$ have constant negative curvature parametrized by tension $T$. We will come back to this result momentarily and discuss the precise relation with the path-integral optimization geometries.\\ Before doing that, we can verify that this negative curvature gives yet another interpretation of the maximization procedure. In fact, we can check that, by definition, when we impose the Neumann boundary condition on $Q$, the extrinsic curvature tensor is proportional to the induced metric
\be
K_{ij}=(K-T)h_{ij}=\frac{T}{d-1}h_{ij},
\ee
where in the second equality we used \eqref{ourEqn}. Consequently, on-shell, the following combination of the extrinsic curvature and the square of its trace  becomes
\be
K^2-K^{ij}K_{ij}=\frac{d}{d-1}T^2.\label{LHS}
\ee
This is nothing but the left-hand side of the ``Hamiltonian" constraint in pure gravity on $AdS_{d+1}$ 
\be
K^2-K^{ij}K_{ij}=R^{(d)}-2\Lambda^{(d+1)},\label{HC}
\ee
where $\Lambda^{(d+1)}$ is the negative cosmological constant of $AdS_{d+1}$ and $R^{(d)}$ is the Ricci scalar curvature of the induced metric on the gravity slice. Plugging \eqref{LHS}, we can check that this constraint reproduces the Ricci scalar $\eqref{Riccid}$. This local Hamiltonian constraint is the contraction of the bulk Einstein's equation\footnote{This can be seen by using Gauss-Codazzi equation, see e.g. \cite{Poisson}.} that is satisfied from the beginning in our construction
\be
n^\mu n^\nu \left(G_{\mu\nu}=-\Lambda^{(d+1)}g_{\mu\nu}\right),
\ee
where $G_{\mu\nu}$ is the Einstein's tensor and $n^\mu$ are the normal vectors to an arbitrary slice of the bulk. Here, since the maximization of the Hartle-Hawking wave functions leads to the Neumann boundary condition this constraint automatically implies that the Ricci scalar curvature of our slices is constant.

Recall that ``Hamiltonian" constraint \eqref{HC} in AdS/CFT usually refers to the ``radial Hamiltonian" where $K_{ij}$ and $K$ are computed locally on some cut-off surface with Ricci scalar $R^{(d)}$. This constraint plays an important role in the recent developments on the relations between $T\bar{T}$-deformations of holographic CFTs and finite cut-off gravity with Dirichlet boundary condition on the cut-off surface \cite{McGough:2016lol,Kraus:2018xrn,Caputa:2020lpa}. Imposing Dirichlet boundary condition for the gravity action with holographic counter-terms allows one to compute the holographic stress-tensor expressed by $K_{ij}$, its trace $K$ and the counter-term contribution. Then, rewriting $K_{ij}$ and $K$ in \eqref{HC} in terms of $T_{ij}$ and $T^i_i$ reproduces trace anomaly equation in a CFT deformed by the $T\bar{T}$-operator with the coupling related to the finite-cut-off radius. This construction was recently generalized to $T\bar{T}$ deformations with time-dependent couplings  \cite{Caputa:2020fbc} (see also \cite{Kruthoff:2020hsi} for discussion on the flow of quantum states under $T\bar{T}$ deformations that is relevant in our context) and studied from the perspective of holographic Tensor Networks as constant $K$ slices of $AdS$. Interestingly, the explicit time-dependent ``folding $T\bar{T}$ deformation" in CFTs, precisely corresponds to the bulk $AdS$ geometry with time-dependent cut-off surfaces identical to the $Q$ in \eqref{surfacep}. We hope that this intriguing connection will shed more light on the operational and circuit-type interpretation of the path-integral optimization and path-integral complexity and we leave it as an interesting future direction.

\subsection{Boundary UV Limit}
Let us finally finish with more evidence for the holographic relation with field theory path-integral optimization.  Firstly, one striking feature of the proposal is that both metrics, Weyl flat metric from path-integral optimization \eqref{curm} (in arbitrary dimensions) and induced metric on $Q$ that maximizes Hartle-Hawking wave function \eqref{diago}, are identical after we relate $w$ and $\tau$. Indeed, they have both constant negative Ricci scalars equal to
\be
R^{(d)}|_{PI}=2\Lambda^{(d+1)}\mu,\qquad R^{(d)}|_{HH}=2\Lambda^{(d+1)}\left(1-\frac{T^2}{(d-1)^2}\right).
\ee
We will analyze various, non-trivial, examples below that confirm this observation. This again leads to the identification \eqref{mutr} between parameters in our proposals. Remember that changing $\mu$ from $\mu=0$ to $\mu=1$ means that we gradually increase the amount of optimization.\footnote{Let us also point that, from the perspective of complexity and gate counting \cite{Czech:2017ryf}, one could think about $\mu$ as a relative ``penalty factor" between unitaries and isometries and we leave this alternative interpretation for future studies (see also \cite{Bhattacharyya:2019kvj}).} In the gravity dual, this corresponds to changing the tension from $T=-(d-1)$ to $T=0$ which tilts the surface $Q$ from the asymptotic boundary $\Sigma$ to the time slice $\tau=0$. 

Secondly, the gravity construction with Hartle-Hawking wave function involves extremizing action \eqref{actgrv}, that can be thought of as the boundary (UV) action with infinite sum of finite cut-off corrections. Indeed we can write $G(\dot{\phi})$ as a series in $\dot{\phi}e^{-\phi}$
\be
G(\dot{\phi})=1+\sum^{\infty}_{k=1}\frac{\Gamma\left(\frac{2k+1}{2}\right)}{\sqrt{\pi}(2k-1)^2\Gamma(k+1)} \,\dot{\phi}^{2k}e^{-2k\phi}.
\ee
In the UV limit, taking only the leading $\dot{\phi}e^{-\phi}$ term, we end up 
\be
e^{d\phi}G(\dot{\phi})=\frac{1}{2}e^{(d-2)\phi(w)}\dot{\phi}^2+e^{d\phi}+O(\dot{\phi}^4),
\ee 
and the gravity action (without the tension term) becomes
\bea
I_G|_{UV}&\simeq &\frac{(d-1)V_x}{2\kappa^2}\int dw \left[e^{(d-2)\phi(w)}\dot{\phi}^2+2e^{d\phi}-2\epsilon^{-d}+O(\dot{\phi}^4)\right]\nn\\
&-&\frac{V_x}{\kappa^2}\frac{\theta_0}{\epsilon^{d-1}},\label{IUVLim}
\eea
where $\theta_0$ is the value of $\arcsin(\dot{\phi}e^{-\phi})$ at $w=0$. Interestingly, we can cancel $\theta_0$ dependence by adding the ``modified Hayward" term (\ref{haymod}), localized on $\Sigma\cap Q$. This in fact will be true in all the Euclidean examples below. This is another argument for its important role in the holographic path-integral complexity. More importantly, \eqref{IUVLim} is precisely the (homogeneous) form of our conjectured action \eqref{HDimA}, therefore the construction with Hartle-Hawking wave function gives a strong justification of the higher dimensional conjecture.\footnote{Giving up the homogeneity assumption $z=f(t)$ and taking $z=f(t,x_i)$ should reproduce the full action in the UV limit.}\footnote{Note that if we add the tension term $I_T$, the total action $I_G+I_T$ in the UV limit leads to a ``UV relation": $\mu=2+\frac{2T}{d-1}$.
We can regard the minimizing $I_G+I_T$ as a partial optimization and this corresponds the changing the cosmological constant $\mu$. However notice that the genuine complexity functional, whose minimization gives the full optimization is obtained by removing the tension term.}\\
Nevertheless, note that the gravity action (\ref{actgrv}) gives the correct holographic finite cut-off corrections to the UV Liouville action (as in (\ref{compigh})) and thus should be thought of as the full answer to the path-integral optimization program.  Having said that, there is still a possibility that both complexity actions, the Liouville and the holographic path-integral complexity, even though apparently different and coming with their own regularization schemes, are actually equivalent. The fact that they lead to the same optimal geometries with constant Ricci scalars may suggest this however, we do not have any definite way to prove/disprove this possibility at the moment and leave it for future investigation.

\section{More Euclidean Examples}\label{Sec:4 MoreEx}
In the following sections we study more examples that will confirm our procedure in various Euclidean setups across dimensions. The algorithm for computing semi-classical Hartle-Hawking wave functions using on-shell gravity action with tension  \eqref{HHwv} is straightforward and we will spare the details of the steps while only showing main results of analogous computations as we did for the vacuum case.

\subsection{Global $AdS_{d+1}$}
We start by analyzing the gravity dual of the vacuum of a CFT on a circle given by global $AdS$ spacetime. We can perform similar analysis to what we did in Poincare coordinates starting now from the global $AdS_{d+1}$ metric
\bea
ds^2&=&\left(1+r^2\right)d\tau^2+\frac{dr^2}{1+r^2}+r^2\,d\Omega^2_{d-1},
\eea
where $\Omega_{d-1}$ stands for the volume element of a $(d-1)$-sphere.\\
The region $M$ in global coordinates is bounded by the cut-off surface $\Sigma$ at $r=r_0=\frac{1}{\epsilon}\to\infty$ and a homogeneous surface $Q$ defined by $r=1/f(\tau)$, so its  induced metric can be written as
\be
ds^2 =\frac{1}{f(\tau)^2}\left[\left(1+f(\tau)^2+\frac{f'(\tau)^2}{1+f(\tau)^2}\right)d\tau^2+d\Omega^2_{d-1}\right]
= e^{2\phi(w)}\left(dw^2+d\Omega^2_{d-1}\right).
\ee 
Again, in the second equality, we introduced coordinate $-\infty<w<0$  and field $\phi(w)$ that render the induced metric on $Q$ conformally flat
\be
\label{eq:EuclideanGlobal: w'}
w'(\tau)=\sqrt{1+f^2+\frac{f'^2}{1+f^2}}= \frac{1+f^2}{\sqrt{1+f^2-\dot{f}^2}},\qquad e^{2\phi}=\frac{1}{f^2}.
\ee
The notation for $f'$ and $\dot{f}$ is the same as in Poincare coordinates (see below \eqref{wprim}).\\
Next, we evaluate the bulk Hartle-Hawking wave function semi-classically as the on-shell bulk action with the tension term on $Q$  \eqref{HHwv}. To compute the classical action, we will need the trace of the extrinsic curvature on $\Sigma$
\be
K|_\Sigma=\frac{d+(d-1)\epsilon^2}{\sqrt{1+\epsilon^2}},
\ee
as well as for surface $Q$ written in terms of $f(\tau)$
\be
K|_Q=-\frac{1}{w'^3}\left[ff''+\frac{f'^2((d-3)f^2+d)+(1+f^2)^2(f^2(d-1)+d)}{(1+f^2)}\right],\label{KQGlf}
\ee
or in terms of $w$ of $\phi(w)$ as
\be
K|_Q=\frac{e^{-2\phi}\left(\ddot{\phi}+(d-1)(\dot{\phi}^2-1)\right)-d}{\sqrt{1-e^{-2\phi}(\dot{\phi}^2-1)}}.\label{KQGl}
\ee
With these ingredients we can evaluate the gravity action with tension term and the answer becomes
\be
I_G+I_T=-\frac{(d-1)V_\Omega L_\tau}{\kappa^2\epsilon^d}\left(1+\epsilon^2\right)-\frac{V_\Omega}{\kappa^2}\int \frac{d\tau}{f^d}w'(\tau)\left[K|_Q-T+\frac{1}{w'(\tau)}\right],
\ee
where $w'(\tau)$ is given by the first equation in \eqref{eq:EuclideanGlobal: w'}, $K|_Q$ by \eqref{KQGlf} and $V_\Omega$ is the volume of the $(d-1)$ sphere.\\
Similarly, in terms of $w$ and $\phi$, after integrating by parts, we derive the main result for the global $AdS_{d+1}$
\bea
I_G+I_T&=&-\frac{(d-1)V_\Omega L_\tau}{\kappa^2 \epsilon^d}\left(1+\epsilon^2\right)+\frac{(d-1)V_\Omega}{\kappa^2}\int dw e^{d\phi}G(\dot{\phi})+\frac{V_\Omega}{\kappa^2}\int dw e^{d\phi}T\nn\\
&-&\frac{V_\Omega}{\kappa^2}\left[e^{(d-1)\phi}\arcsin\left(\frac{\dot{\phi}e^{-\phi}}{\sqrt{1+e^{-2\phi}}}\right)\right]^{0}_{-\infty},\label{IGITGL}
\eea
where now, function $G(\dot{\phi})$, with finite size corrections, becomes
\be
G(\dot{\phi})=\sqrt{1-e^{-2\phi}(\dot{\phi}^2-1)}+\dot{\phi}e^{-\phi}\arcsin\left(\frac{\dot{\phi}e^{-\phi}}{\sqrt{1+e^{-2\phi}}}\right).
\ee
Note that, as we showed in the previous example in Poincare coordinates, the gravity action (without the tension term) that computes the semi-classical Hartle-Hawking wave functions \eqref{HHwv} should be regarded as the CFT path-integral complexity (Liouville action in 2d) together with finite cut-off corrections from the bulk. These holographic finite cut-off corrections will naturally depend on the details of the gravity geometry such as e.g. the finite size in \eqref{IGITGL} or mass of excitations (black hole) that we will see in the following examples. 

\begin{figure}[t!]
  \centering
  \includegraphics[width=7cm]{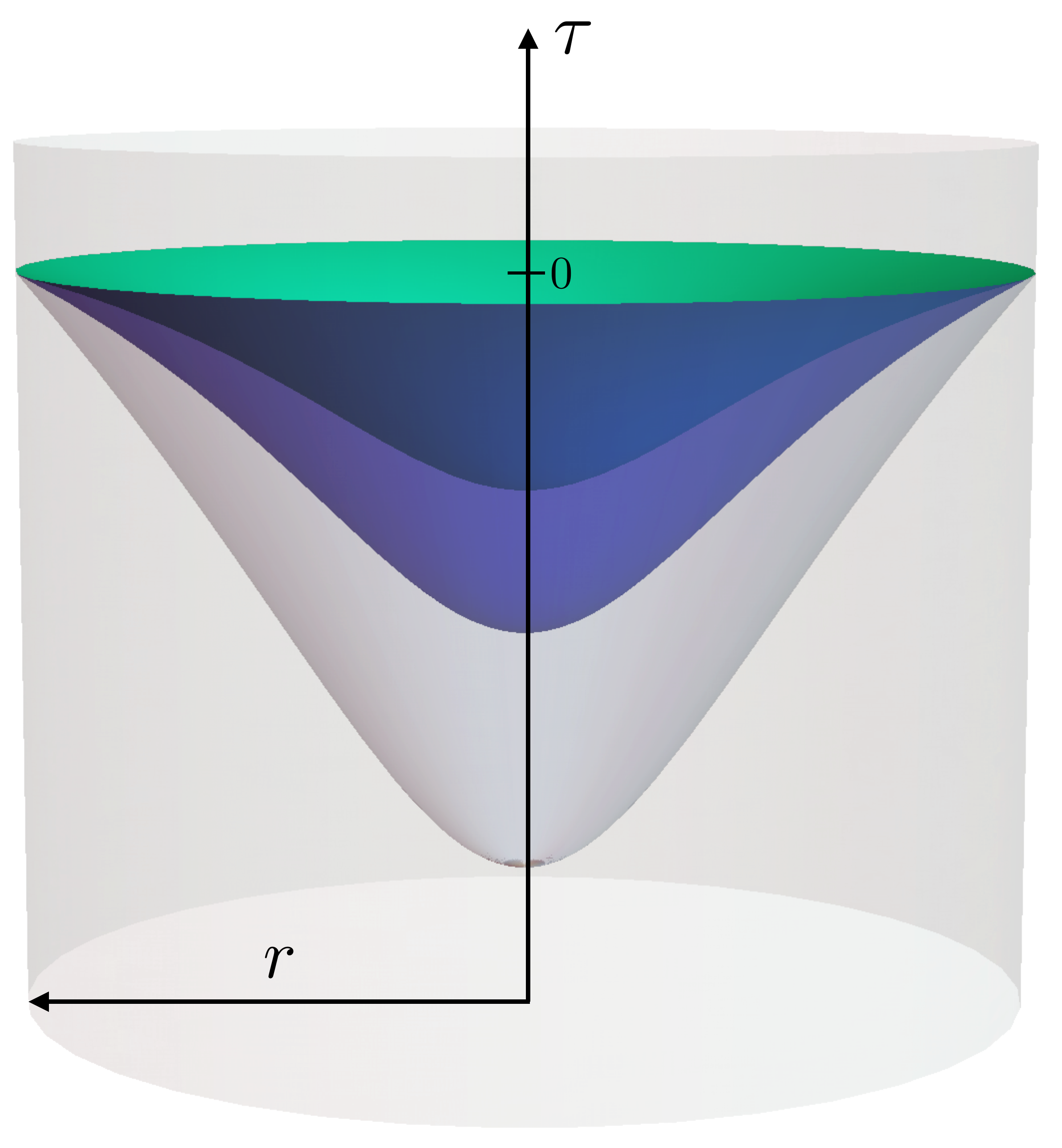}
  \caption{Example slices $Q$ that maximize the Hartle-Hawking wave functions in global $AdS_{d+1}$ for a fixed value of $T$. They interpolate between the maximal, $\tau=0$ slice for $T=0$ (green) and one reaching to $\tau=-\infty$ for $T=-(d-1)$.}
\label{CMCslicesG}
\end{figure}

Maximization of the wave function with respect to $\phi$ again leads to the CMC slices
\be
K|_Q=\frac{d}{d-1}T,
\ee
and we can solve it for $-(d-1)<T\le 0$ by
\be
e^{2\phi(w)}=\frac{1}{f^2(w)}=\frac{1}{\left(1-\frac{T^2}{(d-1)^2}\right)\sinh^2\left(w-c_1\right)}.
\ee
Constant $c_1$ is fixed such that the boundary condition \eqref{bccg}, $f(0)=\epsilon$, is satisfied
\be
c_1\simeq \frac{\epsilon}{\sqrt{1-\frac{T^2}{(d-1)^2}}},
\ee
and in the computations of the on-shell action we will be taking the $\epsilon<<1$ limit first.\\
Again, this result in conformal coordinate $w$ matches the solution that we get from path-integral optimization with the Liouville action (or its generalization to higher dimensions \eqref{HDimA}) on the cylinder.\footnote{Recall that the solution of Liouville equation on the cylinder of circumference $L$ with coordinates $\tau\in (-\epsilon,-\infty)$ and $\sigma=\sigma+L$ written in coordinates $w=\tau+i\sigma$ is given by
\be
e^{2\phi(w,\bar{w})}=\frac{4\pi^2}{L^2\sinh^2\left(\frac{\pi(w+\bar{w})}{L}\right)}=\frac{4\pi^2}{L^2\sinh^2\left(\frac{2\pi \tau}{L}\right)},
\ee
and we have the boundary condition $e^{2\phi(\tau=-\epsilon)}=\frac{1}{\epsilon^2}$ and $e^{2\phi(\tau=-\infty)}\to0$.}\\
Moreover, we can integrate \eqref{eq:EuclideanGlobal: w'} and write the embedding function $f(\tau)$ of the surface $Q$ in original global $AdS_{d+1}$ coordinates 
\be
f(\tau)=\sqrt{1-\frac{T^2}{(d-1)^2}}\frac{\tanh(c_2-\tau)}{\sqrt{\frac{T^2}{(d-1)^2}-\tanh^2(\tau-c_2)}},
\ee
where, from the boundary condition $f(0)=\epsilon$ we can fix $c_2$ to
\be
c_2\simeq \frac{\epsilon|T|}{(d-1)\sqrt{1-\frac{T^2}{(d-1)^2}}}+O(\epsilon^3).\label{c2G}
\ee
To analyze this profile more carefully, it is useful to parametrize $Q$ as $\tau(r)$ and we derive
\be
\tau(r)=\sinh^{-1}\left(\frac{T}{(d-1)\sqrt{(1+r^2)\left(1-\frac{T^2}{(d-1)^2}\right)}}\right)+c_2.
\ee
We can verify that $Q$ has a minimum at $r=0$ where
\be
\tau_{min}=\tanh^{-1}\left(\frac{T}{d-1}\right)+O(\epsilon).
\ee
These surfaces are shown on Fig. \ref{CMCslicesG}. They interpolate between the $\tau=0$ slice for $T=0$ and the boundary surface $\Sigma$ for $T=-(d-1)$.
\\
Finally, for obtaining complexity $\mathcal{C}^{(e)}_T$ \eqref{euclcomp}, we first evaluate the on-shell gravity action (without the tension term). It can be computed in arbitrary dimensions in a closed form written in terms of special functions but, since it is not particularly illuminating, we just show a couple of results for specific dimensions.\\
 On the other hand, the modified Hayward term in general dimensions given by
\be
I'_H=\frac{1}{\kappa^2}\int_\gamma\sqrt{\gamma}\theta_0=\frac{V_\Omega}{\kappa^2\epsilon^{d-1}}\theta_0,\qquad \sin\theta_0=\sqrt{1-\frac{T^2}{(d-1)^2(1+\epsilon^2)}}.
\ee
This way, e.g. in global $AdS_3$ we derive 
\be
\mathcal{C}^{(e)}_T=-\frac{V_xL_\tau}{\kappa^2\epsilon^2}+\frac{V_x}{\kappa^2\epsilon}\left(\theta_0+\frac{|T|}{\sqrt{1-T^2}}\right)-\frac{V_x}{\kappa^2}\left(\frac{ |T|}{1-T^2}+\tanh^{-1}(T)\right)+O(\epsilon).
\ee
In odd $d$, we also get an additional logarithmic divergence from the gravity action part. E.g. in global $AdS_4$ we derive
\bea
\mathcal{C}^{(e)}_T&=& 
- \frac{2 V_\Omega L_\tau}{\kappa^2\epsilon^3}
\left( 1+ \epsilon^2\right) + \frac{V_\Omega}{\kappa^2}
\left[\frac{|T|}{2\sqrt{1-\frac{T^2 }{4}}} \frac{1}{\epsilon^2} -
\text{arctan} \frac{|T|}{2\sqrt{1-\frac{T^2}{4}}} \right] \nn\\
&-&\frac{V_\Omega}{\kappa^2}
\frac{|T|^3}{8\left(1-\frac{T^2}{4}\right)^{3/2}} 
\log \left[ 
\frac{2}{\epsilon} \sqrt{1- \frac{T^2}{4}}\right] + O(\epsilon). 
\eea
From the analytic formula in general $d$ we can verify that the complexity action action is always minimized for $T=0$. The minimal value can be computed in all dimensions and is given by
\be
\mathcal{C}^{(e)}_{T=0}=-\frac{(d-1)V_\Omega L_\tau}{\kappa^2\epsilon^d}\left(1+\epsilon^2\right)+\frac{V_\Omega}{\kappa^2\epsilon^{d-1}}\frac{\pi}{2}.
\ee
The leading divergent term and the modified Hayward term have the same form as in Poincare coordinates (with $V_x$ there and $V_\Omega$ here) and the new finite-size signature enters via the sub-leading contribution (or constant term in $d=2$).
\subsection{Excited states in 3d}
Next, we consider a general family of spacetimes that will allow us to reproduce all the metrics found in the path-integral optimization for 2d CFTs \cite{Caputa:2017urj}. More precisely, we repeat the computation of the semi-classical Hartle-Hawking wave functions \eqref{HHwv} using the on-shell gravity solution given by Euclidean BTZ-type metrics in three dimensions
\ba
ds^2=(r^2-r^2_h)d\tau^2+\frac{dr^2}{r^2-r^2_h}+r^2dx^2,\label{EXStM}
\ea
where $r^2_h=M-1$ can be positive or negative depending on the mass $M$ of the excitation. For $M=0$ or $r^2_h=-1$ we will formally recover the above results in global $AdS_3$ (with $d=2$).\\ We will now compute $\Psi^{(T)}_{HH}$ for region $M$ specified by $\frac{1}{f(\tau)}\le r\le \frac{1}{\ep}$, where $r=1/f(\tau)$ and $r=1/\ep$ describe the surface $Q$ and the asymptotic boundary $\Sigma$, respectively. The $x$ direction can be taken periodic or a real line and will only appear inside the overall factor $V_x$ as before. \\
Following our procedure, for $\Sigma$ at $r=1/\epsilon$ we derive
\be
K_\Sigma=\frac{2-r^2_h\epsilon^2}{\sqrt{1-r^2_h\epsilon^2}}.
\ee
On the other hand, the induced metric on $Q$ is given by
\be
ds^2=\frac{1}{f^2}\left[\left(1-r^2_hf(\tau)^2+\frac{f'^2}{1-r^2_hf(\tau)^2}\right)d\tau^2+dx^2\right]=e^{2\phi}(dw^2+dx^2),
\ee
where we introduced $-\infty<w<0$ such that
\be
w'(\tau)=\sqrt{1-r^2_hf^2+\frac{f'^2}{1-r^2_hf^2}}=\frac{1-r^2_hf^2}{\sqrt{1-r^2_hf^2-\dot{f}^2}}.
\label{eq:EuclideanBTZ-w'}
\ee
The trace of extrinsic curvature in terms of $\tau$ and $f(\tau)$ is now
\be
K|_Q=-\frac{1}{w'^3}\left[ff''+\frac{f'(\tau)^2(2+r^2_h f^2)+(1-r^2_h f^2)^2(2-r^2_h f^2)}{1-r^2_h f^2}\right],
\ee
or in terms of $w$ and $\phi(w)\equiv-\log(f(w))$
\be
K|_Q=\frac{e^{-2\phi}(\ddot{\phi}+\dot{\phi}^2)+r^2_h e^{-2\phi}-2}{\sqrt{1-e^{-2\phi}(\dot{\phi}^2+r^2_h)}} .
\ee
The action that computes the Hartle-Hawking wave function is now given in terms of $f(\tau)$ as
\be
I_G+I_T=-\frac{V_x L_\tau}{\kappa^2\epsilon^2} \left(1-r^2_h\epsilon^2\right)-\frac{V_x}{\kappa^2}\int \frac{d\tau}{f(\tau)^2} w'\left[K|_Q-T+\frac{1}{w'}\right].
\ee
As before, we can integrate by parts and derive the main result for excited states in asymptotically $AdS_{3}$ geometry \eqref{EXStM}
\bea
I_G+I_T=&-&\frac{V_x L_\tau}{\kappa^2\epsilon^2} \left(1-r^2_h\epsilon^2\right)+
\frac{V_x}{\kappa^2}\int dw\,e^{2\phi}G(\dot{\phi})+
\frac{V_x}{\kappa^2}\int dw\,e^{2\phi}T  \\
&-&\frac{V_x}{\kappa^2}
\left[e^{\phi}\arcsin\left(\frac{\dot{\phi}e^{-\phi}}{\sqrt{1-r^2_he^{-2\phi}}}\right)\right]^0_{-\infty}\nn,
\eea
with
\ba
G(\dot{\phi})&=&\sqrt{1-e^{-2\phi}\left(\dot{\phi}^2+r^2_h\right)}
+\dot{\phi}e^{-\phi}\arcsin\left(\frac{\dot{\phi}e^{-\phi}}{\sqrt{1-r^2_he^{-2\phi}}}\right).\label{GEXSt}
\ea
Variation with respect to $\phi$ yields the CMC condition $K|_Q=2T$ and is again equivalent to the Neumann condition \eqref{bdycond}. For negative tension, we can solve it by
\ba
e^{2\phi}=\frac{r^2_h}{\left(1-T^2\right)\sin^2\left(r_h(w-c_1)\right)}\label{solEx},
\ea
where $c_1$ is tuned appropriately to reproduce the boundary condition \eqref{bccg}.\\
This family of solutions precisely matches those in the path-integral optimization \cite{Caputa:2017urj} via the identification (\ref{mutr}). For $r^2_h=-(1-M)=-\alpha^2$ we reproduce excited states from the optimization for primary operators in 2d CFT i.e. conical singularity geometries, including the finite size vacuum.  For $r_h=2\pi/\beta$ and $w$ on the strip, we reproduce the optimal geometry for the thermofield double (TFD) state that was found to describe the time slice of eternal black hole (Einstein-Rosen bridge) \cite{MaE}. We can verify that all these solutions have constant negative curvature 
\be
R^{(2)}=-2(1-T^2),\label{RS2dEX}
\ee
which is just a special case of \eqref{Riccid} with $d=2$ and follows from the the Hamiltonian constraint supplemented by the Neumann boundary condition. We now analyze two of these solutions in more detail.
\subsubsection{Conical Singularities}
Let as first take the solutions with $r^2_h=-\alpha^2$, often referred to as conical singularities
\be
e^{2\phi}=\frac{\alpha^2}{\left(1-T^2\right)\sinh^2\left(\alpha(w-c_1)\right)}.
\ee
In fact this solution has a constant Ricci scalar as \eqref{RS2dEX} with an extra $\delta$-function proportional to $(1-\alpha)$ at $w\to-\infty$. This can be seen by e.g. mapping the 2d surface into a disc.\\
The boundary condition for $\phi$ at $w=0$ implies
\be
c_1\simeq \frac{\epsilon}{\sqrt{1-T^2}}+O(\epsilon^3).
\ee
In terms of $\tau$ the embedding function of $Q$ reads
\be
f(\tau)=\frac{\sqrt{1-T^2}\tanh(\alpha(c_2-\tau))}{\alpha\sqrt{T^2-\tanh^2(\alpha(\tau-c_2))}},
\ee
with
\be
c_2\simeq\frac{\epsilon|T|}{\sqrt{1-T^2}}+O(\epsilon^3).
\ee
It is again convenient to parametrize $Q$ as
\be
\tau(r)=\frac{1}{\alpha}\tanh^{-1}\left(\frac{\alpha T}{\sqrt{r^2(1-T^2)+\alpha^2}}\right)+c_2.
\ee
These rotationally symmetric solutions have minima at $r=0$ 
\be
\tau_{min}=\frac{1}{\alpha}\tanh^{-1}(T)+O(\epsilon),
\ee
and interpolate between $\tau=0$ surface for $T=0$ and a surface with minimum at $\tau=-\infty$ (and $r=0$) for $T\to -1$. For $\alpha=1$ we recover the global $AdS_3$ solution.

For the  Euclidean complexity $\mathcal{C}^{(e)}_T$, the on-shell action without the tension term is equal to 
\be
I_G=-\frac{V_xL_\tau}{\kappa^2\epsilon^2}+\frac{V_x}{\kappa^2}\left(\frac{|T|}{\sqrt{1-T^2}\epsilon}-\frac{\alpha |T|}{1-T^2}-\alpha\tanh^{-1}(T)\right)+O(\epsilon),
\ee
while the modified Hayward term is given by
\be
I'_H=\frac{1}{\kappa^2}\int\sqrt{\gamma}\theta_0=\frac{V_x}{\kappa^2\epsilon}\theta_0,\qquad \theta_0=\sin^{-1}\left(\sqrt{1-\frac{T^2}{1+\alpha^2\epsilon^2}}\right).
\ee
Adding these contributions together, we arrive at
\be
\mathcal{C}^{(e)}_T=-\frac{V_xL_\tau}{\kappa^2\epsilon^2}+\frac{V_x}{\kappa^2\epsilon}\left(\theta_0+\frac{|T|}{\sqrt{1-T^2}}\right)-\frac{V_x\alpha}{\kappa^2}\left(\frac{ |T|}{1-T^2}+\tanh^{-1}(T)\right)+O(\epsilon).
\ee
This action is monotonic and has a minimum at $T=0$ so the $\tau=0$ slices of \eqref{EXStM} have the lowest holographic path-integral complexity. The minimal value is given by
\be
\mathcal{C}^{(e)}_{T=0}=-\frac{V_xL_\tau}{\kappa^2\epsilon^2}+\frac{V_x}{\kappa^2}\frac{\pi}{2\epsilon}+O(\epsilon).
\ee
Again the $\epsilon^{-2}$ divergence originates from the choice of the ``reference metric" at $\Sigma$.
\subsubsection{Thermofield double}
\begin{figure}[t!]
\centering
\begin{minipage}{.55\textwidth}
  \centering
  \includegraphics[width=.8\linewidth]{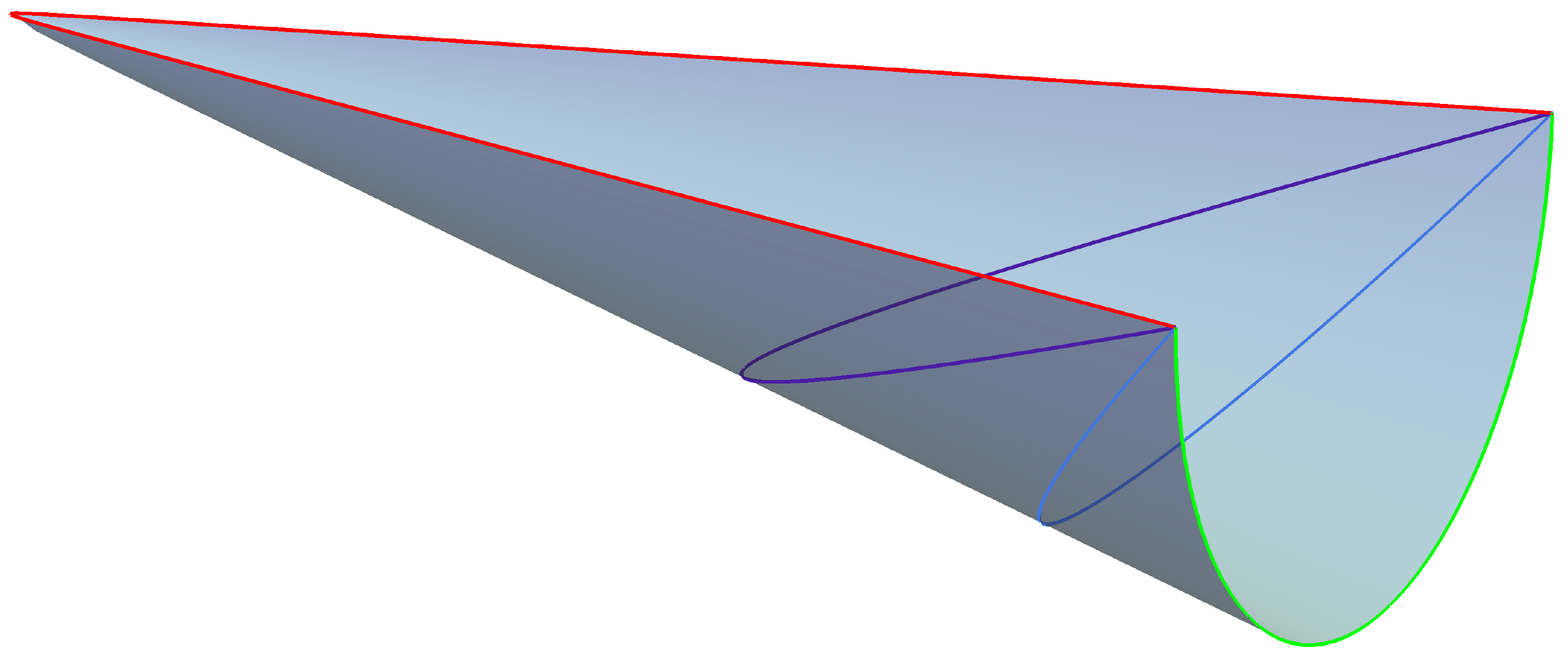}
\end{minipage}%
\begin{minipage}{.5\textwidth}
  \centering
  \includegraphics[width=.6\linewidth]{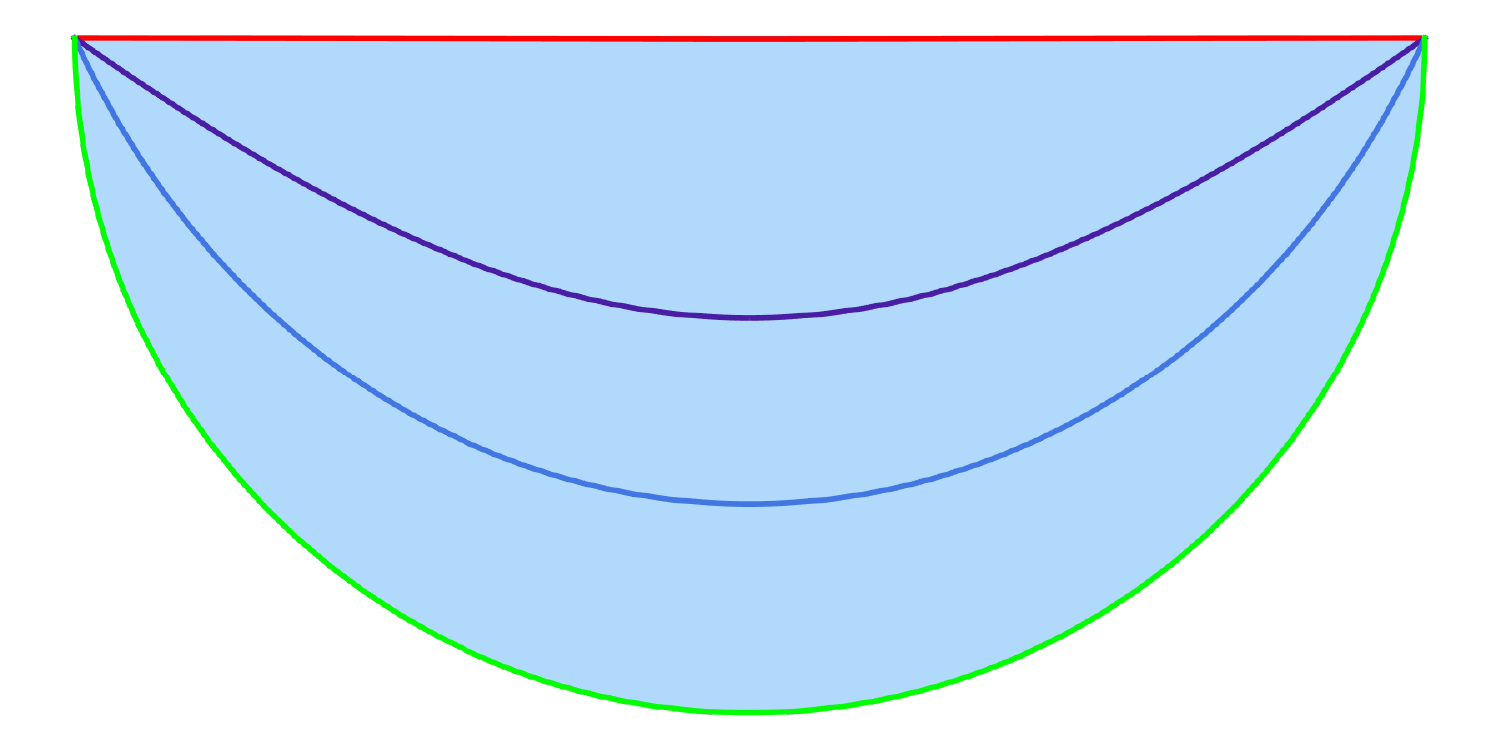}
\end{minipage}
\caption{Solutions \eqref{fTFDt} shown on the cigar geometry (left) and its projection (right) for a few sample values of $T$ interpolating between $T=0$ (red) and $T=-1$ (green). The semi-circles have the range $\tau\in [-\beta/4,\beta/4]$.}
  \label{TFDfigs}
\end{figure}

Next, we can use the solution \eqref{solEx} to analyze the gravity dual of the path-integral optimization for the TFD  state
\be
|\Psi\rangle_{TFD}=\sum_n e^{-\frac{\beta}{2}E_n}|E_n\rangle |E_n\rangle,
\ee
studied in \cite{Caputa:2017urj}. We take $r_h=2\pi/\beta$ and Euclidean time between $\tau\in (-\beta/4,\beta/4)$, while imposing the usual boundary conditions  \eqref{bccg} at both ends. If we then take
\be
w\in\left[-\frac{\beta}{4}+\frac{\epsilon}{\sqrt{1-T^2}},\frac{\beta}{4}-\frac{\epsilon}{\sqrt{1-T^2}}\right],
\ee
we can write the solution as
\be
e^{2\phi}=\frac{4\pi^2}{\left(1-T^2\right)\beta^2\cos^2\left(\frac{2\pi w}{\beta}\right)},
\ee
and it clearly satisfies the boundary condition at each end. This solution is equivalent to the optimized metric from the path-integrals preparing TFD state in 2d CFTs \cite{Caputa:2017urj}\footnote{Note that analogous analysis could be done for the pure state $|\Psi\rangle =e^{-\beta/4 H}|B\rangle$ with $\tau\in[0,\beta/4)$ and $|B\rangle$ representing a boundary state in CFT.}\\
Using \eqref{eq:EuclideanBTZ-w'}, we can derive $f(\tau)$ and surfaces $Q$, plotted on Fig. \ref{TFDfigs}, are given by the embedding function
\be
f(\tau)= \frac{\sqrt{1-T^2}}{r_h\sqrt{1+ T^2 \tan^2\left(r_h\tau\right)}},\label{fTFDt}
\ee
where the range of $\tau$ is now
\be
\tau\in\left[-\frac{\beta}{4}-c_2,\frac{\beta}{4}+c_2\right],
\ee
with cut-off
\be
c_2\simeq \frac{\epsilon T}{\sqrt{1-T^2}}+O(\epsilon).
\ee
Again in the analysis we take $\epsilon\to0$ first.\\
In order to estimate the complexity $\mathcal{C}^{(e)}_T$, we first compute the on-shell action without tension term
\be
I_G=- \frac{V_x \beta}{2\kappa^2 \epsilon^2} + 
\frac{V_x}{\kappa^2} \frac{2 |T|}{\sqrt{1-T^2}\epsilon}
+ O(\epsilon).
\ee
\\
Similarly, the modified Hayward terms from both corners are given by
\be
I'_H=\frac{1}{\kappa^2}\int_{\tau=-\beta/4} \sqrt{\gamma}\theta_0+\frac{1}{\kappa^2}\int_{\tau=\beta/4} \sqrt{\gamma}\theta_0=\frac{2 V_x}{\kappa^2 \epsilon} \theta_0,
\ee
where the inner angles are
\be
\theta_0=\sin^{-1}\left(\sqrt{1-\frac{T^2}{1-\frac{4\pi^2\epsilon^2}{\beta^2}}}\right).
\ee
Adding the two contributions gives the complexity for the holographic TFD setup
\be
\mathcal{C}^{(e)}_T= -\frac{V_x \beta}{2\kappa^2 \epsilon^2} + 
\frac{V_x}{\kappa^2\epsilon} \left(\frac{2 |T|}{\sqrt{1-T^2}}+2\theta_0\right)
+ O(\epsilon).
\ee
We can check that this expression is minimal for $T=0$ and again the $\epsilon^{-2}$ divergence comes from the ``reference metric" at $\Sigma$'s. The minimal value is given by
\be
\mathcal{C}^{(e)}_{T=0}=-\frac{V_x\beta}{2\kappa^2\epsilon^2}+\frac{V_x}{\kappa^2}\frac{\pi}{\epsilon}.
\ee
Note that this result has a similar structure to the vacuum example in Poincare $AdS_3$ and the only dependence on the temperature enters via the range of $\tau$: $L_{\tau}=\beta/2$. This is a similar behaviour to holographic computations of complexity in three dimensions \cite{Susskind,Chapman:2016hwi} and reflects the fact that in three dimensions all of the above metrics are locally $AdS_3$ (see also App. \ref{MapsAdS3} for more details on three-dimensional solutions). We will see that in higher dimensions, black holes lead to a more involved result for holographic path-integral complexity with respect to the vacuum state.
\subsection{Planar Black Holes in Higher Dimensions}
Next, we can study an interesting generalization of the above computations to the Euclidean planar black-hole metric in $d+1$ dimensions
\be
ds^2=\frac{1}{z^2}\left((1-Mz^{d})d\tau^2+\frac{dz^2}{1-M z^{d}}+dx^2_i\right),\label{HDBH}
\ee
where $M^{1/d}=(4\pi)/(d\,\beta)$ and the Euclidean time is periodic $\tau\in [0,\beta]$ such that when we consider the wave function for the TFD state, we focus on $\tau\in[-\frac{\beta}{4},\frac{\beta}{4}]$.\\
In the following, we compute the semi-classical Hartle-Hawking wave function \eqref{HHwv} using the usual algorithm.
For the asymptotic boundary surface $\Sigma$ at $z=\epsilon$  we have the trace of the extrinsic curvature in this geometry
\be
K|_\Sigma=\frac{d\left(2-M\epsilon^{d}\right)}{2\sqrt{1-M\epsilon^{d}}}.
\ee
On the other hand, for surface $Q$ defined by the embedding function $z=f(\tau)$ the induced metric is
\be
ds^2=\frac{1}{f(\tau)^{2}}\left[\left(1-Mf^{d}+\frac{f'(\tau)^2}{1-Mf^{d}}\right)d\tau^2+dx^2_i\right]\equiv e^{2\phi}\left(dw^2+d^2x_i\right),
\ee
where, as before for the TFD in two dimensions, we introduced the conformal coordinate $w$ satisfying
\be
w'(\tau)=\sqrt{1-Mf^{d}+\frac{f'(\tau)^2}{1-Mf^{d}}}=\frac{1-Mf^{d}}{\sqrt{1-Mf^{d}-\dot{f}(w)^2}},
\ee
and its range will be specified below.\\
The trace of the extrinsic curvature on $Q$ is now expressed in terms of $f(\tau)$ as
\be
K|_Q=-\frac{1}{w'^3}\left[ff''+\frac{d(2+Mf^{d})}{2(1-Mf^{d})}f'^2+\frac{d}{2}\left(1-Mf^{d}\right)(2-Mf^{d})\right],
\ee
or in terms of $\phi(w)=-\log(f(w))$ we get a generalization of the previous expressions
\be
K|_Q=\frac{e^{-2\phi}\left(\ddot{\phi}+(d-1)\dot{\phi}^2\right)-d\left(1-\frac{1}{2}Me^{-d\phi}\right)}{\sqrt{1-Me^{-d\phi}-\dot{\phi}^2e^{-2\phi}}}.\label{KQBHHD}
\ee
Then the on-shell action that computes the Hartle-Hawking wave function  \eqref{HHwv} takes the standard form
\bea
I_G+I_T=-\frac{(d-1)V_x L_\tau}{\kappa^2}\left(\frac{1}{\epsilon^{d}}-\frac{dM}{2(d-1)}\right)-\frac{V_x}{\kappa^2}\int \frac{d\tau}{f^d}w' (\tau)\left[K|_Q+\frac{1}{w'(\tau)}-T\right].
\eea
Finally, rewriting it in terms of $w$ and $\phi(w)$ and integrating by parts we arrive at the main result for higher-dimensional black holes \eqref{HDBH}
\bea
I_G+I_T&=&-\frac{(d-1)V_x L_\tau}{\kappa^2}\left(\frac{1}{\epsilon^{d}}-\frac{dM}{2(d-1)}\right)+\frac{(d-1)V_x}{\kappa^2}\int dw e^{d\phi}G(\dot{\phi})+\frac{V_x}{\kappa^2}\int dw e^{d\phi}T\nn\\
&-&\frac{V_x}{\kappa^2}\left[e^{(d-1)\phi}\arcsin\left(\frac{\dot{\phi}e^{-\phi}}{\sqrt{1-Me^{-d\phi}}}\right)\right]_{bdr},
\eea
with a closed formula
\be
G(\dot{\phi})=\frac{1-\frac{d}{2(d-1)}Me^{-d\phi}}{1-Me^{-d\phi}}\sqrt{1-Me^{-d\phi}-\dot{\phi}^2e^{-2\phi}}+\dot{\phi}e^{-\phi}\arcsin\left(\frac{\dot{\phi}e^{-\phi}}{\sqrt{1-Me^{-d\phi}}}\right).
\ee
This generalizes the result from $d=2$ and we can see that only for three-dimensional geometry the pre-factor of the square-root trivializes leading to a simpler equation of motion.\\
Indeed, the variation with respect to $\phi$ that gives the maximization constraint can still be written in a compact form as
\be
\frac{d}{d-1}T-K|_Q=\frac{(d-2)Me^{-d\phi}\left(2e^{-2\phi}(\dot{\phi}^2-\ddot{\phi})+dMe^{-d\phi}\right)}{4(d-1)\left(1-Me^{-d\phi}-\dot{\phi}^2e^{-2\phi}\right)^{3/2}},
\ee
where $K|_Q$ is given by \eqref{KQBHHD}. For $d=2$ the right hand side vanishes so that we reproduce the condition for the CMC slice. Naively, we can expect that CMC slices will no longer maximize the Hartle-Hawking wave functions in this higher-dimensional background. This conclusion is too quick as we can see from analyzing the full Neumann boundary condition.\\
Indeed, we can again consider the Neumann boundary condition \eqref{bdycond} (rewritten  in terms of $\phi(w)$) that in this setup gives two constraints. First, the $(\tau,\tau)$ components 
\be
h_{\tau\tau}\left(T+(d-1)\sqrt{1-Me^{-d\phi}-\dot{\phi}^2e^{-2\phi}}\right)=0,
\ee
and the $(a,b)$ components yield
\be
h_{ab}\left[T-\sqrt{1-Me^{-d\phi}-\dot{\phi}^2e^{-2\phi}}-K|_Q\right]=0.
\ee
If we solve the first one, then the second is equivalent to the trace CMC condition $K|_Q=\frac{d}{d-1}T$. Solving the first constraint is equivalent to the equation
\be
\dot{\phi}^2+Me^{-(d-2)\phi}=\left(1-\frac{T^2}{(d-1)^2}\right)e^{2\phi},\label{BHhig}
\ee
and for $T<0$, $\dot{\phi}$ also solves the second equation.\\
To solve \eqref{BHhig}, for convenience, we first shift
\bea
\phi\to \tilde{\phi}-\frac{1}{2}\log\left(1-\frac{T^2}{(d-1)^2}\right),\qquad \tilde{M}_d=M\left(1-\frac{T^2}{(d-1)^2}\right)^{\frac{d-2}{2}},
\eea
to absorb the constant on the right-hand side and write the equation in terms of $\tilde{\phi}=-\log(F(w))$ as
\be
F'^2+\tilde{M}_d F^{d}-1=0.
\ee
This equation can be solved by standard methods and requires inverting the integral
\be
w+c=\int^F\frac{dF}{\sqrt{1-\tilde{M}_d F^{d}}}=F\,_2F_1\left(\frac{1}{2},\frac{1}{d},\frac{d+1}{d},\tilde{M}_d F^d\right),
\ee
in order to find $F(w)$. This way, in $d=2$ we recover the previous result with $\sin$ function \eqref{solEx}.\\ Interestingly, for $d=4$, we can also obtain analytic result in terms of Jacobi sine Elliptic function $sn(x,-1)$. Namely, for $-3<T<0$, we find the solution of the Neumann boundary condition and $K|_Q=4T/3$ given by
\be
e^{2\phi(w)}=\frac{\sqrt{M}}{\sqrt{1-\frac{T^2}{9}}sn^2\left(M^{1/4}\left(1-\frac{T^2}{9}\right)^{1/4}w,-1\right)}.
\ee
This metric is a new result that on top of constant $K|_Q$ has a constant Ricci scalar \eqref{Riccid} for $d=4$ hence, after the map \eqref{mutr}, it is also a solution of the four dimensional path-integral optimization problem \eqref{HDOE} for the TFD state.\\ 
Let us now fix the appropriate range of $w$. Since the real periodicity of $sn(x,m)$ is $4K(m)$, we can consider the 4d counterpart of the previous TFD solution by taking $w$ in the interval
\be
w\in\left[c_1,\frac{2K(-1)}{M^{1/4}\left(1-\frac{T^2}{9}\right)^{1/4}}-c_1\right],
\ee
where $K(-1)=K(m=-1)$ is a numerical constant and $c_1$ is expressed in terms of the inverse Jacobi $sn^{-1}$ as
\be
c_1=\frac{1}{M^{1/4}\left(1-\frac{T^2}{9}\right)^{1/4}}sn^{-1}\left(\frac{M^{1/4}\epsilon}{\left(1-\frac{T^2}{9}\right)^{1/4}},-1\right)\simeq\frac{\epsilon}{\sqrt{1-\frac{T^2}{9}}}+O(\epsilon^5),
\ee
so that at the end points we satisfy the boundary condition $f(w)=\epsilon$.\\
Finally, this analytic result allows us to evaluate the holographic path-integral complexity from the Hartle-Hawking wave function in this five-dimensional black-hole geometry. The on-shell action without the tension term can now be computed as
\bea
I_G&=&-\frac{3V_x L_\tau}{\kappa^2}\left(\frac{1}{\epsilon^{4}}-\frac{2M}{3}\right)-\frac{V_x}{\kappa^2}\int dw e^{4\phi}\left[\frac{4}{3}T+\frac{1}{w'(\tau)}\right],\nn\\
&=&-\frac{3V_x L_\tau}{\kappa^2}\left(\frac{1}{\epsilon^{4}}-\frac{2M}{3}\right)+\frac{2V_x}{3\kappa^2}\left[\frac{|T|}{\sqrt{1-\frac{T^2}{9}}\epsilon^3}+\frac{|T|\,M^{3/4}K(-1)}{\left(1-\frac{T^2}{9}\right)^{5/4}}\right]+O(\epsilon),
\eea
where $L_\tau=\beta/2=\pi/(2M^{1/4})$.\\
Moreover, the modified Hayward term again consists of two equal contributions from each inner-angle
\be
I'_H=2\frac{V_x}{\kappa^2\epsilon^{3}}\theta_0,\qquad \theta_0=\sin^{-1}\left(\sqrt{1-\frac{T^2}{9\left(1-M\epsilon^4\right)}}\right).
\ee
Adding two contributions we can verify that the Euclidean complexity $\mathcal{C}^{(e)}_T$ is minimized for $T=0$. Its minimal value is given by
\be
\mathcal{C}^{(e)}_{T=0}=\frac{V_x}{\kappa^2}\left(-\frac{\beta}{2\epsilon^4}+\frac{\pi}{\epsilon^3}+\frac{\pi^4}{3\beta^3}\right)+O(\epsilon).
\ee
Interestingly, unlike in $d=2$, we obtain a constant ($\epsilon^0$), $\beta$-dependent term that increases the value of complexity for large temperatures. In fact, as also noted in \cite{Chapman:2016hwi}, the constant term is proportional to the entropy.\footnote{We thank the anonymous referee for pointing this to us.} \\
For other dimensions, inverting the hypergeometric function would have to be done numerically and we leave this analysis as an interesting future problem.\\
It is also useful to comment on the relation between the on-shell values of our Euclidean complexity actions and the complexity equals volume \cite{Susskind} computations. Since, on-shell, the extrinsic curvature is proportional to $T$, our complexity action contains the contribution from the pure volume of $Q$ as well as from the bulk and the Hayward term. However, once we further minimize over $T$, the contributions become more difficult to disentangle. Nevertheless, e.g. similarity with \cite{Chapman:2016hwi} may suggest that the final result contains (parts of) the CV answer and we leave determining these details and connections with holographic complexity conjectures as an interesting future problem.
\section{JT Gravity}\label{Sec:5 JT}
In this section we will test our construction in the context of Euclidean JT gravity dual to the SYK model \cite{SYKa,SYKb,SYKc,SYKd,Jensen:2016pah,Harlow:2018tqv}. Most of the computations presented above goes through in this two dimensional gravity model as well. However, it turns out that in order to naturally generalize our discussion, it is more advantageous to introduce the tension term on surface $Q$ by coupling to the dilaton\footnote{From the three dimensional origin of JT action such term naturally arises by adding a brane with tension in 3d in which $\sqrt{h}$ yields the dilaton field.} field $\Phi$, such that the action computing the Hartle-Hawking wave functions \eqref{HHwv} in JT will be given by
\begin{eqnarray}
I_{JT}+I_{T_\Phi}&=&-\Phi_0\chi(M)-\int_{M}\sqrt{g}\Phi(R+2)-2\int_{\partial M}\sqrt{h}\Phi K+2T_\Phi\int_Q \sqrt{h}\Phi, \label{JTAction}
\end{eqnarray}
with $\chi(M)$ being the Euler characteristic of the region $M$.\\
On-shell solutions of this theory (again we work in the probe limit and have in mind solutions of pure JT gravity without $T_\Phi$) have constant negative curvature $R=-2$ and dilaton equation of motion is set by the vanishing energy-momentum tensor. As an example solution, we can take the 2d Schwarzschild metric with dilaton profile
\be
ds^2=(r^2-r^2_h)d\tau^2+\frac{dr^2}{r^2-r^2_h},\qquad \Phi(r)=\Phi_b r.
\ee
We will take $r_h=2\pi/\beta$ and be interested in $\tau\in[-\beta/4,\beta/4]$.
By analogy to higher dimensions, we consider region $M$ bounded by  $\Sigma$ at $r=\frac{1}{\epsilon}\to \infty$ and $Q$ specified by $r=1/f(\tau)$. The induced metric on $Q$ becomes
\be
ds^2=e^{2\phi}dw^2,\qquad e^{\phi}=f^{-1},
\ee
where $w'$ is given by the same expression as in \eqref{eq:EuclideanBTZ-w'}.\\
Following the same steps as before, we can derive the on-shell action that computes the semi-classical Hartle-Hawking wave function
\bea
I_{JT}+I_{T_\Phi}=-\frac{2L_\tau\Phi_b}{\epsilon^2}-2\Phi_0\int dw e^\phi\left[ K|_Q+\frac{1}{w'}\right] -2\Phi_b\int dw e^{2\phi} \left[K|_Q-T_\Phi\right],
\eea
with $w'$ as in \eqref{eq:EuclideanBTZ-w'} and 
\be
K|_Q=\frac{\ddot{\phi}e^{-2\phi}-1}{\sqrt{1-e^{-2\phi}(\dot{\phi}^2+r^2_h)}}.\label{KQJTG}
\ee
Similarly, the action written in terms of $w$ and $\phi(w)$ in the first-derivative form can be obtained as
\bea
I_{JT}+I_{T_\Phi}&=&-\frac{2L_\tau\Phi_b}{\epsilon^2}+2\Phi_b\int_Q e^{2\phi}G(\dot{\phi})+2\Phi_b\int_Q e^{2\phi}T_\Phi\nn\\
&-&2\left[\left(\Phi_0+\Phi_be^\phi\right)\arcsin\left(\frac{\dot{\phi}e^{-\phi}}{\sqrt{1-r^2_he^{-2\phi}}}\right)\right]_{bdr},\label{JTHHA}
\eea
with monotonic a function of $\dot{\phi}$
\begin{eqnarray}
G(\dot{\phi})&=&\frac{\sqrt{1-e^{-2\phi}(\dot{\phi}^2+r^2_h)}}{1-r^2_he^{-2\phi}}
+\dot{\phi}e^{-\phi}\arcsin\left(\frac{\dot{\phi}e^{-\phi}}{\sqrt{1-r^2_he^{-2\phi}}}\right).
\end{eqnarray}
With the intuition from previous examples we can again consider the UV limit. Interestingly, for small $\dot{\phi}$ as well as small $r_h$, we get the following action in the leading order
\be
I_{JT}|_{UV}=\Phi_b\int\left(\dot{\phi}^2+2e^{2\phi}\right)+...
\ee
With the usual replacement of the reparametrization function $e^\phi \sim \tilde{w}'(w)$, this action precisely describes the Schwarzian sector of the SYK model (compare e.g. \cite{SYKd,Jensen:2016pah}).

The saddle point equation from the maximization of the wave-function is again equivalent to the Neumann boundary condition. More explicitly, varying the action with respect to $\phi$ yields the equation
\be
K|_Q-T_\Phi=-\frac{\left(1-\dot{\phi}^2e^{-2\phi}\right)\left(r^2_h+\left(\dot{\phi}^2-\ddot{\phi}\right)\right)e^{-2\phi}}{2\left(1-e^{-2\phi}(r^2_h+\dot{\phi}^2)\right)^{3/2}},\label{MaxEqJT}
\ee
where $K|_Q$ is given by \eqref{KQJTG}.\\
We can verify that for $-1<T_\Phi<0$, this equation is solved by
\be
e^{2\phi}=\frac{r^2_h}{(1-T^2_\Phi)\cos^2\left(r_hw\right)},
\ee
with
\be
w\in\left[-\frac{\beta}{4}+\frac{\epsilon}{\sqrt{1-T^2}},\frac{\beta}{4}-\frac{\epsilon}{\sqrt{1-T^2}}\right].
\ee
Similarly to the higher-dimensional planar black holes, the right hand side of \eqref{MaxEqJT} vanishes on-shell and the surface is again described by the CMC slice $K|_Q=T_\Phi$. This solution is a natural generalization of its higher dimensional TFD cousins.\\

Finally, we can compute holographic path-integral complexity in the Euclidean JT model. First, we evaluate the on-shell JT gravity action without tension $T_\Phi$ term 
\bea
I_{JT}&=&-\frac{\beta\Phi_b}{\epsilon^2}-2\Phi_0\int dw e^\phi\left[ T_\Phi+\frac{1}{w'}\right] -2\Phi_bT_\Phi\int dw e^{2\phi} ,\nn\\
&=&-\frac{\beta\Phi_b}{\epsilon^2}+\frac{4|T_\Phi|\Phi_b}{\sqrt{1-T^2_\Phi}\epsilon}-4\Phi_0\left(\pi-\arccos(T_\Phi)\right).
\eea
On the other hand, the internal angles $\theta_0$ that appear in the modified Hayward terms are now related to $T_\Phi$ by
\be
\theta_0=\arcsin\left(\sqrt{1-T^2_\Phi}\right).\label{t0JT}
\ee
 From the form of the JT action \eqref{JTAction}, it is then natural to generalize the modified Hayward term from section \ref{Sec:Hayward} to include the coupling to $\Phi_0$  as well as the dilaton $\Phi(r)$ at each corner\footnote{See also e.g. \cite{Harlow:2018tqv} where standard Hayward/corner terms in JT gravity were included in the same form and \cite{Iliesiu:2020zld} for related analysis.}
\be
I'_H=2\sum_i\theta^i_0\left(\Phi_0+\Phi(r)\right)|_{C_i}=2\left(\Phi_0+\frac{\Phi_b}{\epsilon}\right)2\theta_0,
\ee
where $C_i$ stands for the $i$-th corner in the wedge region $M$. In fact this term plays the same role as its higher-dimensional counterparts and, when added to the JT action, cancels the boundary term in \eqref{JTHHA}.\\
This way, the holographic path-integral complexity action in JT model becomes
\bea
\mathcal{C}^{(e)}_T&=&-\frac{\beta\Phi_b}{\epsilon^2}+\frac{2\Phi_b}{\epsilon}\left(\frac{2|T_\Phi|}{\sqrt{1-T^2_\Phi}}+2\theta_0\right)\nn\\
&-&4\Phi_0\left(\pi-\cos^{-1}(T_\Phi)-\theta_0\right).
\eea
We can then check that after inserting \eqref{t0JT} the second line of this expression, as well as dependence on $\Phi_0$, vanishes. Therefore, analogously to the TFD example in the previous sections, the complexity is mimimized at $T_\Phi=0$. The minimum value is given by
\be
\mathcal{C}^{(e)}_{T=0}=-\frac{\beta\Phi_b}{\epsilon^2}+\frac{2\pi\Phi_b}{\epsilon}.
\ee
With this example we finish the analysis of the Euclidean Hartle-Hawking wave functions and confirm our procedure, including the estimation of holographic path-integral complexity for $T\neq 0$ slices, across dimensions. In the next section, we will proceed to analyze gravitational path-integrals in Lorentzian geometries.
\section{Lorentzian Spacetimes}\label{Sec:6 Lor}
The main important motivation for generalizing the above construction to Lorentzian signature has to do with understanding how general, time-dependent, holographic geometries emerge from CFT. An interesting clue in this direction was given in  \cite{Takayanagi:2018pml} that proposed to interpret general slices of holographic spacetimes as different quantum circuits constructed from path-integrals on these geometries (Euclidean or Lorentzian) that prepare quantum states in CFTs (see also \cite{Milsted:2018yur,Milsted:2018san}). Nevertheless, in order to test this proposal, the optimization of Lorentzian path-integrals in CFTs requires a much better understanding. In this section, with the gravitational construction at hand, we will first perform the Lorentzian computations in $AdS$ and then use them to shed some top-down light on how one could advance the optimization of path-integrals in Lorentzian CFTs.

Moreover, the advantage of the Hartle-Hawking approach is that we can apply it also to gravity on a de Sitter (dS) spacetime. Up to date, there have been various attempts to formulate holographic duals of de Sitter gravity, the so called dS/CFT correspondence \cite{Strominger:2001pn,Maldacena:2002vr}, in which the dual CFTs are usually argued to be associated with the future spacelike boundary (refer also to other approaches e.g. \cite{MT,Dong:2018cuv}). However, comparing AdS/CFT, understanding of dS/CFT holography has been highly limited and it is fair to say that the validity of the holographic duality with $dS$ itself is under debate at present. Therefore, working under the assumption that the dual theory is associated with the spacelike boundary region $\Sigma$, we hope that the Hartle-Hawking approach may provide a new way to probe how holography on de Sitter spaces looks like. In this section, we will also analyze Hartle-Hawking wave functions in de Sitter spacetimes and extract new lessons for possible path-integral optimization in dS/CFT.
 
Conceptually, the standard Lorentzian path-integral in gravity is computing transition amplitudes between two geometries: some reference metric $g^{(1)}$ and a different metric $g^{(2)}$ (see e.g. \cite{Loll:2019rdj} for interesting applications in quantum gravity). If we naturally extend our proposal to Lorentzian signature, the gravitational path-integral will be evaluating transition amplitudes between a geometry on a timelike boundary  $\Sigma$ of $AdS_{d+1}$ (or spacelike boundary $\Sigma$ for $dS_{d+1}$) and a timelike or spacelike surface $Q$ with tension $T$ up to which we integrate the bulk gravity action
\be
\Psi^{(T)}_{HH}[g^{(2)},g^{(1)}]=\langle g^{(2)}|g^{(1)}\rangle^{(T)}=\int [Dg_{\mu\nu}] e^{i\,(I_G+I_T)}\delta(g|_Q-g^{(2)})\delta(g|_\Sigma-g^{(1)}),
\ee
where $I_G$ is the Lorentzian Einstein-Hilbert action with the Gibbons-Hawking term and $I_T$ is the tension term on $Q$. With a slight abuse of language, we will still refer to these transition amplitudes as Hartle-Hawking wave functions.

To make progress, we will again only be able to compute the Lorentzian Hartle-Hawking wave function semi-classically by the gravity on-shell action with the tension term. Namely, after introducing the conformal coordinate $w$ that will put the metric on $Q$ (specified by embedding homogeneous function $f(t)$) in the Weyl Minkowski (for timelike $Q$) or Weyl flat (for spacelike $Q$) form, we will evaluate the Hartle-Hawking wave functional of the Weyl factor $\phi(w)=-\log(f(w))$ as
\be
\Psi^{T}_{HH}[\phi]\sim e^{i\left(I_G+I_{T}\right)}|_{on-shell}.\label{HHwvL}
\ee
Then, the ``maximization" will be performed by finding the extremum of the classical action with respect to $\phi$ with analogous boundary condition to \eqref{bccg}.\\
Last but not least, we propose a generalization of the holographic path-integral complexity from Hartle-Hawking wave functions to Lorentzian signature given by
\be
\mathcal{C}^{(l)}_{T}=-\left(I_G+I'_H\right),  \label{cltlo}
\ee
where $I_G$ is the Lorentzian gravity action without the tension term and $I'_H$ stands for an appropriate modified Lorentzian Hayward term that, as in the Euclidean case (\ref{euclcomp}), will be careful computed for spacelike and timelike surfaces $Q$ below. The overall minus sign in (\ref{cltlo}) is chosen such that it is consistent with the holographic analysis in \cite{Takayanagi:2018pml}.

Before we proceed with explicit examples, let us also review the argument with the Hamiltonian constraint in gravity and the curvature of the slices that maximize the Hartle-Hawking wave function. In Lorentzian signature, we will be interested in slices of solutions to Einstein's equations in $d+1$ dimensions with negative ($AdS_{d+1}$) or positive $(dS_{d+1})$ cosmological constants
\be
G_{\mu\nu}=-\Lambda^{(d+1)}_{AdS/dS}g_{\mu\nu}.
\ee
We will then consider arbitrary non-null slices $Q$ with normal vector $n^{\mu}$ such that $n_{\mu}n^{\mu}=\varepsilon$, where $\varepsilon=+1$ corresponds to timelike and $\varepsilon=-1$ to spacelike slices respectively. We can again project Einstein's equations to the normal components and derive a local ``Hamiltonian" constraint (see e.g. \cite{Poisson})
\be
\varepsilon\left(K^2-K^{ij}K_{ij}\right)=R^{(d)}-2\Lambda^{(d+1)}_{AdS/dS}.
\ee
As will see below, the maximization of the Hartle-Hawking wave functions will again be equivalent to imposing Neumann boundary condition on $Q$, that together with its trace being constant, implies \eqref{LHS}. This way, from the Hamiltonian constraint, we can derive that Ricci scalar curvature of slices $Q$ is constant and given by
\be
R^{(d)}=2\Lambda^{(d+1)}_{AdS/dS}+\varepsilon \frac{d}{d-1}T^2.\label{HCL}
\ee 
We will verify this formula in the examples below and indeed see that timelike CMC slices of both $AdS_{d+1}$ and $dS_{d+1}$ are positively curved whereas spacelike CMC slices in these geometries have constant negative curvatures. We will then speculate on possible extensions of the path-integral optimization to Lorentzian CFTs that could give rise to continuous tensor networks with both of these constant curvatures.

\subsection{Semi-classical wave functions in $AdS_{d+1}$}
In this section we evaluate the semi-classical Hartle-Hawking wave function \eqref{HHwvL} in $AdS_{d+1}$ using the on-shell Lorentzian action
\be
I_G+I_T=\frac{1}{2\kappa^2}\int_{M} \sqrt{-g}\left(R-2\Lambda\right)+\frac{1}{\kappa^2}\int_{Q} \sqrt{|h|}(K-T)+\frac{1}{\kappa^2}\int_{\Sigma} \sqrt{|h|}K.\label{ActL}
\ee
For simplicity, we analyze the $d+1$-dimensional $AdS$ in Poincare coordinates dual to the vacuum of Lorentzian $CFT_d$
\be
ds^2=l^2\frac{dz^2-dt^2+dx^2_i}{z^2},\label{AdSdLP1}
\ee
where, in order to compare the results with those in de-Sitter spacetime in the next section, we keep the radius $l$ of $AdS_{d+1}$ explicit.

As in the Euclidean computation, we focus on the asymptotic region $M$ of $AdS_{d+1}$ between $\Sigma$: at $z=\epsilon$, that is timelike, and a spacelike or timelike surface $Q$ in the bulk defined by radial coordinate as a function of the boundary time $t$: $z=f(t)$. In M, we compute the bulk action together with boundary terms and tension \eqref{ActL} that will give the Lorentzian semi-classical Hartle-Hawking wave function $\Psi^{(T)}_{HH}[\phi]$ in \eqref{HHwvL}.\footnote{Again we keep the boundary condition on $\Sigma$ implicit.} 

To evaluate the result, we will need the trace of the extrinsic curvature at $\Sigma$ given by 
\be 
K|_\Sigma=\frac{d}{l}.
\ee
Then, we will have to pay a special attention to the signature of the induced metric on $Q$. More precisely, the induced metric on $Q$: $z=f(t)$ is given by
\be
ds^2=l^2\frac{-(1-f'(t)^2)dt^2+dx^2_i}{f^2(t)},\label{IndMQLA}
\ee
and in what follows, we will distinguish two cases: timelike when $|f'(t)|<1$ and spacelike when $|f'(t)|>1$, and analyze them separately below.
\subsubsection{Timelike $Q$}
In the timelike case with $|f'(t)|<1$, we write the metric \eqref{IndMQLA} on Q as
\be
ds^2=\frac{l^2}{f(w)^2}\left(-dw^2+dx^2_i\right)\equiv e^{2\phi(w)}\left(-dw^2+dx^2_i\right),
\ee
where we defined coordinate $w$ with the range $-\infty<w\le0$, related to $t$ by
\be
w'(t)=\sqrt{1-f'(t)^2}=\frac{1}{\sqrt{1+\dot{f}(w)^2}}.\label{wpAdSLtLQ}
\ee
The trace of the extrinsic curvature on timelike $Q$ is then given by
\be
K|_Q=\frac{f(t)f''(t)-d(1-f'(t)^2)}{l(1-f'(t)^2)^{3/2}},
\ee
or in terms of $w$ and $\phi(w)$
\be
K|_Q=-\frac{le^{-2\phi}(\ddot{\phi}+(d-1)\dot{\phi}^2)+\frac{d}{l}}{\sqrt{1+l^2\dot{\phi}^2e^{-2\phi}}}.\label{KLAdSP}
\ee
With these ingredients, the Lorentzian on-shell action that computes the Hartle-Hawking wave function (``transition amplitude")\eqref{HHwvL} between timelike $\Sigma$ and timelike $Q$ can be written as
\be
I_G+I_T=\frac{(d-1)l^{d-1}}{\kappa^2}\frac{V_x L_t}{\epsilon^{d}}+\frac{V_xl^d}{\kappa^2}\int \frac{dt}{f(t)^{d}}w'(t)(K|_Q+\frac{1}{lw'(t)}-T),
\ee
with all the expressions spelled in terms of $t$ and $f(t)$ above and $V_x$ and $L_t$ defined as in the Euclidean examples. We can again rewrite this action in terms of $w$ and $\phi(w)$ as
\bea
I_G+I_T=\frac{(d-1)l^{d-1}}{\kappa^2}\frac{V_x L_t}{\epsilon^{d}}+\frac{V_x}{\kappa^2}\int dwe^{d\phi(w)}\left[-\frac{le^{-2\phi}(\ddot{\phi}+(d-2)\dot{\phi}^2)+\frac{d-1}{l}}{\sqrt{1+l^2\dot{\phi}^2e^{-2\phi}}}-T\right].
\eea
Integrating by parts, allows to write the action in the first derivative form and we arrive at 
\bea
I_G+I_T&=&\frac{(d-1)l^{d-1}}{\kappa^2}\frac{V_x L_t}{\epsilon^{d}}+\frac{(d-1)V_x}{l\kappa^2}\int dwe^{d\phi(w)}G(\dot{\phi})-\frac{V_x}{\kappa^2}\int dwe^{d\phi(w)}T\nn\\
&-&\frac{V_x}{\kappa^2}\left[e^{(d-1)\phi}\sinh^{-1}\left(l\dot{\phi}e^{-\phi}\right)\right]^0_{-\infty},\label{IGTLAdS}
\eea
with
\be
G(\dot{\phi})=l\dot{\phi}e^{-\phi}\sinh^{-1}\left(l\dot{\phi}e^{-\phi}\right)-\sqrt{1+l^2\dot{\phi}^2e^{-2\phi}}.
\ee
This is the main result for timelike surfaces $Q$ in Lorentzian $AdS_{d+1}$.

\begin{figure}[b!]
  \centering
  \includegraphics[width=8cm]{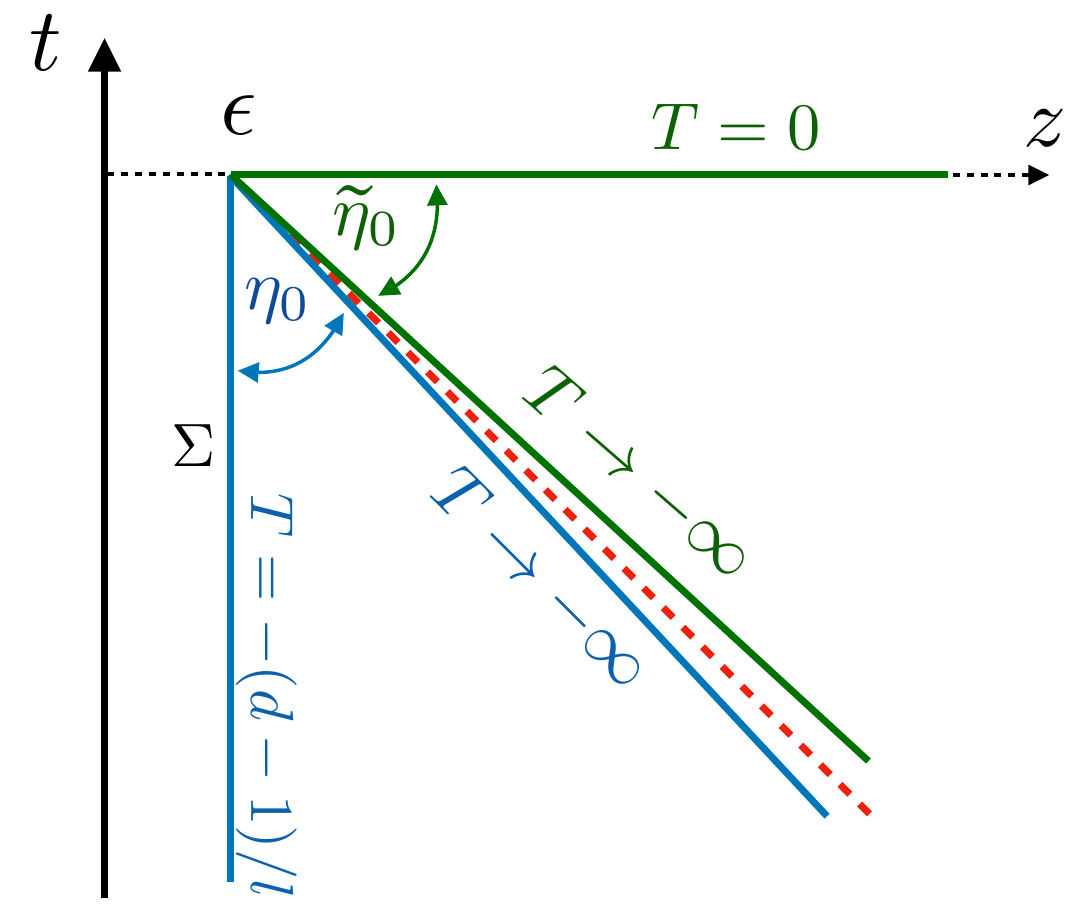}
  \caption{Solutions of the maximization in Lorentzian $AdS_{d+1}$.  Timelike half-planes \eqref{AdSLTL}, shown in blue, interpolate between the boundary at $T=-(d-1)/l$ and null plane, dashed red,  in the limit of $T\to-\infty$. The counterpart of the inner angle (``rapidity") $\eta_0$ given by \eqref{etaTLADS} is shown in blue. Spacelike half planes \eqref{AdSLSL}, shown in green, interpolate between the null sheet for $T\to-\infty$ and the $t=0$ slice at $T=0$. Their inner angle $\tilde{\eta}_0$, given by \eqref{etaSLADS}, is shown in green.}
\label{SolLorAdS}
\end{figure}

To gain some intuition on what to expect from the Lorentzian path-integral optimization, we can again use this result to extract the CFT action in the UV limit. Expanding the answer for small $\dot{\phi}$ we derive 
\be
e^{d\phi}G(\dot{\phi})\simeq \frac{1}{2}l^2e^{(d-2)\phi}\left(\dot{\phi}^2-\frac{2}{l^2}e^{2\phi}+O(\dot{\phi}^4)\right).
\ee
In two-dimensions, this limit reproduces the standard Lorentzian Liouville action $S^{(l)}_{L}$, which takes the following form\footnote{If we include the tension term also, then  $I_G+I_T$ gives the cosmological constant $\mu=2+\frac{2T}{d-1}$, where we set $l=1$. Note that for $T<-(d-1)$ which is assumed in the present case, we find $\mu<0$. This is 
the reason why we can have the dS$_d$ solutions with positive Ricci scalar. However for the evaluation of complexity (\ref{cltlo}) we always need to remove the tension term contribution as we did in Euclidean case, which leads to $\mu=2$.}
 \ba
S^{(l)}_L=\frac{c}{24\pi}\int d^2x\left[(\de_t\phi)^2-(\de_x\phi)^2-\mu e^{2\phi}\right]. \label{ILact}
\ea
The Euclidean continuation of this action is the standard Liouville action (\ref{LVac}). We will come back to this action at the end of this section.\\
The equation of motion from the maximization of the Hartle-Hawking wave function with \eqref{IGTLAdS} becomes
\be
-\frac{le^{-2\phi}(\ddot{\phi}+(d-1)\dot{\phi}^2)+\frac{d}{l}}{\sqrt{1+l^2\dot{\phi}^2e^{-2\phi}}}=\frac{d}{d-1}T.
\ee
Comparing with \eqref{KLAdSP}, this is again the constraint for the CMC slices in Lorentzian geometry \eqref{AdSdLP1}
\be
K|_Q=\frac{d}{d-1}T.
\ee
The solution with correct boundary condition, $f(0)=\epsilon$, is obtained for the negative range of the tension parameter $-\infty<T<-(d-1)/l$ and yields the timelike metric on $Q$
\be
ds^2=\frac{l^2}{\left(-\sqrt{\frac{l^2T^2}{(d-1)^2}-1}w+\epsilon\right)^2}\left(-dw^2+dx^2_i\right).\label{SolutionAdST}
\ee
These are the Lorentzian d-dimensional  de Sitter slices of $AdS_{d+1}$ geometry \eqref{AdSdLP1}. Indeed, we can verify that the Ricci scalar of this metric is constant positive
\be
R^{(d)}=\frac{d(d-1)}{l^2}\left(\frac{l^2T^2}{(d-1)^2}-1\right)=-2\Lambda^{(d+1)}_{AdS}\left(\frac{l^2T^2}{(d-1)^2}-1\right),
\ee
which is consistent with the general result \eqref{HCL} from the gravity Hamiltonian constraint.\\
After integrating \eqref{wpAdSLtLQ}, we can find the embedding function $f(t)$ of the timelike $Q$ in original time coordinate
\be
z=f(t)=\sqrt{\frac{l^2T^2}{(d-1)^2}-1}\frac{d-1}{lT}t+\epsilon.\label{AdSLTL}
\ee
These are half-planes that interpolate between the timelike boundary $\Sigma$ for $T\to-(d-1)/l$ and the null plane in the limit $T\to -\infty$, see blue lines on Fig. \ref{SolLorAdS}. The inner angle (i.e. ``rapidity") between $\Sigma$ and $Q$ is denoted by $\eta_0$ in blue. We can also verify that they satisfy the full Neumann boundary condition.
\\

Up to now, the above results are natural generalizations of the Euclidean derivation so in the next step, in analogy with section \ref{Sec:Hayward}, we would like to define the Lorentzian holographic path-integral complexity $\mathcal{C}^{(l)}_T$ and investigate which slices $Q$ are favoured by this candidate for a relative measure of complexity.

Similarly to the Euclidean $AdS_{d+1}$, the dependence on tension $T$ in the on-shell action with the tension term evaluated on \eqref{AdSLTL} cancels. Therefore, as before, in the definition of Lorentzian complexity we will only take gravity action (without tension term) evaluated on region $M$. For the solution \eqref{AdSLTL} the answer becomes
\be
I_G=\frac{(d-1)l^{d-1}}{\kappa^2}\frac{V_x L_t}{\epsilon^{d}}+\frac{V_xl^{d-1}}{\kappa^2\epsilon^{d-1}}\frac{ lT}{(d-1)\sqrt{\frac{l^2T^2}{(d-1)^2}-1}}.
\ee
Again, for the correct variational principle we should add an appropriate Hayward term in the definition of complexity. For consistency with the notion of relative complexity, we will again require that the Hayward term does not contribute to the complexity action when $Q\to \Sigma$. Moreover, mimicking the Euclidean construction, we will require that the Lorentzian modified Hayward term should cancel the boundary term in \eqref{IGTLAdS}. These two conditions uniquely single out the candidate in terms of  the ``inner" rapidity parameter $\eta_0$ defined by the two normal vectors $n_\Sigma$ and $n_Q$ (normal to $\Sigma$ and $Q$ respectively) at the AdS boundary follows
\be
I'_H=\frac{1}{\kappa^2}\int_\gamma\sqrt{\gamma}\,\eta_0,\qquad n_\Sigma \cdot n_Q=-\cosh\eta_0.
\ee
For the solution \eqref{AdSLTL}, this modified Hayward term is the following function of the tension parameter
\be
I'_H=\frac{V_xl^{d-1}}{\kappa^2 \epsilon^{d-1}}\eta_0,\qquad \sinh\eta_0=\sqrt{\frac{l^2T^2}{(d-1)^2}-1}.\label{etaTLADS}
\ee
Finally, with these two ingredients we define the Lorentzian holographic path-integral complexity as minus the sum of the two above contributions
\be
\mathcal{C}^{(l)}_{T}=-\left(I_G+I'_H\right).
\ee
The overall minus sign is fixed so that it is consistent with the previous holographic analysis in \cite{Takayanagi:2018pml}. 
We will test this proposal and its implications in the Lorentzian examples below.

In our first example, the Lorentzian holographic path-integral complexity becomes
\be
\mathcal{C}^{(l)}_{T}=-\frac{(d-1)l^{d-1}}{\kappa^2}\frac{V_x L_t}{\epsilon^{d}}-\frac{V_xl^{d-1}}{\kappa^2 \epsilon^{d-1}}\left(\eta_0-\coth\eta_0\right). \label{timeAdScompx}
\ee
Interestingly, this action is minimized in the limit $\eta_0\to\infty$ or $T\to-\infty$. This is the light sheet surface and, in fact, it was argued to be naturally related to geometry on which Lorentzian path-integral corresponds to MERA-type quantum circuits \cite{Milsted:2018san,Takayanagi:2018pml}. Notice also that this optimization limit is equivalent to maximizing the absolute value of the phase of the Hartle-Hawking wave function at zero tension $T=0$. This can be regarded as a Lorentzian counterpart of  the optimization in the Euclidean case, where we maximize the amplitude of the Hartle-Hawking wave function.\\
Observe that again we chose the modified Hayward term so that in the limit of $Q\to\Sigma$ action vanishes. This is now obtained in the limit
\be
\eta_0\to \coth^{-1}\left(\frac{(d-1)L_t}{\epsilon}\right),
\ee
that generalizes the Euclidean limit \eqref{ForLim}.

\subsubsection{Spacelike $Q$}
Now we briefly revisit the above analysis for the spacelike surfaces $Q$. In this case we have instead the spacelike induced metric on $Q$ given by
\be
ds^2=\frac{l^2}{f(w)^2}\left(dw^2+dx^2_i\right)\equiv e^{2\phi(w)}\left(dw^2+dx^2_i\right),
\ee
where we defined coordinate $w$ with range $-\infty<w<0$ that puts the induced metric in a conformally flat form and is expressed by relations
\be 
w'(t)=\sqrt{f'(t)^2-1}=\frac{1}{\sqrt{\dot{f}(w)^2-1}}.\label{wpAdSLSLQ}
\ee
The only new ingredient that we need to evaluate the action is the trace of the extrinsic curvature of $Q$. It is given  in terms of the embedding function as
\be
K|_Q=-\frac{f''(t) f(t)+d(f'(t)^2-1)}{l\left(f'(t)^2-1\right)^{3/2}},
\ee
or equivalently in terms of $w$ and $\phi(w)=\log(l/ f(w))$ as
\be
K|_Q=-\frac{le^{-2\phi}(\ddot{\phi}+(d-1)\dot{\phi}^2)-\frac{d}{l}}{\sqrt{l^2\dot{\phi}^2e^{-2\phi}-1}}.\label{KAdSSLKPQ}
\ee
The on-shell action that computes \eqref{HHwvL} is given in terms of w and $\phi(w)$ as
\bea
I_G+I_T=\frac{(d-1)l^{d-1}}{\kappa^2}\frac{V_x L_t}{\epsilon^{d}}-\frac{V_x}{\kappa^2}\int dw e^{d\phi(w)}\left[\frac{le^{-2\phi}(\ddot{\phi}+(d-2)\dot{\phi}^2)-\frac{d-1}{l}}{\sqrt{l^2\dot{\phi}^2e^{-2\phi}-1}}+T\right].\label{ALSL}
\eea
Finally, after integrating by parts, we derive the main result for spacelike $Q$
\bea
I_G+I_T&=&\frac{(d-1)l^{d-1}}{\kappa^2}\frac{V_x L_t}{\epsilon^{d}}-\frac{(d-1)V_x}{l\kappa^2}\int dwe^{d\phi(w)}G(\dot{\phi})-\frac{V_x}{\kappa^2}\int dwe^{d\phi(w)}T\nn\\
&-&\frac{V_x}{\kappa^2}\left[e^{(d-1)\phi}\sinh^{-1}\left(i\,l\dot{\phi}e^{-\phi}\right)\right]^0_{-\infty},\label{IGSLAdS}
\eea
with function $G(\dot{\phi})$ now given by
\be
G(\dot{\phi})=\sqrt{l^2\dot{\phi}^2e^{-2\phi}-1}-l\dot{\phi}e^{-\phi}\sinh^{-1}\left(i\,l\dot{\phi}e^{-\phi}\right).\label{GdotAdSS}
\ee
Observe that since these surfaces cannot reach the boundary surface $\Sigma$, we cannot perform the UV expansion $l\dot{\phi}e^{-\phi}<<1$ for this action.\\
The equation of motion from \eqref{ALSL}  is now given by
\be
-\frac{le^{-2\phi}(\ddot{\phi}+(d-1)\dot{\phi}^2)-\frac{d}{l}}{\sqrt{l^2\dot{\phi}^2e^{-2\phi}-1}}=\frac{d}{d-1}T,
\ee
which is again the trace of the Neumann boundary condition $K|_Q=\frac{d}{d-1}T$ requiring spacelike slices $Q$ to be CMC. For the range of the tension parameter $-\infty<T<0$ we can solve it for $\phi(w)$ that gives the metric on the spacelike $Q$
\be
ds^2=\frac{l^2}{\left(-\sqrt{1+\frac{l^2T^2}{(d-1)^2}}w+\epsilon\right)^2}\left(dw^2+dx^2_i\right).\label{SolutionAdSS}
\ee
These in turn are the $d$-dimensional hyperbolic slices of the Lorentzian $AdS_{d+1}$ \eqref{AdSdLP1} and their constant negative Ricci curvature is given by
\be
R^{(d)}=-\frac{d(d-1)}{l^2}\left(1+\frac{l^2T^2}{(d-1)^2}\right)=2\Lambda^{(d+1)}_{AdS}\left(1+\frac{l^2T^2}{(d-1)^2}\right),
\ee
which can be computed explicitly or using \eqref{HCL}.\\
Integrating \eqref{wpAdSLSLQ}, we can find the embedding function expressed in original coordinates 
\be
z=f(t)=\sqrt{1+\frac{l^2T^2}{(d-1)^2}}\frac{d-1}{lT}t+\epsilon.\label{AdSLSL}
\ee
These are half-planes that interpolate between the null sheet in the limit of $T\to-\infty$ and the $t=0$ slice for $T=0$, shown by green lines on Fig. \ref{SolLorAdS}.

Let us now discuss the holographic path-integral complexity \eqref{cltlo} for spacelike $Q$. Firstly, the action without tension term in this case is given in terms of the negative tension $T$ as
\be
I_G=\frac{(d-1)l^{d-1}}{\kappa^2}\frac{V_x L_t}{\epsilon^{d}}+\frac{V_x}{\kappa^2\epsilon^{d-1}}\frac{T}{(d-1)\sqrt{1+\frac{l^2T^2}{(d-1)^2}}}.
\ee
Secondly, we add the modified Hayward term expressed by the rapidity $\tilde{\eta}_0$ defined in terms of normal vectors $n_\Sigma$ to timelike $\Sigma$ and $n_Q$ to spacelike $Q$ (see Fig. \ref{SolLorAdS}) as
\be
I'_H=\frac{1}{\kappa^2}\int_\gamma \sqrt{\gamma}\,\tilde{\eta}_0,\qquad n_\Sigma \cdot n_Q=-\sinh\ti{\eta}_0.\label{MHTAdSS}
\ee
For the solution \eqref{AdSLSL} this modified Hayward term for spacelike $Q$ becomes
\be
I'_H=\frac{V_xl^{d-1}}{\kappa^2 \epsilon^{d-1}}\ti{\eta}_0\qquad \sinh\ti{\eta}_0=\frac{l|T|}{d-1}. \label{etaSLADS}
\ee
Let us point that for the spacelike $Q$ this modified Hayward term was chosen slightly different than for timelike $Q$.\footnote{Perhaps it would then be more appropriate to denote it as $\tilde{I}'_H$ but we avoid that for the sake of simplicity of our formulas and hope that it will be clear from the context which term we use.}. Indeed the on-shell value of the boundary term in \eqref{IGSLAdS} is given by
\be
-\frac{V_x}{\kappa^2}\left[e^{(d-1)\phi}\sinh^{-1}\left(i\,l\dot{\phi}e^{-\phi}\right)\right]^0_{-\infty}=-\frac{V_xl^{d-1}}{\kappa^2\epsilon^{d-1}}\left(\sinh^{-1}\left(\frac{l|T|}{d-1}\right)+\frac{i\pi}{2}\right). \label{boundaryadstim}
\ee
Therefore, if we insisted on canceling it, we would rather have to add the modified Hayward term \eqref{MHTAdSS} with rapidity $\eta_0$ related to $\tilde{\eta}_0$ by $\eta_0=\ti{\eta}_0+\frac{\pi}{2}i$. From the perspective of our holographic complexity this would lead to complex on-shell values of $\mathcal{C}^{(l)}_T$ with imaginary part independent on the tension $T$. This complexity would clearly give the same solution for optimal $T$ but we would need to take the absolute value for the final answer to be real. Instead, since we do not consider the UV limit for spacelike $Q$ where the above cancelation was a consistency check, for spacelike slices we propose to evaluate real-valued holographic path-integral complexity in terms of rapidity $\tilde{\eta}_0$. We will study this proposal below as well as in the de Sitter example and leave its test to more complicated time-dependent geometries for future works. 

Finally, the relative holographic path-integral complexity between timelike $\Sigma$ and spacelike $Q$ for the solution \eqref{AdSLSL} is given by
\be
\mathcal{C}^{(l)}_T=-(I_G+I'_H)=-\frac{(d-1)l^{d-1}}{\kappa^2}\frac{V_x L_t}{\epsilon^{d}}-\frac{V_xl^{d-1}}{\kappa^2 \epsilon^{d-1}}\left(\ti{\eta}_0-\tanh\ti{\eta}_0\right).  \label{spadscoj}
\ee
This action is minimized for $\ti{\eta}_0\to\infty$ or $T\to-\infty$, where the surface $Q$ becomes light-like so the  optimal slice coincides with the time-like case (\ref{timeAdScompx}). Notice also that the absolute phase of the Hartle-Hawking wave function is maximized in this limit.

Let us also point the value of $\mathcal{C}^{(l)}_T$ for $\tilde{\eta}_0=0$ or $T=0$ that is given by
\be
\mathcal{C}^{(l)}_0=-\frac{(d-1)l^{d-1}}{\kappa^2}\frac{V_x L_t}{\epsilon^{d}}.\label{CT0SAdS}
\ee
This result is different that Euclidean counterpart \eqref{euccompvacc}. Even though naively one could expect that both  relative complexities are concerned with slice $Q$ with the same induced metric, \eqref{CT0SAdS} hints on a very different nature of our Euclidean and Lorentzian complexities. We leave better understanding of this nature in the language of Euclidean and Lorentzian continuous tensor networks as an interesting future problem.

\subsection{Semi-classical wave-functions in $dS_{d+1}$}
In this section we will evaluate semi-classical Hartle-Hawking wave functions \eqref{HHwvL} for 
$d+1$ dimensional de Sitter spacetimes, using the gravity action
\be
I_G+I_T=\frac{1}{2\kappa^2}\int_{M} \sqrt{-g}\left(R-2\Lambda\right)+\frac{1}{\kappa^2}\int_{Q} \sqrt{|h|}(K-T)+\frac{1}{\kappa^2}\int_{\Sigma} \sqrt{|h|}K.\label{ActLdS}
\ee
This time, we consider a classical solution given by the de Sitter spacetime in the Poincare coordinates
\be
ds^2=l^2\frac{-dt^2+dx^2+dy^2_i}{t^2},\label{dSPoincareM}
\ee
where we again keep the radius $l$ of $dS_{d+1}$ explicit.  This metric solves vacuum Einstein's equations with
\be
R=\frac{d(d+1)}{l^2},\qquad \Lambda=\Lambda^{(d+1)}_{dS}=\frac{d(d-1)}{2l^2}.
\ee 
Generalization of our construction to de Sitter spacetimes will involve region $M$ between the spacelike boundary cut-off surface $\Sigma$ given by $t=-\epsilon$, that we assume is again associated with the dual theory on the fixed flat metric that specifies the state dual to \eqref{dSPoincareM},\footnote{This seems like a natural assumption but it is possible that holography for dS works in a completely different manner as the AdS counterpart.} and a surface $Q$ defined by
 \be
 t=h(x)<0.
 \ee
The Hwartle-Hawking wavefunction will again implicitly depend on the fixed flat metric on $\Sigma$ and will be an explicit functional of $h(x)$ (or $\phi$) with respect to which we will extremize it.\\ 
 Our choice (convention) is such that $M$ will be in the negative $t$ and negative $x$ region of the geometry (see Fig. \ref{SolLordS}).\\ 
The trace of extrinsic curvature of $\Sigma$ will always be given by $K_\Sigma=-d/l$. For surfaces $Q$ we again assume the translational invariance in $y_i$ directions and write their infinite volumes as $V_y$. Similarly to $AdS_{d+1}$, we will now consider the two cases of timelike and spacelike $Q$'s separately.   

\subsubsection{Timelike $Q$}
 For the timelike surface $Q$ we get the induced metric 
 \be
 ds^2=l^2\frac{-(h'^2-1)dx^2+dy^2_i}{h^2}\equiv e^{2\phi}\left(-dw^2+dy^2_i\right),
 \ee
 where we introduced conformal coordinate $w$ with range $-\infty<w<0$. In this example, it is related to $x$ by derivatives
 \be
 w'(x)=\sqrt{h'^2-1}=\frac{1}{\sqrt{\dot{h}^2-1}},\label{wpdSLTLQ}
 \ee
where as in previous examples we denote $h'=\partial_x h(x)$ and $\dot{h}=\partial_w h(w)$.

\begin{figure}[b!]
  \centering
  \includegraphics[width=7cm]{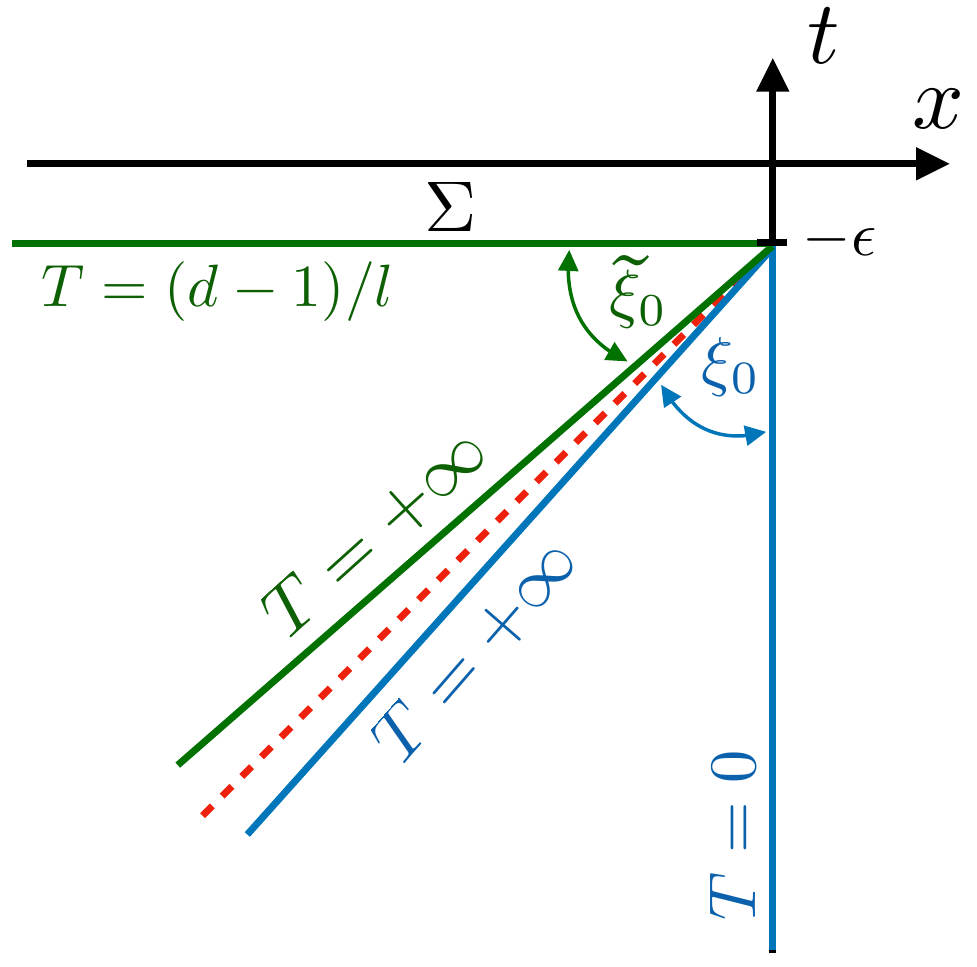}
  \caption{Solutions in Lorentzian $dS_{d+1}$.  Timelike half-planes \eqref{dSLTL}, shown in blue, interpolate between the  the timelike plane $x=0$ at $T=0$ and null plane, dashed red, in the limit of $T\to+\infty$. Their inner angle $\xi_0$ given by \eqref{etaTLdS} is shown in blue. Spacelike half planes \eqref{dSLSL}, shown in green, interpolate between the spacelike boundary $\Sigma$ at $T=(d-1)/l$ and the null sheet for $T\to+\infty$. Their inner angle $\tilde{\xi}_0$, given by \eqref{etaSLdS}, is shown in green.}
\label{SolLordS}
\end{figure}

The relevant trace of extrinsic curvature on $Q$ can then be computed in terms of $h(x)$ or $\phi(w)$ as
 \be
 K|_Q=\frac{h''h+d(h'^2-1)}{l(h'^2-1)^{3/2}}=\frac{le^{-2\phi}(\ddot{\phi}+(d-1)\dot{\phi}^2)-\frac{d}{l}}{\sqrt{l^2e^{-2\phi}\dot{\phi}^2-1}}.
 \ee
Observe that in accordance with \cite{Maldacena:2002vr}, this result can be formally obtained from its $AdS$ counterpart \eqref{KLAdSP} by continuation $l_{dS}=-il_{AdS}$.\\
 With these ingredients, we have the on-shell action in the Lorentzian semi-classical Hartle-Hawking wave function  \eqref{HHwvL} in de Sitter geometry \eqref{dSPoincareM}
 \bea
 I_G+I_T&=&-\frac{(d-1)V_y L_x l^{d-1}}{\kappa^2\epsilon^{d}}+\frac{V_y}{\kappa^2}\int dx\frac{l^d}{h^{d}} w'(x)\left(K|_Q-\frac{1}{lw'(x)}-T\right).
 \eea
 More explicitly, we can write the action in terms of $w$ and $\phi(w)$ as
  \be
 I_G+I_T=-\frac{(d-1)V_y L_x l^{d-1}}{\kappa^2\epsilon^{d}}+\frac{V_y}{\kappa^2}\int dw e^{d\phi} \left[\frac{le^{-2\phi}(\ddot{\phi}+(d-2)\dot{\phi}^2)-\frac{d-1}{l}}{\sqrt{l^2e^{-2\phi}\dot{\phi}^2-1}}-T\right],
  \ee
and comparing to the AdS counterpart, we have a formal relation $l_{dS}=-il_{AdS}$ for the action of $Q$.\\
As for the spacelike $AdS_{d+1}$, we can rewrite it in the first derivative form as follows
\bea
I_G+I_T&=&-\frac{(d-1)V_y L_x l^{d-1}}{\kappa^2\epsilon^{d}}+\frac{(d-1)V_y}{l\kappa^2}\int dwe^{d\phi(w)}G(\dot{\phi})-\frac{V_y}{\kappa^2}\int dwe^{d\phi(w)}T\nn\\
&+&\frac{V_y}{\kappa^2}\left[e^{(d-1)\phi}\sinh^{-1}\left(i\,l\dot{\phi}e^{-\phi}\right)\right]^0_{-\infty},\label{ActiondSSL}
\eea
with the function
\be
G(\dot{\phi})=\sqrt{l^2\dot{\phi}^2e^{-2\phi}-1}-l\dot{\phi}e^{-\phi}\sinh^{-1}\left(i\,l\dot{\phi}e^{-\phi}\right).
\ee
The form of $G(\dot{\phi})$ is analogous to the one for spacelike surfaces in Lorentzian $AdS_{d+1}$ \eqref{GdotAdSS} and the reason that we cannot expand it for $\dot{\phi}e^{-\phi}$ is a manifestation of the fact that timelike slices in Lorentzian de Sitter geometry will not reach the spacelike boundary $\Sigma$ (see Fig. \ref{SolLordS}). 

The maximization of the Hartle-Hawking wave function again leads to the CMC constraint in de Sitter geometry
\be
\frac{le^{-2\phi}(\ddot{\phi}+(d-1)\dot{\phi}^2)-\frac{d}{l}}{\sqrt{l^2e^{-2\phi}\dot{\phi}^2-1}}=\frac{d}{d-1}T,
\ee
and in equivalent to imposing the Neumann boundary condition on $Q$.
   \\

This equation with boundary condition $h(0)=-\epsilon$ can be solved for the range of tension parameter $0<T<+\infty$ by $\phi(w)$ that corresponds to the induced metric on $Q$ 
 \be
 ds^2=\frac{l^2}{\left(\sqrt{1+\frac{l^2T^2}{(d-1)^2}}w-\epsilon\right)^2}\left(-dw^2+dy^2_i\right).\label{SolutiondST}
 \ee
 These timelike geometries are the $d$-dimensional de Sitter slices of the $(d+1)$-dimensional de Sitter geometry \eqref{dSPoincareM}. Their constant positive Ricci scalar curvature is given by
\be
R^{(d)}=\frac{d(d-1)}{l^2}\left(1+\frac{l^2T^2}{(d-1)^2}\right)=\Lambda^{(d+1)}_{dS}\left(1+\frac{l^2T^2}{(d-1)^2}\right),
\ee
which is indeed the result \eqref{HCL} for timelike slices in $dS_{d+1}$ spacetime.\\
Integrating \eqref{wpdSLTLQ}, and finding $w(x)$ allows us to write  the embedding function for this surface given by
 \be
 t=h(x)=\sqrt{1+\frac{l^2T^2}{(d-1)^2}}\frac{d-1}{lT}x-\epsilon.
 \label{dSLTL}
 \ee
 These are the half-planes shown in Fig. \ref{SolLordS} in blue and we can verify that they satisfy the full Neumann boundary condition. Up to this point the results are naturally in parallel with the computation in Lorentzian $AdS_{d+1}$.

Finally, following steps from the previous section, we compute the Lorentzian holographic path-integral complexity in this de Sitter setup. As we discussed, the rules for holography in de Sitter spacetimes may be completely different than those in standard AdS/CFT therefore using the same proposal for the holographic path-integral complexity seems far from obvious. Nevertheless, given the universality of definition \eqref{cltlo}, we assume/conjecture its validity beyond AdS/CFT, straightforwardly evaluate it and examine the consistency of the result.

The first ingredient is the on-shell gravity action computed without the tension term in region $M$ of de Sitter geometry and for solution \eqref{dSLTL} it becomes
\be
I_G=-\frac{(d-1)V_y L_x l^{d-1}}{\kappa^2\epsilon^{d}}+\frac{V_y l^{d-1}}{\kappa^2\epsilon^{d-1}}\frac{lT}{(d-1)\sqrt{1+\frac{l^2T^2}{(d-1)^2}}}.
\ee
Then, naturally adapting the computation for spacelike slices in $AdS_{d+1}$, the modified Hayward term for timelike $Q$ in de Sitter spacetime is given by 
\be
I'_H=-\frac{V_yl^{d-1}}{\epsilon^{d-1}}\xi_0,\qquad \sinh\xi_0=\frac{lT}{d-1},\label{etaTLdS}
\ee
where the angle between two surfaces $\Sigma$ and $Q$ is given by 
$n_\Sigma\cdot n_{Q}=-\sinh\xi_0$.\\
Note that the boundary term in the action \eqref{ActiondSSL} is given on-shell by
\be
\frac{V_y}{\kappa^2}\left[e^{(d-1)\phi}\sinh^{-1}\left(i\,l\dot{\phi}e^{-\phi}\right)\right]^0_{-\infty}=\frac{V_yl^{d-1}}{\kappa^2\epsilon^{d-1}}\left(\sinh^{-1}\left(\frac{lT}{d-1}\right)+\frac{i\pi}{2}\right).  \label{dstimpa}
\ee
and again we chose the angle $\xi_0$ so that we end up with real value of holographic path-integral complexity.\\
Adding the them  up we arrive at the following result in de Sitter spacetime
\be
I_G+I'_H=-\frac{(d-1)V_y L_x l^{d-1}}{\kappa^2\epsilon^{d}}+\frac{V_y l^{d-1}}{\kappa^2\epsilon^{d-1}}
(\tanh\xi_0-\xi_0).  \label{actiondsorp}
\ee
If we continue to employ the definition of gravitational complexity (\ref{cltlo}) in the Lorentzian signature,
we obtain $\mathcal{C}^{(l)}_T=-(I_G+I'_H)$ which is a minus (\ref{actiondsorp}). Interestingly, 
this complexity  $\mathcal{C}^{(l)}_T[dS]$ for the time-like surface in  dS$_{d+1}$ is equal to $-\mathcal{C}^{(l)}_T[AdS]$ of the complexity for the space-like surface in AdS$_{d+1}$ given by (\ref{spadscoj}) if we relate $\xi_0$ to $\ti{\eta}_0$.
This might suggest that the optimization 
in the de Sitter case is opposite to the AdS case i.e. the maximization of the complexity, which leads to
the limit $\xi_0\to \infty$. This is equivalent to our claim, common to the AdS case, that the optimization 
corresponds to the maximization of the absolute value of phase factor of Hartle-Hawing wave function.\\
Similarly to \eqref{CT0SAdS}, the holographic path-integral complexity in $dS_{d+1}$ for timelike $Q$ would give only the divergent contribution as $\xi_0\to 0$ or $T=0$. Interpretation of this result in terms of continuous tensor networks in CFTs dual to de Sitter spacetime is certainly an interesting future problem too. 

\subsubsection{Spacelike $Q$}
The same computation can be repeated for spacelike surface $Q$: $t=h(x)$ on which we get the induced metric
 \be
 ds^2=l^2\frac{(1-h'^2)dx^2+dy^2_i}{h^2}\equiv e^{2\phi}\left(dw^2+dy^2_i\right),
 \ee
 where we introduced the conformal coordinate $w$ with range $-\infty<w<0$ related to $x$ via derivatives
 \be
 w'(x)=\sqrt{1-h'^2}=\frac{1}{\sqrt{1+\dot{h}^2}}.\label{wpdSLSLQ}
 \ee
The trace of the extrinsic curvature for spacelike $Q$ in de Sitter spacetime \eqref{dSPoincareM} is now given in terms of $h(x)$ and $\phi(w)$ as
\be
K|_Q=\frac{-hh''+d(1-h'^2)}{l(1-h'^2)^{3/2}}=\frac{le^{-2\phi}(\ddot{\phi}+(d-1)\dot{\phi}^2)+\frac{d}{l}}{\sqrt{1+l^2e^{-2\phi}\dot{\phi}^2}},
\ee
where as before $h'=\partial_xh(x)$ and $\dot{\phi}=\partial_w\phi(w)$. Again, this result can be formally obtained from its $AdS$ counterpart \eqref{KAdSSLKPQ} by continuation $l_{dS}=-il_{AdS}$ \cite{Maldacena:2002vr}.\\
The analog of the gravity action with tension term that computes the Hartle-Hawking wave function \eqref{HHwvL} between spacelike $\Sigma$ and spacelike $Q$ is then
\be
 I_G+I_T=-\frac{(d-1)V_y L_x l^{d-1}}{\kappa^2\epsilon^{d}}+\frac{V_y}{\kappa^2}\int dw e^{d\phi} \left[\frac{le^{-2\phi}(\ddot{\phi}+(d-2)\dot{\phi}^2)+\frac{d-1}{l}}{\sqrt{1+l^2e^{-2\phi}\dot{\phi}^2}}-T\right].
  \ee
Comparing with spacelike $AdS$ action, we again get the formal continuation $l_{dS}\to-il_{AdS}$ for the action on $Q$.\\
As for timelike $AdS_{d+1}$, integrating by parts, yields the action in the first derivative form
\bea
I_G+I_T&=&
-\frac{(d-1)V_y L_x l^{d-1}}{\kappa^2\epsilon^{d}}+\frac{(d-1)V_y}{l\kappa^2}\int dw e^{d\phi}G(\dot{\phi})-\frac{V_y}{\kappa^2}\int dw e^{2\phi}T\nn\\
&+&\frac{V_y}{\kappa^2}\left[e^{(d-1)\phi(w)}\sinh^{-1}\left(l\dot{\phi}e^{-\phi}\right)\right]^0_{-\infty},\label{ActiondSSL2}
\eea
where the monotonic function $G(\dot{\phi})$ is now
\be
G(\dot{\phi})=\sqrt{1+l^2e^{-2\phi}\dot{\phi}^2}-l\dot{\phi}e^{-\phi}\sinh^{-1}\left(l\dot{\phi}e^{-\phi}\right).
\ee
This time, since spacelike surfaces $Q$ are again connected to the spacelike boundary $\Sigma$, we can consider the analog of the ``UV limit" and expand $G(\dot{\phi})$ for small $\dot{\phi}$. The combination that appears in the action for $Q$ becomes in this limit
\be
e^{d\phi}G(\dot{\phi})\simeq \frac{1}{2}l^2e^{(d-2)\phi}\left[-\dot{\phi}^2+\frac{2}{l^2}\
e^{2\phi}+O(\dot{\phi}^4)\right].
\ee

Interestingly, in two dimensions ($d=2$), this action corresponds to the following form of the timelike Liouville theory
action \cite{TL1,TL2,TL3} on an Euclidean flat space
\ba
S_{TL}=\frac{c}{24\pi}\int d^2x\left[-(\de_\tau\phi)^2-(\de_x\phi)^2+\mu e^{2\phi}\right]. \label{ILactTL}
\ea
This observation offers a hint on how a holographic dual of de Sitter space looks like. The dual CFT  seems to lead to the above timelike Liouville theory rather than the conventional Liouville theory.\footnote{If we add the tension term,
then the UV limit of the total action $I_G+I_T$ gives the cosmological constant $\mu=2-\frac{2T}{d-1}$, where we set $l=1$.  For this range of the tension we have $\mu<0$ and this is the reason why the dS$_{d}$ solutions for the surface $Q$ are obtained.} We will comment more on this action in the next section.

The equation from the maximization of $I_G+I_T$ is again given by the CMC condition
\be
\frac{le^{-2\phi}(\ddot{\phi}+(d-1)\dot{\phi}^2)+\frac{d}{l}}{\sqrt{1+l^2e^{-2\phi}\dot{\phi}^2}}=\frac{d}{d-1}T.
\ee
We can solve it with the boundary condition $h(0)=-\epsilon$ for positive $(d-1)/l<T<+\infty$ and we find $\phi(w)$ as well as the metric on spacelike $Q$ given by
 \be
 ds^2=\frac{l^2}{\left(\sqrt{\frac{T^2l^2}{(d-1)^2}-1}\,w-\epsilon\right)^2}\left(dw^2+dy^2_i\right).  \label{hypplo}
 \ee
These are the $d$-dimensional hyperbolic slices of the $(d+1)$-dimensional de Sitter space \eqref{dSPoincareM}.
Their constant negative curvature is computed as 
\be
R^{(d)}=-\frac{d(d-1)}{l^2}\left(\frac{l^2T^2}{(d-1)^2}-1\right)=-\Lambda^{(d+1)}_{dS}\left(\frac{l^2T^2}{(d-1)^2}-1\right),
\ee
and agrees with \eqref{HCL} for $d$-dimensional spacelke slices of the $dS_{d+1}$ spacetime.
 
Integrating $w'(x)$ in \eqref{wpdSLSLQ} allows us to rewrite the embedding function $h(x)$ for spacelike $Q$ as
 \be
 t=h(x)=\sqrt{\frac{l^2T^2}{(d-1)^2}-1}\frac{d-1}{lT}x-\epsilon,
 \label{dSLSL}
 \ee
 that represents half-planes, shown in Fig. \ref{SolLordS} in green, that interpolate between the null-sheet and the spacelike boundary $\Sigma$. Again both $h(x)$ and \eqref{hypplo} solve the full Neumann boundary conditions on the spacelike $Q$.
 
 Finally, we finish by computing the Lorentzian path-integral complexity in this example. First, the on-shell gravity action without the tension term evaluated on the solution \eqref{dSLSL} is given by
\be
I_G=-\frac{(d-1)V_y L_x l^{d-1}}{\kappa^2\epsilon^{d}}+\frac{V_y l^{d-1}}{\kappa^2\epsilon^{d-1}}\frac{lT}{(d-1)\sqrt{\frac{l^2T^2}{(d-1)^2}-1}}.
\ee
By analogy with the timelike $Q$ in $AdS_{d+1}$, we can now evaluate the modified Hayward term  expressed in terms of rapidity $\tilde{\xi}_0$ 
\be
I'_H=-\frac{V_yl^{d-1}}{\kappa^2\epsilon^{d-1}}\ti{\xi}_0,\qquad \sinh\ti{\xi}_0=\sqrt{\frac{l^2T^2}{(d-1)^2}-1},\label{etaSLdS}
\ee
where the $\tilde{\xi}_0$ is defined in terms of unit normal vectors $n_\Sigma$ and $n_Q$ to spacelike surfaces $\Sigma$ and $Q$ as $n_\Sigma\cdot n_{Q}=-\cosh\ti{\xi}_0$.\\
Observe that this modified Hayward term plays the same role as its counterpart for timelike $Q$ in Lorentzian $AdS_{d+1}$ and precisely cancels the on-shell boundary term in \eqref{ActiondSSL2} that reads
\be
\frac{V_y}{\kappa^2}\left[e^{(d-1)\phi(w)}\sinh^{-1}\left(l\dot{\phi}e^{-\phi}\right)\right]^0_{-\infty}=\frac{V_yl^{d-1}}{\kappa^2\epsilon^{d-1}}\sinh^{-1}\left(\sqrt{\frac{l^2T^2}{(d-1)^2}-1}\right).
\ee

Adding the two contributions we arrive a the following result 
\be
I_G+I'_H=-\frac{(d-1)V_y L_x l^{d-1}}{\kappa^2\epsilon^{d}}+\frac{V_y l^{d-1}}{\kappa^2\epsilon^{d-1}}
(\coth\ti{\xi}_0-\ti{\xi}_0).  \label{actiondsorpp}
\ee
As in the time-like surface case,  if we employ the definition of gravitational complexity 
(\ref{cltlo}) in the Lorentzian signature,
we obtain $\mathcal{C}^{(l)}_T=-(I_G+I'_H)$ which is minus of (\ref{actiondsorpp}). Interestingly, 
this complexity  $\mathcal{C}^{(l)}_T[dS]$ for the space-like surface $Q$ in  dS$_{d+1}$ is equal to $-\mathcal{C}^{(l)}_T[AdS]$ of the complexity for the time-like surface in AdS$_{d+1}$ given by (\ref{timeAdScompx})
if we relate 
$\ti{\xi}_0$ to $\eta_0$.
This again may imply that the optimization in the de Sitter case is opposite to the AdS case i.e. the maximization of the complexity, which leads to the limit $\ti{\xi}_0\to \infty$. In this limit, the surface $Q$ becomes light-like. This is equivalent to the claim common to the AdS case that the optimization corresponds to the maximization of the absolute value of phase factor of Hartle-Hawing wave function.

As before, the Hayward term \eqref{etaSLdS} was chosen such that the holographic path-integral complexity vanishes in the limit of $Q\to\Sigma$ corresponding to $\tilde{\xi}_0\to 0$. In this example, this limit should be taken carefully as
\be
\tilde{\xi}_0\to  \coth^{-1}\left(\frac{(d-1)L_x}{\epsilon}\right).
\ee
This concludes the Lorentzian examples and in the next section we will employ these results in order to draw new lessons for path-integral optimization in Lorentzian CFTs.
\subsection{Towards Lorentzian Path-Integral Optimization}
After getting some intuition and motivation from the gravity computations in the two sections above, we are now ready revisit the idea of extracting geometry from Lorentzian path-integrals in CFT.  In this section we will discuss one possible approach to Lorentzian path-integrals and their optimization that naturally leads to Lorentzian AdS geometries. Furthermore, we will speculate on the timelike Liouville action derived above that gives rise to de Sitter metrics. For simplicity we focus on two spacetime dimensions.

While Euclidean path-integrals define states in quantum field theories (QFT), their Lorentzian conterparts serve as natural description of transition amplitudes. Indeed given a QFT Hamiltonian
\be
H=\int dx \,\mathcal{H}(\varphi(x),\pi(x)), 
\ee
and a quantum field $\varphi(t,x)$ which has eigenstates
\be
\hat{\varphi}(t,x)\ket{\varphi(x),t}=\varphi(t,x)\ket{\varphi(x),t},
\ee
we define a transition amplitude between two field values $\varphi_i$, $i=1,2$ at two given times $t_i$, using Lorentzian path-integrals
\bea
\langle \varphi_f(x),t_f| \varphi_i(x),t_i\rangle&=&\langle \varphi_f(x)|e^{i\hat{H}(t_f-t_i)} |\varphi_i(x)\rangle\nn\\
&=&\int^{\varphi(t_f,x)=\varphi_f(x)}_{\varphi(t_i,x)=\varphi_i(x)}[D\varphi][D\pi]e^{i\int^{t_f}_{t_i}dt dx\left[\pi \dot{\varphi}-\mathcal{H}(\varphi,\pi)\right]}\nn\\
&=&\int^{\varphi(t_f,x)=\varphi_f(x)}_{\varphi(t_i,x)=\varphi_i(x)}[D\varphi]e^{i\int^{t_f}_{t_i}dtdx\mathcal{L}},
\eea
where $\mathcal{L}$ is the Lorentzian Lagrangian of the theory.

Now, we can naturally generalize the optimization procedure for the transition amplitudes by performing the above path-integral of a unitary CFT on a general two dimensional Lorentzian background
\be
ds^2=e^{2\phi(x,t)}\eta_{\mu\nu}dx^\mu dx^\nu=e^{2\phi(x,t)}\left(-dt^2+dx^2\right),
\ee
and, by analogy to the ratio of the the wave functions \eqref{pathinf}, define the ratio of transition amplitudes by
\be
\frac{\langle \varphi_f(x),t_f| \varphi_i(x),t_i\rangle_{e^{2\phi}\eta}}{\langle \varphi_f(x),t_f| \varphi_i(x),t_i\rangle_\eta}\equiv e^{-i S^{(l)}_L[\phi,\eta]},  \label{ratiopi}
\ee
where $S^{(l)}_L$ is the Liouville action on a Lorentzian space, explicitly given by  (\ref{ILact}) 
as explained in App. \ref{LorentzianDet}.

We could then regard $-S^{(l)}_L$ with $\mu=1$ defined this way as Lorentzian path-integral complexity that, when optimized, determines a continuous tensor network built from generally both spacelike and timelike tensors. The reason why we put a minus sign is because when we consider the flat metric $e^\phi=1/\ep$, which corresponds to the original path-integral before the optimization, $S^{(l)}_L$ gets negatively divergent.

The next step, i.e. optimization of this action, boils down to solving the Liouville equation of motion 
\be
4\partial_+\partial_-\phi=\mu e^{2\phi},  \label{liveqp}
\ee
where $x_\pm=x\pm t$, which is equivalent to the condition for constant curvature of the optimal surface $R=-2\mu$.  The most general solution to (\ref{liveqp}) is given by
\be
e^{2\phi(x^+,x^-)}=\frac{4A'(x^+)B'(x^-)}{\mu(1-A(x^+)B(x^-))^2},
\ee
and this would be the family of networks achievable with this procedure.

As in the Euclidean AdS case, the Lorentzian Hartle-Hawking result receives finite cut-off corrections. Indeed, the analysis in the previous subsection shows that the optimization in gravity which minimizes  $\mathcal{C}^{(l)}_T$ in Lorentzian AdS is given by the limit $\eta_0\to \infty$, where the surface $Q$ becomes light-like i.e. $e^{2\phi}\to 0$. The light-like surface in the Lorentzian AdS has been argued to describe the MERA network \cite{Milsted:2018yur,Milsted:2018san,Takayanagi:2018pml}. Thus, the analysis of path-integral optimization using the Hartle-Hawking wave function suggests that MERA-type networks may be the most optimized quantum circuits which produce the CFT ground states.

Next we move on to the path-integral optimization in the de Sitter gravity dual. 
In this case we expect that the dual Euclidean CFT 
is no longer unitary as long as we keep the standard notion of 
dS/CFT \cite{Strominger:2001pn,Maldacena:2002vr}, where the dual CFT is expected to live on the space-like future boundary of de Sitter spacetime. Therefore, field theoretic calculations of transition amplitudes like (\ref{ratiopi}) are not available. However, the gravity analysis in de Sitter suggests that the path-integral complexity functional is 
given by a finite cut-off corrected version of the timelike Liouville theory $S_{TL}$ \eqref{ILactTL} on a Euclidean flat space. This was derived from the gravity action for a spacelike surface $Q$ situated closely to the dS boundary $\Sigma$.
The equation of motion of $S_{TL}$ reads
\ba
(\de^2_\tau+\de^2_x)\phi=-\mu e^{2\phi},  \label{hyepwqa}
\ea
and its solutions are surfaces with constant curvature $R=2\mu$. If we believe the dS/CFT, this implies that there exists a class of (perhaps non-unitary) Euclidean CFTs which leads to the timelike Liouville theory when we perform  a path-integral on a curved manifold to calculate a partition function.

In the gravity calculations, solutions for spacelike surfaces $Q$ in de Sitter (for a non-zero tension $T$) are given by hyperbolic spaces as in (\ref{hypplo}). This arises as solutions to (\ref{hyepwqa}) for $\mu<0$. Again, the actual complexity computed from the gravity dual includes the finite cut-off corrections. The gravity calculations in the previous subsection leads to an interesting result that holographic path-integral complexity in dS$_{d+1}$ is given by the minus of the complexity in AdS$_{d+1}$. Following the general optimization idea for Lorentzian CFTs, we 
maximize the absolute value of the phase of Hartle-Hawking wave function.
This leads to the optimized solution  given by the light-like limit $\xi_0\to \infty$ of the surface $Q$. 
In this case, the complexity for dS$_{d+1}$, defined as in the Lorentzian AdS case, is maximized rather than minimized. This might be either because our definition of complexity in the dS case is not appropriate or 
because the Euclidean CFT dual to a de Sitter space is ill-behaved.  
We would like to leave further considerations of the dS/CFT case, which include
tensor network interpretations of this null surface in de Sitter space, as a future problem. 

\section{Tension as an Emergent Time}\label{Sec:Tension}
In this last section we would like to present an interesting observation about slices $Q$ that optimize Hartle-Hawking wave functions for a fixed value of the tension parameter $T$. Namely, in all the examples studied above, surfaces $Q$ provide natural foliations of the $d+1$ dimensional $AdS$ and $dS$ spacetimes, once we regard their induced metric as a function of $T$. Recall that we originally introduced the tension parameter to construct intermediate configurations of the surface $Q$, between the (most unoptimal) boundary $\Sigma$ and the fully optimized solution reached in the $T\to 0$ limit (e.g. in the Euclidean cases). As we will see below, this suggests that the tension parameter, which quantifies how much we optimize the path-integral, plays a role of an emergent time. We believe that this interesting interpretation may shed an important new light on the emergence of time from tensor networks. Below we summarize few mathematical structures that are revealed in our analysis. For simplicity, we list examples from Poincare coordinates and set $l=1$.

\subsection{Euclidean $AdS$ }
The Euclidean Poincare AdS$_{d+1}$ $(=H_{d+1})$ metric can be written in a form 
\ba
 ds^2=d\rho_{H}^2+e^{2\phi(\rho_H,w)}(dw^2+dx_i^2),
\ea
with the Weyl factor
\be
e^{2\phi(\rho_H,w)}=\frac{\cosh^2\rho_{H}}{w^2},
\ee
where $\phi$ is the solution of the path-integral optimization on a Euclidean flat space with the tension $T$ \eqref{solta}.
Then the radial coordinate can be identified with the tension parameter ($-(d-1)<T\leq 0$)
\be
\cosh\rho_{H}=\frac{1}{\s{1-\frac{T^2}{(d-1)^2}}},
\ee
or equivalently
\be
|T|=(d-1)\tanh\rho_H.
\ee
This is in fact a standard relation from AdS/BCFT \cite{Ta}.
\subsection{Lorentzian $AdS$ }
Analogously, the Lorentzian Poincare AdS$_{d+1}$ can be written as a union of the following two spacetimes
\ba
 ds^2_{(1)}=d\rho_{A1}^2+e^{2\phi(\rho_{A1},w)}\left(-dw^2+dx_i^2\right), \label{laco}
\ea
with the Weyl factor
\be
 e^{2\phi(\rho_{A1},w)}=\frac{\sinh^2\rho_{A1}}{w^2},
\ee
as well as 
\ba
 ds^2_{(2)}=-d\rho_{A2}^2+e^{2\phi(\rho_{A2},w)}\left(dw^2+dx_i^2\right), \label{lact}
\ea
with the Weyl factor
\be
e^{2\phi(\rho_{A2},w)}=\frac{\sin^2\rho_{A2}}{w^2}.
\ee
Again, we can interpret the above $\phi$ as solutions of the path-integral optimization on a Lorentzian 
flat space with non-trivial tension parameters \eqref{SolutionAdST} and \eqref{SolutionAdSS} respectively (see Fig.\ref{SolLorAdS} for these slices).

Indeed, the radial coordinate in (\ref{laco})  can be identified with the tension parameter as
\be
\coth\rho_{A1}=\f{|T|}{d-1},
\ee
for $-\infty<T<-(d-1)$. Similarly, the radial coordinate in (\ref{lact})  can be identified as
\be
\cot\rho_{A2}=\f{|T|}{d-1},
\ee
for $-\infty<T\leq 0$.

It is worth stressing that the coordinate $\rho_{A2}$ is time like. This implies a real time can emerge from
the Euclidean CFT path-integral on a hyperbolic plane, where the time estimates the amount of cut off scale of the Euclidean CFT.

\subsection{Lorentzian $dS$ }
Finally, the Poincare $dS_{d+1}$ can be written equivalently as a union of the following two spacetimes
\ba
 ds^2_{(1)}=d\rho_{D1}^2+e^{2\phi(\rho_{D1},w)}\left(-dw^2+dx_i^2\right), \label{ldco}
\ea
with the Weyl factor
\be
e^{2\phi(\rho_{D1},w)}=\frac{\sin^2\rho_{D1}}{w^2},
\ee
as well as 
\ba
 ds^2_{(2)}=-d\rho_{D2}^2+e^{2\phi(\rho_{D2},w)}\left(dw^2+dx_i^2\right),\label{ldct}
\ea
with the Weyl factor
\be
e^{2\phi(\rho_{D2},w)}=\frac{\sinh^2\rho_{D2}}{w^2}.
\ee
Notice that the above  $\phi$ is identical to the solution of the path-integral optimization on a Lorentzian 
flat space with the tension parameter, related to the time-like Liouville theory.\\
The two solutions above correspond to \eqref{SolutiondST} and \eqref{hypplo} respectively (see Fig.\ref{SolLordS} for these slices).. This way, the radial coordinate in (\ref{ldco})  can be identified with the tension parameter as
\be
\cot\rho_{D1}=\frac{T}{d-1},
\ee
for $0\leq T<\infty$. On the other hand, the radial coordinate in (\ref{ldct})  can be identified as
\be
\coth\rho_{D2}=\frac{T}{d-1},
\ee
for $(d-1)<T<\infty$. Notice that $\rho_{D2}$ provides a time-like coordinate and this may provide a mechanism of emergence of real time from an Euclidean CFT.

All these suggestive ways of re-writing of $AdS_{d+1}$ as well as $dS_{d+1}$ appear to be hinting on the physical interpretation of $T$ as time from tensor networks. If this intuition is correct, it would not only further support status of path-integral optimization as a tool to extract holographic geometry from CFT states but also help us to understand more general slices of fully dynamical holographic spacetimes \cite{Takayanagi:2018pml}. Unfortunately, at present, we do not have a precise understanding of the microscopic mechanism behind this emergence of time. For example, we still do not know how the metric component in the tension direction (i.e. $g_{TT}$) would emerge from CFTs and hope return to this important problem in the future works.

\section{Conclusions }\label{Sec:7 Conclusions}

In this work, we further developed the recent proposal for interpreting the path-integral optimization \cite{Boruch:2020wax} from the perspective of the AdS/CFT correspondence. The key role in our construction is played by the semi-classical Hartle-Hawking wave functions, computed by gravity action in a wedge region from the asymptotic boundary $\Sigma$ up to a surface $Q$ on which we included an extra tension term. For Euclidean AdS geometries, we maximized  Hartle-Hawking wave-functions that lead to the condition which is equivalent to Neumann boundary condition imposed on $Q$. By solving this constraint we fixed the shape of $Q$, i.e. induced metric on $Q$. 

We analyzed various important examples in $AdS_3$ as well as higher-dimensional holographic geometries and found that these ``maximal" metrics on $Q$ coincide with those previously found by the path-integral optimization procedure in CFTs \cite{Caputa:2017urj}. Moreover, we found that the gravity action provides a complete form of the 
path-integral complexity functional which includes finite cut-off corrections. In particular, in three dimensions, 
the gravity action yields a finite cut-off corrected version of the Liouville action on a Euclidean space. On the other hand, in higher dimensions, we managed to reproduce the conjectured path-integral complexity functional after taking the UV limit (removing finite cut-off corrections). This way, along the lines of \cite{Takayanagi:2018pml}, we see that the minimization of the full path-integral complexity, equivalent to the maximization of Hartle-Hawking wave function, leads to the optimal non-unitary quantum circuit which prepares CFT states. 

In the latter half of this paper, we generalized the analysis of Hartle-Hawking wave function to Lorentzian AdS and dS spacetimes. The guiding principle was that the optimization is realized by maximizing the absolute value of the phase of the Hartle-Hawking wave function, which also computes relative holographic path-integral complexity in the Lorentzian construction. For the Lorentzian AdS$_{3}$, we found that the complexity functional obtained from the gravity calculation can be regarded as a Liouville action on a Lorentzian space plus finite cut-off corrections. Notably, we found that the optimization of the gravitational complexity functional leads to a light-like metric on the surface  $Q$. This implies that the most optimal unitary quantum circuit which realizes the CFT vacuum is the MERA-like tensor network (as also noticed in \cite{Milsted:2018san,Takayanagi:2018pml}). 

It is important to stress that we find two definitions of complexity: the complexity based on the Euclidean path-integral
as well as complexity based on the Lorentzian path-integral, which are denoted by  $\mathcal{C}^{(e)}_T$ and  $\mathcal{C}^{(l)}_T$, respectively.
They are different in that when we consider e.g. a CFT vacuum, we can easily see  $\mathcal{C}^{(e)}_T\neq \mathcal{C}^{(l)}_T$, by comparing the Euclidean result (\ref{euccompvacc}) and the Lorentzian result (\ref{timeAdScompx}) in the $\eta_0\to \infty$ limit. It is also useful to note that the leading divergence $\sim O(\ep^{-d})$ 
of (\ref{euccompvacc}) agrees with that of  (\ref{timeAdScompx}) and that this negative divergence arises because 
we consider a difference of complexity between a given path-integral and the original un-optimized path-integral 
corresponding to uniform $e^\phi=1/\ep$. It is curious to note that these two different definitions of complexity look ``analogous" to what is happening in the holographic counterparts \cite{Susskind}. Indeed, the ``complexity = volume" proposal behaves similarly as the Euclidean path-integral complexity $\mathcal{C}^{(e)}_T$, while the ``complexity = action" proposal resembles the Lorentzian path-integral complexity $\mathcal{C}^{(l)}_T$. It is still though an open problem to make a precise connection (if at all) between this work and the two leading holographic proposals.

Finally we studied Hartle-Hawking wave functions in the Lorentzian de Sitter space. For dS$_3$, the gravitational  complexity functional leads to the action of timelike Liouville theory plus finite cut-off corrections on a Euclidean space. Its optimization again points to a light-like metric on the optimal surface $Q$. Even though our understanding of the dS/CFT correspondence  is much more limited comparing to the AdS/CFT, it is at least clear that one of the most important aspects of dS/CFT is an emergence of time from a Euclidean CFT. We expect that such a CFT may be exotic and non-unitary since gravity duals of standard unitary and well-behaved CFTs should be the Euclidean AdS geometries (or equally hyperbolic spaces). Our result suggests that partition function of such exotic CFT on a curved space is described by a timelike Liouville theory. Moreover, the path-integral optimization using the  Hartle-Hawking wave function at least shows an emergence of a light cone from Euclidean CFTs. 

At the same time, we also noted that the full AdS and dS spacetimes can be regarded as foliations of the surface $Q$ for various values of the tension $T$. In this context, the tension parameter plays the role of an emergent time in gravity duals. We are certain that our analysis provides important hints for understanding the emergent time in dS/CFT and we would like to come back to this deep problem with more understandings in future works.

There are many remaining open problems and interesting directions for further exploration in this program. To list a few, it will be very interesting to better understand and develop the Lorentzian setups that include matter and non-trivial dynamics as well as those that deal with de Sitter geometries. The connection between our setup with CMC slices and holographic tensor networks from $T\bar{T}$-deformations \cite{Caputa:2020fbc} also deserves further study, and may offer a way to definite finite cut-off corrections to the Liouville complexity action in order to match the correct holographic  path-integral complexity. Understanding the role of the tension term (possibly its relation to the York time complexity proposals \cite{York:1972sj,Belin:2018bpg}) and, more generally, the emergence of time from tensor networks remains another interesting open problem. Last but not least, exploring quantum aspects of the Hartle-Hawking wave functions and their connection to tensor networks is one of the most interesting directions that we hope to address in the near future.
 
\section*{ Acknowledgements } 
We are grateful to Shira Chapman, Sumit Das, Diptarka Das, Maciej Kolanowski, Masamichi Miyaji, Pratik Nandy, Dimitris Patramanis, Onkar Parrikar, Yoshiki Sato and Joan Simon for useful discussions and comments on the draft. TT is supported by Grant-in-Aid for JSPS Fellows No.~19F19813.
TT is supported by the Simons Foundation through the ``It from Qubit'' collaboration.  
TT is supported by Inamori Research Institute for Science and 
World Premier International Research Center Initiative (WPI Initiative) 
from the Japan Ministry of Education, Culture, Sports, Science and Technology (MEXT). 
TT is supported by JSPS Grant-in-Aid for Scientific Research (A) No.~16H02182. 
TT is also supported by JSPS Grant-in-Aid for Challenging Research (Exploratory) 18K18766. PC, DG and JB are supported by NAWA “Polish Returns 2019” and NCN Sonata Bis 9 grants.

\begin{appendix}

\section{Lorentzian determinant}\label{LorentzianDet}
In this section, following \cite{Mavromatos:1989nf}  we  directly evaluate the Jacobian of transformation of path-integral measure  after placing it on Lorentzian metric
\be
ds^2=e^{2\phi(x,t)}\eta_{\mu\nu}dx^\mu dx^\nu=e^{2\phi(x,t)}\left(-dt^2+dx^2\right).\label{Mink}
\ee
We will need the curvature of this metric is given by\footnote{More generally, for Lorentzian reference metric $ds^2=e^{2\phi}\hat{g}_{ij}dx^i dx^j$ we have
\be
R=e^{-2\phi}\left(\hat{R}-2\Delta\phi\right).
\ee}
\be
R=-2e^{-2\phi}\left(-\partial^2_t+\partial^2_x\right)\phi=-2e^{-2\phi}\Delta\phi.
\ee
More precisely, we compute the Jacobian of the transformation $g=e^{2\phi}\eta$ 
\be
J=\frac{\delta g(x)}{\delta \eta(x')}=e^{2\phi(x)}\delta^{(2)}(x-x'),
\ee
under which the path-integral measure changes as
\be
[D\varphi]_{e^{2\phi}\eta}=[D\varphi]_{\eta}\det(J)\equiv [D\varphi]_{\eta} e^{S_{eff}[\phi,\eta]}.
\ee
The effective action is then given by
\be
S_{eff}[\phi,\eta]=\ln(\det(J)).
\ee
In order to compute it, we take the variation with respect to $\phi$
\be
\delta S_{eff}=\delta\ln(\det(J))\equiv \delta tr(\ln J)=2\int d^2x\delta^{(2)}(0)\delta\phi(x).
\ee
The delta function $\delta^{(2)}(0)$ is divergent and can be regularized by e.g. the Lorentzian heat-kernel method
\be
\delta^{(2)}(0)=K_L(\epsilon,x,x)\simeq i\sqrt{g}\left(\frac{1}{4\pi\epsilon}+\frac{1}{24\pi}R+ O(\epsilon)\right).
\ee
This way we have
\be
\delta S_{eff}=2i\int d^2x\sqrt{g}\left(\frac{1}{4\pi \epsilon}+\frac{1}{24\pi}e^{-2\phi}\left(\hat{R}-2\Delta\phi\right)\right)\delta\phi.
\ee
For the metric \eqref{Mink}, after integrating over $\phi$, integrating by parts and shifting $\phi$ to set the coefficient of the potential to 1 we arrive at
\be
S_{eff}[\phi,\eta]=\frac{i}{24\pi}\int d^2x\left(-(\partial_t\phi)^2+(\partial_x\phi)^2+e^{2\phi}\right)+O(\epsilon).
\ee
This effective action is related to the Lorentzian Liouville action that we discussed in the main text by
\be
S_{eff}[\phi,\eta]=-iS^{(l)}_L[\phi,\eta],
\ee
where $S^{(l)}_L$ is the standard Liouville action on a Lorentzian flat space (\ref{ILact}).

\section{Inhomogeneous slices in $AdS_3$ }\label{InhomSlices}
In this section we discuss more general, inhomogeneous, surfaces $Q$ and solutions of Neumann boundary condition on slices of $AdS_3$ in Poincare coordinates
\be
ds^2=\frac{d\eta^2+dZd\bar{Z}}{\eta^2},
\ee
where we introduced complex coordinates $Z=\tau+ix$.
\subsection{Action and solutions}
We will now consider a general surface $Q$ defined by $\eta=f(Z,\bar{Z})$ and evaluate the Hartle-Hawking wave function in region $M$ from the boundary $\Sigma$  at $z=\epsilon$ up to the inhomogeneous $Q$ with induced metric $h_{ij}$  given by
\be
ds^2=\frac{1}{f^2}\left[(\partial f)^2dZ^2+(1+2\partial f\bar{\partial}f)dZd\bar{Z}+(\bar{\partial}f)^2d\bar{Z}^2\right].\label{hijIH}
\ee
It will be useful to write its determinant as
\be
\sqrt{h}=\frac{W}{2f^2},\qquad W\equiv\sqrt{1+4\partial f\bar{\partial}f}.
\ee
The unit normal vector to $Q$ and projectors are then given by
\be
n^{\mu}=\frac{z}{W}\{1,-2\bar{\partial}f,-2\partial f\},\qquad e^{\mu}_Z=\{\partial f,1,0\},\qquad e^{\mu}_{\bar{Z}}=\{\bar{\partial}f,0,1\},
\ee
and the trace of extrinsic curvature is 
\be
K=-2\left[f\partial\left(\frac{\bar{\partial}f}{W}\right)+f \bar{\partial}\left(\frac{\partial f}{W}\right)+\frac{1}{W}\right].\label{KCompl}
\ee
With this data, we can write the semi-classical action in $M$ as before
\be
I_G+I_T=-\frac{1}{\kappa^2}\frac{V_xL_\tau}{\epsilon^2}-\frac{1}{\kappa^2}\int \frac{d^2x}{f^2}W\left[K|_Q-T+\frac{1}{W}\right].
\ee
We argued, on general grounds that slices minimizing this action should be solutions of Neumann boundary condition (see App. \ref{INHeom}) as well as slices of constant Ricci scalar. Let us analyze this in more detail.\\
The full extrinsic curvature tensor defined by
\be
K_{ij}=e^\mu_ie^\nu_j\nabla_\mu n_\nu,
\ee
can be expressed as
\be
K_{ij}=A_{ij}+\frac{1}{2}K h_{ij},\label{KandA}
\ee
with $h^{ij}A_{ij}=0$. More explicitly we find
\be
A_{ij}=-\frac{1}{fW}\left(H_{ij}-\frac{1}{2}H h_{ij}\right),
\ee
given in terms of the Hessian matrix of $f$ and its trace
\be
H_{ij}=\partial_i\partial_j f,\qquad H=h^{ij}H_{ij}=W^2\Box f,
\ee
where $\Box$ is the Laplace-Beltrami operator of $h_{ij}$.\\
Now we want to focus on the Neumann boundary condition and its trace in $3d$
\be
K_{ij}-Kh_{ij}=-Th_{ij},\qquad K=2T.
\ee
If we just solve the equation for the trace, then we still have remaining constraints
\be
K_{ij}=(K-T)h_{ij}=Th_{ij}.
\ee
Using \eqref{KandA} this is equivalent to vanishing $A_{ij}=0$\footnote{This is also true in higher dim. is we set $K=\frac{d}{d-1}T$ to the Neumann boundary condition}
that, in the $3d$ case, becomes
\be
H_{ij}=\frac{1}{2}H h_{ij}=-(1+TW)f h_{ij},\label{Aij0}
\ee
where in the second equality we used the relation
\be
H=-f(2+KW),
\ee
that can be checked by explicit computation as well as $K=2T$.\\
From \eqref{Aij0} we can immediately find a class of linear solutions with vanishing Hessian
\be
f(Z,\bar{Z})=a Z+b\bar{Z}+d,
\ee
and fixing the boundary condition yields
\be
d=\epsilon,\qquad a=b=\frac{\sqrt{1-T^2}}{2T},\label{BCIH}
\ee
that satisfies \eqref{Aij0}  $W=-1/T$ (with negative $T$) and corresponds to the solution \eqref{surfacep}. Note that for this solution it is very important that $T\neq 0$. \\
More generally, we can take a surface (see. e.g. \cite{Akal:2020wfl})
\be
(z-\alpha)^2+(Z-a)(\bar{Z}-\bar{a})=\beta^2,
\ee
with arbitrary real (or purely imaginary) coefficient $\beta$ and complex $a=a_1+ia_2$, that corresponds to
\be
z=f(Z,\bar{Z})=\alpha\pm\sqrt{\beta^2-|Z-a|^2}.
\ee
We can verify that $A_{ij}=0$ as well as 
\be
K_{ij}=\frac{1}{2}Kh_{ij},\qquad K=\pm\frac{2\alpha}{\beta}=2T,\qquad R^{(2)}=-2\left(1-\frac{\alpha^2}{\beta^2}\right)=-2\left(1-T^2\right).
\ee
Now, in order to impose our boundary condition on this general solution, we have to first take the formal ``planar" limit of of coefficients $(\alpha,\beta,a,\bar{a})\to\infty$ with their ratios fixed. Indeed this limit corresponds to the surface parametrized by
\be
0=(z-\alpha)^2+(Z-a)(\bar{Z}-\bar{a})-\beta^2\simeq -2\alpha z-a\bar{Z}-\bar{a}Z+\alpha^2-\beta^2+a\bar{a},
\ee
or embedding function
\be
 z=-\frac{a}{2\alpha}\bar{Z}-\frac{\bar{a}}{2\alpha}Z+\frac{\alpha^2+a\bar{a}-\beta^2}{2\alpha}.
\ee
As before \eqref{BCIH}, the boundary condition requires
\be
-\frac{a}{2\alpha}=-\frac{\bar{a}}{2\alpha}=\frac{\sqrt{1-T^2}}{2T},\qquad \frac{\alpha^2+a\bar{a}-\beta^2}{2\alpha}=\epsilon.
\ee
Notice in particular that the boundary condition requires $\alpha^2+a\bar{a}-\beta^2\simeq 2\alpha \epsilon$.
\subsection{Uniformized metric}
In the construction for homogeneous solutions, it was important to choose coordinate $w$ that was uniformizing the metric into the conformally flat one. In two dimensions, we can use this trick to write an arbitrary metric this way. Indeed we can write the induced metric on the general inhomogeneous slice \eqref{hijIH} as
\be
ds^2=e^{2\Phi(\omega,\bar{\omega})}d\omega d\bar{\omega}=e^{2\phi(Z,\bar{Z})}\left(dZ+\mu d\bar{Z}\right)\left(d\bar{Z}+\bar{\mu} dZ\right),\label{CFMet}
\ee
with coordinates $(\omega(Z,\bar{Z}),\bar{\omega}(Z,\bar{Z}))$ specified by
\be
\Phi(\omega,\bar{\omega})=\phi(Z,\bar{Z})-\frac{1}{2}\log(\partial \omega\bar{\partial}\bar{\omega}),\qquad \mu= \frac{\bar{\partial}\omega}{\partial \omega},\qquad \bar{\mu}=\frac{\partial \bar{\omega}}{\bar{\partial}\bar{\omega}},
\ee
where in our example we can take
\be
e^{2\phi(Z,\bar{Z})}=\frac{(1+W)^2}{4f^2},\qquad \mu=\frac{4(\bar{\partial} f)^2}{(1+W)^2},\qquad \bar{\mu}=\frac{4(\partial f)^2}{(1+W)^2}.
\ee
The last two identifications allow us to find coordinates $(\omega,\bar{\omega})$ once we solve the Beltrami equations with $\mu$ and its complex conjugate $\bar{\mu}=\mu^*$
\be
(\bar{\partial}-\mu\partial)\omega=0,\qquad (\partial-\bar{\mu}\bar{\partial})\bar{\omega}=0.
\ee
In general, the Ricci scalar of the conformally flat metric \eqref{CFMet} is given by
\be
R^{(2)}=-8e^{-2\Phi}\partial_\omega\partial_{\bar{\omega}}\Phi,
\ee
where
\be
 \partial_\omega=\frac{\partial-\bar{\mu}\bar{\partial}}{\partial\omega(1-\mu\bar{\mu})}, \qquad \partial_{\bar{\omega}}=\frac{\bar{\partial}-\mu\partial}{\bar{\partial}\bar{\omega}(1-\mu\bar{\mu})},
\ee
and requiring that $R^{(2)}=-2(1-T^2)$ would be solved by (Liouville equation)
\be
e^{2\Phi(\omega,\bar{\omega})}=\frac{4A'(\omega)B'(\bar{\omega})}{(1-T^2)\left(1-A(\omega)B(\bar{\omega})\right)^2}.
\ee
Constraint of constant Ricci scalar is clearly weaker than the full Neumann boundary condition. Indeed all our homogeneous (often referred to as mini-superspace) examples of metrics on $Q$ embedded in $AdS_3$ geometries correspond to $A=\exp(2p \omega)$, $B=\exp(2p \bar{\omega})$ and $\omega=w+ix$ so that
\be
e^{2\phi(w)}=\frac{16p^2e^{4pw}}{(1-T^2)\left(1-e^{4pw}\right)^2}=\frac{4p^2}{(1-T^2)\sinh^2(2pw)},
\ee
and all satisfy
\be
R^{(2)}=K^2-K^{ij}K_{ij}-2,
\ee
with $R^{(2)}=-2(1-T^2)$. However the true minimal solutions of $K=2T$ and full Neumann boundary condition are obtained by setting $p\to 0$ for the vacuum, $p=\alpha/2$ for conical singularities (and $p=1/2$ global $AdS_3$) as well as $p=ir_h/2$ for the BTZ geometry.
\subsection{Inhomogeneous equations of motion}\label{INHeom}
We can consider a general variation of the gravity action with respect to the metric on $Q$
\be
\int_Q d^{2}x \sqrt{h} (K h^{ij}-K^{ij}-T h^{ij}) \delta h_{ij} =0,
\ee
where in the most general case in Poincare $AdS_3$, with surfaces specified by $z=f(\tau,x)$, we have
\be
ds^2=h_{ij}dx^idx^j=\frac{1}{f^2}\left[(1+f_\tau^2)d\tau^2+2f_\tau f_x d\tau dx+(1+f_x^2)dx^2\right].
\ee
We then have
\be
\delta (e^{2\phi} \hat{h}_{ij}) = h_{ij} 2 \delta \phi
+ e^{2\phi} \delta \hat{h}_{ij} ,\qquad e^{2\phi}=\frac{1}{f^2},
\ee
which gives 
\be
\int_Q d^{2}x \sqrt{h} \left[
(K  - 2 T) 2 \delta \phi + 
(K h^{ij}-K^{ij}-T h^{ij})e^{2\phi} \delta \hat{h}_{ij} \right] =0 .
\ee
Then, since
\be 
\delta \hat{h}_{\tau \tau} = 2 f_\tau \delta f_\tau ,
\qquad 
\delta \hat{h}_{x x} = 2 f_x \delta f_x ,
\qquad
\delta \hat{h}_{\tau x} = f_x \delta f_\tau 
+ f_\tau \delta f_x,
\ee
after integrating by parts we can write the terms coming from extrinsic curvature as 
\begin{align}
\mathcal{K} \delta f \equiv &- \frac{2}{f} \sqrt{h} K \delta f - 
\partial_\tau [2 e^{2\phi}\sqrt{h}(K h^{\tau \tau}-K^{\tau \tau}) f_{\tau} ] \delta f \nn\\
&- 
\partial_x [2 e^{2\phi}\sqrt{h}(K h^{xx}-K^{xx}) f_{x} ] \delta f
- 
\partial_\tau [2 e^{2\phi}\sqrt{h}(K h^{\tau x}-K^{\tau x}) f_{x} ] \delta f\nn \\
&-
\partial_x [2 e^{2\phi}\sqrt{h}(K h^{\tau x}-K^{\tau x}) f_{\tau} ] \delta f ,
\end{align}
while the terms proportional to the tension are 
\begin{align}
\mathcal{T} \delta f \equiv &\frac{4}{f} T \sqrt{h} \delta f +
\partial_\tau [2 T \sqrt{h} \hat{h}^{\tau \tau} f_\tau]
\delta f 
+
\partial_x [2 T \sqrt{h} \hat{h}^{xx} f_x]
\delta f  \nn\\
& +
\partial_\tau [2 T \sqrt{h} \hat{h}^{\tau x} f_x]
\delta f +
\partial_x [2 T \sqrt{h} \hat{h}^{\tau x} f_\tau]
\delta f .
\end{align}
Evaluating these terms we can check that
\be 
(\mathcal{K} + \mathcal{T}) \delta f =0 \qquad 
\Leftrightarrow \qquad \textup{EOM} = 0. 
\ee
\section{Properties of the gravitational action with modified Hayward term}\label{GRaction}
We can check the co-cycle properties of the gravity action (without tension term) with the modified Hayward term that depends on the internal angle $\theta^{(i)}_0$ between $\Sigma$ and $Q_i$. The action is given by
\begin{align}
\kappa^2 I [\Sigma,Q_i] =& -\frac{1}{2} \int_{\Sigma} ^{Q_i} d^{d+1} x 
\sqrt{g}
(R-2\Lambda)  - \int_{Q_i} d^{d}x \sqrt{h} K 
 -   \int_{\Sigma} d^{d}x \sqrt{h} K +\int_\gamma\sqrt{\gamma}\theta^{(i)}_0,
\end{align}
and the action between two surfaces $Q_1$ and $Q_2$ (with Neumann boundary condition on both) is defined as
\be
\tilde{I}[Q_1,Q_2] \equiv I[\Sigma,Q_2]-I[\Sigma,Q_1].
\ee
Then we can write the following combinations for three surfaces ordered such that $0<\theta^{(1)}_0<\theta^{(2)}_0<\theta^{(3)}_0<\pi/2$
\be
\kappa^2\tilde{I}[Q_1,Q_2]=-\frac{1}{2} \int_{Q_1} ^{Q_2} 
\sqrt{g}
(R-2\Lambda) - \int_{Q_2}  \sqrt{h}K
+  \int_{Q_1} \sqrt{h}K+\int_{\gamma}\sqrt{\gamma}\left(\theta^{(2)}_0-\theta^{(1)}_0\right),
\ee
\be
\kappa^2\tilde{I}[Q_2,Q_1]=-\frac{1}{2} \int_{Q_2} ^{Q_1}  
\sqrt{g}
(R-2\Lambda) - \int_{Q_1}\sqrt{h}K
+  \int_{Q_2} \sqrt{h}K+\int_{\gamma}\sqrt{\gamma}\left(\theta^{(1)}_0-\theta^{(2)}_0\right),
\ee
\be
\kappa^2\tilde{I}[Q_2,Q_3]=-\frac{1}{2} \int_{Q_2} ^{Q_3}  
\sqrt{g}
(R-2\Lambda) - \int_{Q_3}  \sqrt{h}K
+  \int_{Q_2}  \sqrt{h}K+\int_{\gamma}\sqrt{\gamma}\left(\theta^{(3)}_0-\theta^{(2)}_0\right),
\ee
\be
\kappa^2\tilde{I}[Q_1,Q_3]=-\frac{1}{2} \int_{Q_1} ^{Q_3} 
\sqrt{g}
(R-2\Lambda) - \int_{Q_3}  \sqrt{h}K
+  \int_{Q_1} \sqrt{h}K+\int_{\gamma}\sqrt{\gamma}\left(\theta^{(3)}_0-\theta^{(1)}_0\right).
\ee
From these definitions, changing the range of the bulk integrals (with minus sign) we get
\be
\tilde{I}[Q_1,Q_2]=-\tilde{I}[Q_2,Q_1].
\ee
Moreover, by adding the first and the third expression we can easily show that it is equal to the fourth, i.e.
\be
\tilde{I}[Q_1,Q_2]+\tilde{I}[Q_2,Q_3]=\tilde{I}[Q_1,Q_3].
\ee
These are precisely the co-cycle properties that we saw in the CFT definition of complexity using the improved Liouville action as well as its higher-dimensional generalization.
\section{Maps in $AdS_3$}\label{MapsAdS3}
In 3 dimensions we can use the solution in Poincare coordinates to obtain surface $Q$ in different metrics simply by rewriting the surface equation in new coordinates. The most general asymptotically $AdS$ metrics with flat boundary that we can consider are the Euclidean Banados geometries given by 
\be 
ds^2 = \frac{dz^2}{z^2} + \frac{(dw + z^2 \Bar{T}(\Bar{w}) d \Bar{w} )
(d\Bar{w} + z^2 T(w) dw)}{z^2} ,
\ee
where $w=x+ i t$, $\Bar{w}=x-i t$, and
\be
T(w) =  \frac{3 A''(w)^2 - 2 A'(w) A'''(w)}{4 A'(w)^2}
,\qquad \bar{T} (\Bar{w}) = \frac{3 B''(\Bar{w})^2 - 2 B'(\Bar{w}) B'''(\Bar{w})}{4 B'(\Bar{w})^2} .
\ee
This geometry can be mapped to Poincare coordinates
\be 
ds^2 = \frac{d\eta^2 + dv d\Bar{v}}{\eta^2} ,
\ee
via relations
\begin{align}
\eta &= \frac{4z (A'(w) B'(\Bar{w}))^{3/2}}{4 A'(w) B'(\Bar{w}) + z^2 A''(w) B''(w) } ,\nn
\\
v &= A(w) - \frac{2 z^2 A'(w)^2 B''(\Bar{w})}{4 A'(w) B'(\Bar{w}) + z^2 A''(w) B''(w)},\nn
\\
\bar{v} &= B(\Bar{w}) - \frac{2 z^2 B'(\bar{w})^2 A''(w)}{4 A'(w) B'(\Bar{w}) + z^2 A''(w) B''(w)} .
\end{align}
Applying these maps to the surface \footnote{Note that we absorbed the cutoff into $\tau$-coordinate redefinition.}
\be
\eta = \frac{\sqrt{1-T^2}}{T} \tau ,
\ee
and subsequently solving for $z(t,x)$ the resulting equation
\bea 
\frac{4z (A'(w) B'(\Bar{w}))^{3/2}}{4 A'(w) B'(\Bar{w}) + z^2 A''(w) B''(w) } = 
&-&\frac{ i \sqrt{1-T^2}}{2 T} \left(
A(w) - B(\Bar{w})\right.\nn\\
 &+& \left.\frac{2z^2 (B'(\bar{w})^2 A''(w) - A'(w)^2 B''(\Bar{w}) )}{4 A'(w) B'(\Bar{w}) + z^2 A''(w) B''(w) }
\right),
\label{eq:Banados-Poincare_surfaceEq}
\eea
we can find the profile for $Q$ in these 3d metrics. The solution for $z(t,x)$ in the general Banados geometry is complicated, so in practice it is better to insert particular $A(w),B(\Bar{w})$ before solving the above equation. In the following subsection, we present how this approach allows us to find non-trivial inhomogeneous solution in the spinning BTZ metric.
\subsection{Spinning BTZ}
To apply the above result we first need to rewrite the Euclidean spinning BTZ metric in Banados form. The spinning BTZ metric is given as
\be
ds^2=f(r)^2d\tau^2+f(r)^{-2}dr^2+r^2(d\phi+i\omega(r)d\tau)^2 ,\label{SpinBTZm}
\ee
with
\be
f(r)^2=\frac{(r^2-r^2_+)(r^2-r^2_-)}{r^2},\qquad \omega(r)=\frac{r_+r_-}{r^2}.
\ee
We can rewrite it in Banados coordinates by introducing $w=\varphi + i\tau$, $\Bar{w}=\varphi-i\tau$ , and 
\be 
z^2 = \frac{r^2-L_0-\Bar{L}_0+ \sqrt{(L_0 + \Bar{L}_0 -r^2)^2 - 4 L_0 \Bar{L}_0}}{2 L_0 \Bar{L}_0} ,
\ee
with expectation values of the stress-tensor components
\be 
T(w) = L_0 = \frac{(r_+ + r_-)^2}{4} , \qquad \Bar{T}(\Bar{w}) = \Bar{L}_0 = \frac{(r_+ - r_-)^2}{4} .
\ee
We can check that these stress-energy tensors correspond to maps
\be 
A(w) = e^{(r_+ + r_-) w} , \qquad B(\Bar{w}) = e^{(r_+ - r_-)\Bar{w}} .
\ee
Inserting these maps into equation \eqref{eq:Banados-Poincare_surfaceEq} together with $z$-coordinate reparametrization, we can solve the resulting equation for $r(\tau,\varphi)$:
\be
r^2 = \frac{1}{\tilde{f}(\tau,\varphi)^2} = r_+ ^2 \left(
1 + \frac{ T^2 \left( 1-\frac{r_- ^2}{r_+ ^2} \right)}
{(1- T^2) \sin^2\left[r_+\left(\tau - \frac{ir_-}{r_+} \varphi\right)\right]}
\right) .
\ee
Setting $r_- =0$ we reproduce the previous thermofield double solution. 
Further we can verify that the inhomogeneous function $\tilde{f}(\tau,\phi)$ introduced in this way indeed satisfies appropriate equations of motion for $Q$ surface in spinning BTZ metric.\\
From the boundary perspective, the Euclidean BTZ geometry is described by a CFT on a torus with Euclidean time $\tau$ and angle $\varphi$ with identifications
\be
(\tau,\varphi)\sim(\tau,\varphi+2\pi)\sim(\tau+\beta,\varphi+\theta),
\ee
where $\beta=1/T$ is the inverse temperature and $\theta$ is the twist angle, written as
\be
\beta=\frac{2\pi r_+}{r^2_+-r^2_-},\qquad \Omega_E=-\frac{ir_-}{r_+},\qquad \theta=-\beta\Omega_E.
\ee
It is also useful to define
\be
\beta_\pm=\beta(1\mp i\Omega_E)=\beta\pm i\theta=\frac{2\pi}{r_+\pm r_-}.
\ee
The standard boundary conditions for the charged thermofield double state (with momentum charge) are then imposed at
\be
\tau=-\Omega_E\varphi,\qquad \tau=-\Omega_E\varphi+\frac{\pi}{r_+}=-\Omega_E\varphi+\frac{\beta_+\beta_-}{2\beta}.
\ee
On the other hand, we could start from the vacuum solution of the Liouville equation
\be
ds^2=\frac{4dzd\bar{z}}{\mu(z+\bar{z})^2},
\ee
and map it with
\be
z=A(w)=e^{\beta_+ w},\qquad \bar{z}=B(\bar{w})=e^{\beta_- \bar{w}}
\ee
with $w=x_1+ix_2$, $\bar{w}=x_1-ix_2$. This way the new solution of the Liouville equation becomes
\be
ds^2=\frac{4\pi^2dwd\bar{w}}{\mu\beta_+\beta_-\cos^2\left[r_+\left(x_1-\frac{ir_-}{r_+}x_2\right)\right]},\label{LSBTZ}
\ee
and the boundary conditions are imposed at
\be
x_1=\frac{ir_-}{r_+}x_2\pm\left(\frac{\pi}{2r_+}-\frac{\epsilon}{\sqrt{\mu}}\sqrt{1-\frac{r^2_-}{r^2_+}}\right).
\ee
Note that in these coordinates metric is automatically diagonal. On the other hand, the induced metric in the spinning BTZ geometry \eqref{SpinBTZm} on the CMC surface $r=1/\tilde{f}(\tau,\varphi)$ can be brought to \eqref{LSBTZ} by mapping
\begin{align}
\tan(r_+ \tau - i r_- \theta) =|T|\cot(r_+ x_1 - i r_- x_2),\qquad r_+ \theta + i r_- \tau=r_+ x_2 + i r_- x_1.
\end{align}

\end{appendix}

\end{document}